# IoT Data Processing for Smart City and Semantic Web Applications

Thesis submitted in partial
fulfillment of the requirements for
the degree of

*Master of Science in* ***Electronics and Communication Engineering*** *by Research*

by
Shubham Mante
2019702003
shubham.mante@research.iiit.ac.in

International Institute of Information
Technology Hyderabad - 500 032, INDIA
March 2023

International Institute of Information Technology
Hyderabad, India

## CERTIFICATE

It is certified that the work contained in this thesis, titled **"IoT Data Processing for Smart City and Semantic Web Applications"** by Shubham Mante, has been carried out under my supervision and is not submitted elsewhere for a degree.

| | |
|---|---|
| Date | Adviser: Dr. Aftab M Hussain |

To my family and friends

# Acknowledgments

I would like to express sincere gratitude to my research advisor Dr. Aftab M. Hussain, for his constant guidance throughout my masters. He has always been an ideal teacher, mentor, and thesis supervisor, giving advice and encouragement with the perfect blend of insights and humour. He always guided me to overcome my weaknesses and improve my writing and presentation skills, which helped me during my masters and will also help me in the future. I would also like to thank Dr. Sachin Chaudhari, Dr. Deepak Gangadharan and Mrs. Anuradha Vattem for their guidance and for providing me the opportunity to be a part of the Smart Campus IIIT-H project. I sincerely thank Prof. Nathalie Hernandez and Thierry Monteil [presently affiliated to IRIT, Toulouse, France] for allowing me to work on a collaborative research project remotely during the pandemic. I would like to thank Prof. Nathalie for her support and valuable inputs that helped me gain profound knowledge about the Semantic Web and its applications in the Internet of Things (IoT) field.

A big thanks to Vrushali Arute, Kuntal Chatterjee, Gaurav Zalariya, Ashutosh Pathy, Souradeep Deb, Debtanu Gupta, Vivek Kumar, and Nayan Vats for making my campus life wonderful and memorable. Your support during the campus placements was very helpful in releasing the exams and interview pressure. I like to express my heartfelt thanks to Suhas Vaddhiparthy, Ganesh CGS, Ruthwik Muppala and my lab mates from Smart City Research Center and PATRIoT-FleCS lab for all the discussions and fun that we had while working together on various projects.

A special thanks to Surabhi Gupta for continuously supporting and helping me throughout my masters' journey. Finally, I would like to thank my dearest family members for their encouragement and support during all the ups and downs of my life.

## Abstract

The world has been experiencing rapid urbanization over the last few decades, putting a strain on existing city infrastructure such as waste management, water supply management, public transport and electricity consumption. We are also seeing increasing pollution levels in cities threatening the environment, natural resources and health conditions. However, we must realize that the real growth lies in urbanization as it provides many opportunities to individuals for better employment, healthcare and better education. Cities have now become one of the major contributors to the world's gross domestic product; therefore, urbanization should always be seen as an opportunity. However, it is imperative to limit the ill effects of rapid urbanization through integrated action plans to enable the development of growing cities. This gave rise to the concept of a smart city in which all available information associated with a city will be utilized systematically for better city management.

Needless to say, data associated to the city plays a vital role in understanding the city's current state and quickly addressing the issues discussed earlier. The systematic collection, processing and analysis of data allow for the incorporation of data intelligence and helps inform decisions on issues that matter to the lives of city residents. City management officials have been collecting data for years; however, their department-centric nature has created data silo issues and made it difficult to manage and share, thus hindering the data analysis process. Considering these issues in existing data systems, this work proposes a novel Smart City data management system architecture to systematically collect, store, exchange and analyze data collected from various sensing nodes depicting smart city applications.

The proposed system architecture is divided in subsystems and is discussed in individual chapters. The first chapter introduces and gives overview to the reader of the complete system architecture. The second chapter discusses the data monitoring system (DMS) and data lake system (DLS) based on the oneM2M standards. DMS employs oneM2M as a middleware layer to achieve interoperability, and DLS uses a multi-tenant architecture with multiple logical databases, enabling efficient and reliable data management. The third chapter discusses energy monitoring and electric vehicle charging systems developed to illustrate the applicability of the oneM2M standards. The fourth chapter discusses the Data Exchange System (DES) based on the Indian Urban Data Exchange (IUDX) framework. DES uses IUDX's standard data schema and open APIs to avoid data silos and enable secure data sharing. The fifth chapter discusses the 5D-IoT framework that provides uniform data quality assessment of sensor data with meaningful data descriptions.

# Contents

# List of Figures

# List of Tables

*Chapter 1*

# Introduction

## 1.1 Motivation

A report published in 2018 by the United Nations Department of Economic and Social Affairs [1], states that the urban population is projected to grow from 55% of the world's population in 2018 to 68% of the world's population by 2050, which means almost 2.5 billion more people living in urban areas by 2050. While urbanization of such magnitude vastly improves the Gross Domestic Product (GDP) of any nation and the quality of life of its citizens, it also rapidly increases the consumption of resources such as water, electricity, public transport and healthcare facilities, which puts a strain on the city's infrastructure. Hence there is a need for better utilization of resources with integrated action plans for urban development with technological enhancements. The concept of a Smart City conforms to this need, and the Internet of Things (IoT) plays a vital role in various technological fields [2–15] which includes the Smart City development [16]. Accordingly, metropolitan cities around the world are transforming into smart cities. Parameters such as traffic, weather, and pollution are being monitored in Aarhus (Denmark) [17]. The trajectories of taxis [18] and air quality [19] are being monitored in Beijing (China). Bike sharing data is being monitored in New York and Chicago (USA) [20] while water table level, air quality, etc., are being monitored in Barcelona (Spain) [21]. The monitored data can be used to detect and reduce air pollution, undertake measures to conserve groundwater and other natural resources and reduce traffic congestion. However, data monitoring systems deployed at the smart city scale generate huge datasets, complicating data aggregation, storage, secure sharing, and analytics. Therefore, well-defined standardized system architecture to perform these tasks is essential in any smart city. Many architectures targeting smart city applications have been proposed in recent years, but most of them only discuss similar functionality and architecture designs at an abstract level. So there is still a gap between what a smart city system looks like and how it should be implemented properly.

## 1.2 Related Works

Gomes et al. [22] proposed a five-layered architecture that utilizes the Pentaho platform for data inte- gration and analytics. The architecture contains a Kettle tool that provides the Extraction, Transformation, and Loading (ETL) engine to cleanse the captured data and store it in a uniform format. The subsequent layers include an Apache Cassandra database and an R environment for statistical computing. However, the proposed architecture did not present a systematic way to share the data with external users and preserve the data privacy, thus, hindering privacy-enabled data sharing. Liu et al. [23] developed an ETL tool as part of a data management framework that additionally categorizes data based on sensitivity and adopts different processing (ex: anonymization for private data) and publishing strategies for each level. The ETL tool provides additional features such as data quality and privacy, compared to the architecture in [22]; however, the authors did not discuss maintaining semantic consistency for heterogeneous data sources during the data sharing. Thus, making it difficult for data users to understand the data, which may delay the data usage process. Cheng et al. [24] proposed a live City Data and Analytics Platform (CiDAP). The platform employs Aeron, an open source IoT broker, to collect data from multiple sources, a NoSQL database for storage, and a REpresentational State Transfer (REST) based Application Programming Interface (API) to grant access to external applications. Though the CiDAP platform provides REST API-based access feature in addition to features proposed in [22] and [23], it provides limited security and also lacks the provision of support for anomaly detection [25]. Moreover, the architectures were commonly lacking in the inclusion of a global standard and thus failed to propose an interoperable system architecture at a smart city level.

To resolve this problem, Datta et al. [26] proposed an architecture based on sensor markup language (SenML) to standardize sensor metadata in an oneM2M-based architecture. oneM2M is the first global architecture for M2M and IoT [27]. SenML provides a structured way to encode sensor measurement and additional attributes of a typical IoT device, whereas oneM2M acts as a horizontal service layer that provides common service functions (CSFs) such as data and device management, discovery, security, and more. Additionally, serialized metadata representation in SenML allows parallel parsing, resulting in efficient oneM2M servers. Thus the architecture proposed in [26] provides interoperability among the Smart City components, which was lacking in many of the architectures proposed in the past; however, it misses extending the architecture further to incorporate an important concept of linked data. The linked data is a method for publishing structured data using vocabularies like schema.org that can be connected together and interpreted by machines [1]. It helps users query data from different sources seamlessly, which is very important in the Smart City scenario where several data sources publish data simultaneously. Therefore, with a similar approach of using oneM2M standards, Jeong et al. [28] propose a city data hub by incorporating linked data concepts leveraging Next Generation Service Interfaces- Linked Data (NGSI-LD) compatible data models and APIs. Thus, the data hub facilitates the exchange and sharing of structured information by defining a modular data platform architecture that provides data interoperability. However, the city data hub neither provides a detailed description of the data models required for heterogeneous IoT verticals nor provides any guidelines. Furthermore, it lacks in dealing with data quality aspect too, which is vital in the smart city deployments. Thus, a common framework that outlines all smart city aspects while identifying key elements and expected outcomes and organizing them under a single system architecture was missing.

[1]https://wordlift.io/blog/en/entity/linked-data/

## 1.3 Contributions of this work

This work proposes an interoperable data processing system architecture for Smart City and Semantic Web applications. In the chapter, oneM2M standards for Smart City Development, a data monitoring system (DMS) consisting of a common service layer based on oneM2M standards and a data lake system (DLS) with multi-tenant database architecture to store the continuously monitored historical data into a common place are proposed. Further in the chaptaer Indian Urban Data Exchange, a Next Generation Service Interfaces for Linked Data (NGSI-LD)[2] based data exchange system (DES) based on the IUDX framework for seamless access to the historical data are proposed. Once the data is successfully stored then there arise a need to provide the correct data to the data users to enable a correct analysis otherwise an erroneous data may lead to produce false results. Thus, in the chapter, 5D-IoT framework, a data quality assessment layer based on Semantic Web principles is proposed. By incorporating all the necessary aspects, the proposed architecture provides a complete solution for developing a smart city system. Furthermore, the work presents a proof of concept implementation in the form of Smart Campus, IIIT-H system depicting Smart City applications. Fig. 1.1 illustrates the proposed architecture for IoT data processing for smart city and semantic web applications and the following subsections introduce the reader to each subsystem.

### 1.3.1 oneM2M standards for Smart City Development

Developing a smart city system requires the cooperation of many government departments and private partners to work seamlessly across departmental and operational boundaries. However, each department works independently and generates department-centric data resulting in a data silo thereby hindering the development of a smart city system. Therefore, this data silo is a significant issue that governments and system integrators must overcome by changing existing data management policies. This data silo issue can be overcome by incorporating standard-oriented methodologies during the system development process. oneM2M [29] is one such open and internationally recognized standard that provides a software middleware known as the Common Service Layer, which contains a standardized set of tools called Common Service Functions (CSFs) for building interoperable smart city systems. These CSFs define functionalities such as detailed data structure, data security and privacy aspects, device management and

[2]https://en.wikipedia.org/wiki/NGSI-LD

APIs to enable third-party application developers to develop applications using data from several smart systems easily.

Thus, a real-time data monitoring system (DMS) based on oneM2M standards is proposed in the first part of this work. The system can seamlessly accumulate data from various sensor networks independent of underlying communication technologies. Later, a Data Lake System (DLS) is proposed as an integration to DMS, leveraging the oneM2M subscription technique. The DLS consists of multi-tenant architecture [30], enabling efficient and reliable data management.

### 1.3.2 Indian Urban Data Exchange

Once the data is accumulated and stored in a commonplace, it becomes necessary to serve it to the users in a controlled and efficient manner. It can be achieved by incorporating a data exchange technique to allow data access in a unified, standard format and enable data sharing between multiple users within the city's ecosystem. Thus, in the second part of this work, a Data Exchange System (DES) is proposed based on the IUDX [31] framework. The DES incorporates a linked data technique that allows the system to publish the accumulated data by DMS in a structured format using standardized data models, thus enabling data interoperability. It also incorporates the Open Authorization (OAuth) 2.0 framework [32] to provide controlled data access to users thereby enabling data privacy. Furthermore, the DES incorporates the NGSI-LD-based APIs to ease the data retrieval.

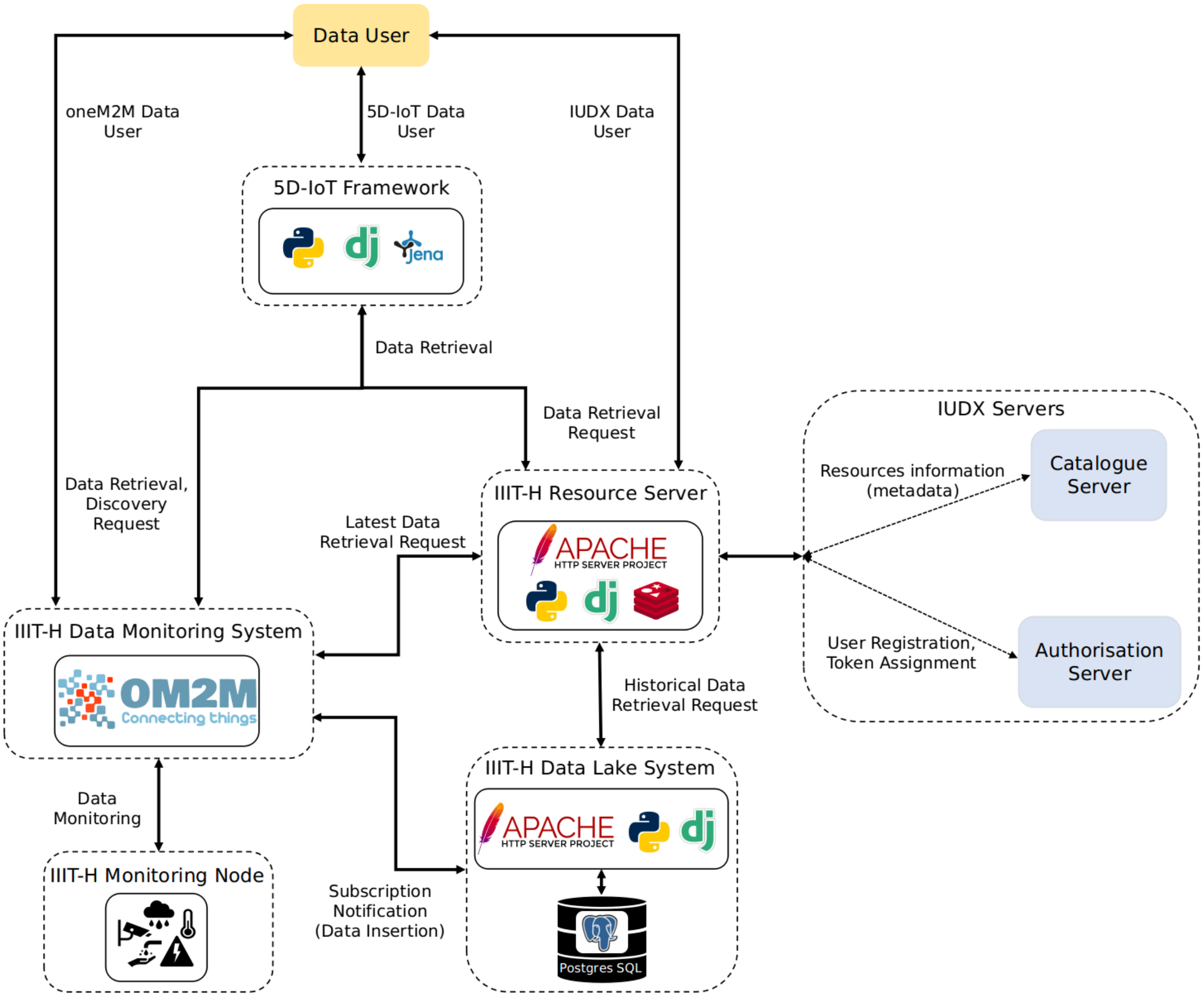

Figure 1.1: Illustration of the proposed architecture for IoT data processing for smart city and semantic web applications.

DES consists of a Catalogue Server (CS), an Authorization Server (AS) and a Resource Server (RS), where the RS is proposed and developed in this work and the CS and AS are developed and maintained by the IUDX organization. The CS maintains the information of the resource groups and resource items. A resource group is a collection of all the resource items that are associated with it, where a resource item refers to an IoT node. The AS manages the processes such as user registration, access policy creation/deletion, and generation of a Javascript Object Notation (JSON) Web Token (JWT) [33], an open standard that defines the method of securely transmitting information as a JSON object in a compact format. The RS manages processes such as JWT verification, data retrieval from DLS or DMS, and provision of data in a standardized format to the data requester. Multiple APIs are proposed as part of RS for data retrieval. These APIs are; the metadata API to return the information associated with an IoT node, the latest data API to return the latest data instance transmitted by the IoT node, and the temporal data API to return the historical data stored in the DLS.

### 1.3.3 5D-IoT Framework

The data exchange enables secure and seamless data retrieval; however, IoT systems have further issues such as additional data transmission delays, duplicate data generation, and erroneous data transmission. The data users need to understand the entire IoT deployment and factors that affect the data accuracy to filter the inaccurate data before using it. However, if this has to be done at a smart city level with numerous applications, it delays the data usage process. Therefore, it becomes necessary to resolve this issue to ease the accurate data quality understanding and usage. Thus, in the third part of this work, a five-layered 5-D IoT framework is proposed to add data quality assessment results to data before providing the data to users.

The first layer of the framework enriches the raw data from heterogeneous IoT sources and converts it into graph data based on the Resource Description Framework (RDF) [34] defined by semantic web principles. The subsequent three layers represent the novel approach of the data quality assessment using Shapes Constrained Language (SHACL) [35] shapes, which assesses duplicate data transmission, transmission delays, and inaccurate data generation, respectively. The assessment results are then added to the data graph using the SHACL inference rule and the proposed IoT Data Quality Assessment (IDQA) vocabulary. Finally, the fifth layer consists of the RDF Triple Store, which stores the processed data and makes it available as a Simple Protocol and RDF Query Language (SPARQL) [36] endpoint to the users.

## 1.4 Thesis Organization

Chapter 2 initially discusses the need for a global standard for smart city applications and how oneM2M standards can be helpful for smart city development. It later introduces the reader to the Smart Campus system, IIIT-H, by briefly explaining various monitoring applications depicting smart city applications. Then it discusses the proposed DMS and its implementation using OM2M [37], an open- source platform based on oneM2M standards. Further, the proposed architecture and implementation of DLS are discussed. Finally, the results of the performance assessment of the DMS and DLS are presented to show the added value of the proposed architecture.

Chapter 3 discusses the two applications, LoRaWAN Enabled Smart Energy Meters and Integration of OCPP based EV charger and oneM2M platform, that are developed in this work. The two IoT applications are developed in this work to present the capability of oneM2M standards to integrate diverse applications independent of the underlying communication technologies.

Chapter 4 discusses the need for a data exchange framework such as IUDX and its integration with the DMS and DLS. It later discusses CS, AS and RS in detail, along with the data models and APIs that are proposed and developed in this work. Finally, the performance results are added to evaluate the feasibility and sustainability of the proposed DES.

Chapter 5 discusses the need for a data quality assessment framework and how the 5D-IoT framework is able to provide the assessment results using Semantic Web principles. It then discusses each assessment layer in detail, followed by a proof-of-concept implementation of the framework. Finally, evaluation results of the framework are added by evaluating the real-time monitoring data received from the smart city system, IIIT-H. The results show the added value of the framework.

Chapter 6 concludes this work and provides some ideas that can be incorporated into the proposed system in the future.

*Chapter 2*

# oneM2M standards for Smart City Development

Chapter 1 introduced the reader to the problems of current smart city systems and how the proposed architecture can solve them. This chapter discusses how oneM2M standards address the need for global standards for smart city applications. A real-time data monitoring system architecture based on the oneM2M standards is proposed to collect data using a common service layer (CSL) independent of the underlying communication technologies. Due to horizontal CSL, oneM2M allows interoperability across multiple IoT verticals and thus provides interoperability across various domains. The chapter then discusses the need to store data collected from sensing nodes and proposes a DLS based on multi-tenant database architecture. It helps in the efficient virtualization of hardware resources available in the database. A typical smart city scenario consists of numerous IoT sensor networks acting as data tenants. Separate logical databases help achieve better data isolation and efficient data management in these sensor networks. Further, a detailed description of the developed DMS and DLS is provided to fully familiarize the reader with the proof-of-concept implementation, followed by performance evaluation results that illustrate the added value of the proposed architecture.

## 2.1 Background and Related Works

With the increasing use of IoT devices, we are seeing an increase in data generation. Various manufacturers produce devices that transmit data using numerous communication protocols and data models. Thus with such great diversity in the data collection process, it becomes difficult to use the data to build applications later. Hence, standardization of device specifications, data models, and communication protocols plays a vital role in achieving interoperability that ultimately helps to develop large-scale, cost-effective solutions. Therefore, in 2012, oneM2M was established as the global partnership project for IoT interoperability that consists of the world's leading countries in information and communication technology standardization development, namely, the USA, Japan, China, Europe, India, and Korea. This organization aims to provide a global technical standard for interoperability of M2M technologies by incorporating standardization in IoT systems architecture, API specifications, security and access control, device communication, data management and data semantics. oneM2M standards body develops technical specifications related to a common service layer that acts as a middleware to provide a horizontal layer between devices, communication protocols, and cloud and business applications, ultimately enabling IoT interoperability. Figure 2.1 illustrates the multi-layered IoT stack and the position of oneM2M common service layer

oneM2M currently has more than 200 participating partners, including companies such as Samsung, Intel, Adobe, IBM, etc. Recently, oneM2M has already been adopted at a national level in India to develop the country's 100 smart cities plan [38]. An IoT system based on the oneM2M standards breaks down the silo boundaries by enabling the interworking between any authorized IoT application and another, independent of underlying communication technologies. The standardized features of the oneM2M standards facilitate its use to develop interoperable solutions in various IoT use-cases such as smart buildings, smart cities, and Industry 4.0 by increasing reusability, reducing fragmentation and the deployment costs of the IoT systems.

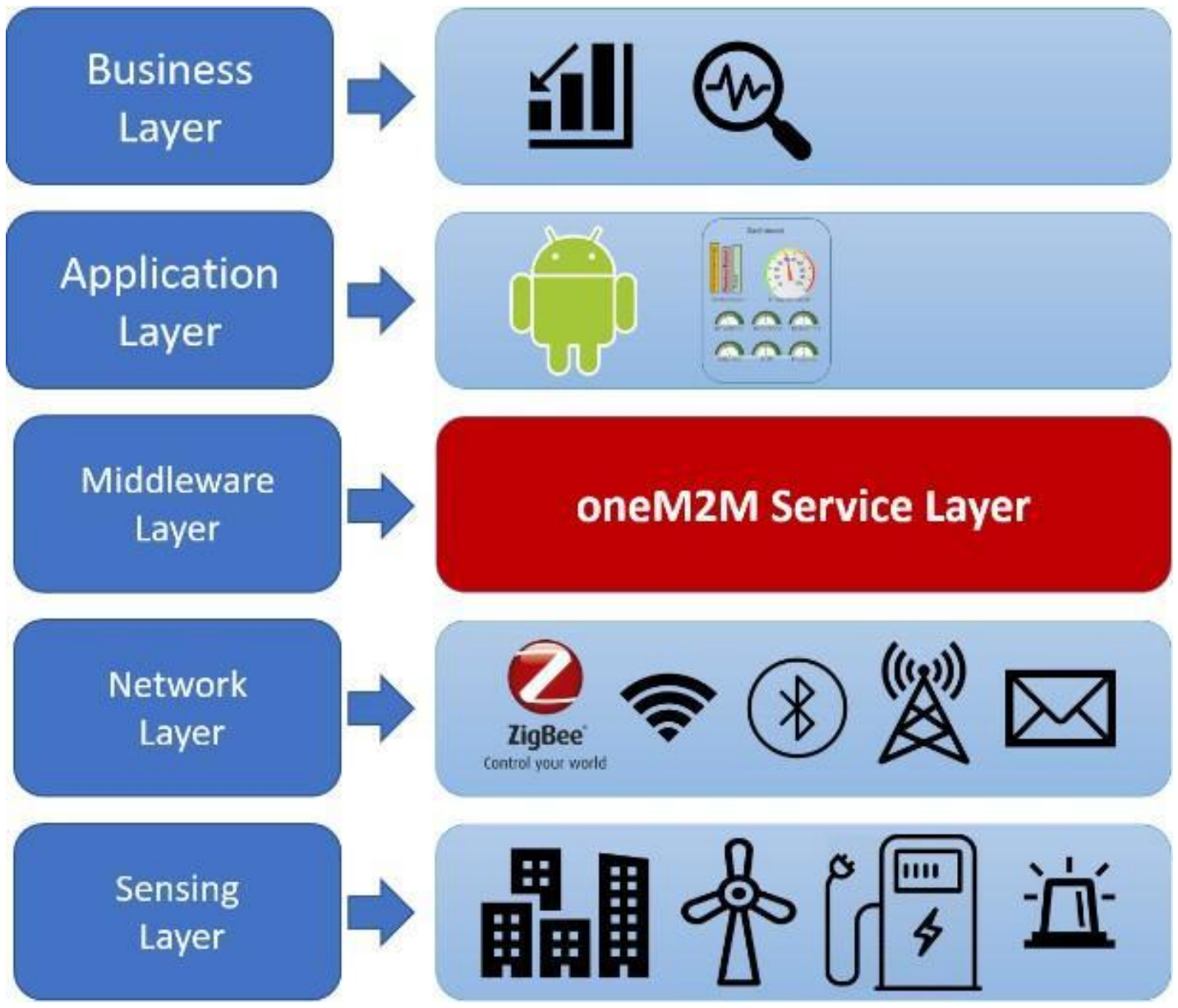

Figure 2.1: Illustration of layers in an IoT stack and the position of oneM2M common service layer.

## 2.2 Smart City Living Lab, IIIT-H

Smart City Living Lab is a part of the Smart City Research Centre at the International Institute of Information Technology, Hyderabad (IIIT-H). It is a setup with the support from the Indian Ministry of Electronics and Information Technology (MEITY), Smart City Mission India, and the Telangana

State Government. The living lab is supported by European Business & Technology Centre (EBTC) and Amsterdam Arena.

The main objective of Living Lab is to develop and test smart city solutions that can then be deployed in cities across India. Thus, multiple IoT nodes equipped with various sensors are deployed across the campus to monitor and collect real-time data of various parameters in a specified area. This data can then be used to take smart and intelligent decisions to solve various problems like energy saving, maintaining water quality, proper distribution of water and so on. Thus, in the first part of this work, a data monitoring system is proposed and developed which is explained in the next section.

### 2.2.1 Data Monitoring System (DMS)

At IIIT-H, with the help of a open-source platform based on oneM2M standards, a centralized real-time DMS has been proposed and developed to collect data belong to various verticals. Each vertical has its nodes equipped with various sensors, which are deployed at multiple locations on campus. The different verticals deployed at the IIIT-H campus are described below.

- **Air Quality Monitoring (AQM):** As mentioned in [39], several air quality parameters such as Particulate Matter (PM2.5, and PM10), Temperature, Relative Humidity and CO concentration are monitored through densely deployed sensor nodes to increase the spatio-temporal resolution of air quality data.
- **Crowd Monitoring (CM):** The vertical monitors crowds of people to check the total number of mask violations to avoid any covid-outbreak situation inside the campus. Videos obtained from the cameras deployed on the campus are processed by a Crowd Monitoring System (CMS), which extracts information of current people count, the total number of safe distance violations, and a total number of mask violations for analysis purposes.
- **Energy Monitoring (EM):** To understand the usage statistics and to provide faster resolutions to power outages, several energy parameters such as the individual phase currents and voltages, average power factor, frequency, energy consump- tion, apparent and real power are monitored by nodes deployed in this vertical [40].
- **Smart Rooms Monitoring (SRM):** To build a state-aware intelligent space and to build a dynamic fault detection system, several parameters like occupancy, energy consumption are monitored to dynamically adjust the air conditioning, lighting, and ventilation.
- **Solar Monitoring (SM):** To efficiently analyse the solar energy generated by solar panels, parame- ters such as energy generated in a day, signed active power, instantaneous frequency, output power factor, voltage, and current are monitored.
- **Water Monitoring (WM):** In this vertical [41], the parameters associated with water quality such as total dissolved solids, water temperature are being monitored. Similarly, the water quantity parameters like water flow and water pressure are monitored to understand the consumption and to avoid health related problems caused by poor quality water.

- **Weather Monitoring (WE):** For predicting weather status and to understand the current climate condition, multiple weather monitoring stations are deployed to monitor parameters such as solar radiation, temperature, relative humidity, wind direction, wind speed, gust speed and dew point.

In addition to the hardware, which includes sensors and micro-controllers deployed to collect data from each node, the whole software side that collects data from each node is implemented with the help of the OM2M platform, an open-source implementation of oneM2M standards.

#### 2.2.1.1 OM2M

It provides a horizontal service layer as defined in oneM2M standards. OM2M is a Java language-based implementation running on top of an Open Service Gateway Initiative (OSGi) [42] Equinox run-time, which makes it highly extensible via plugins. Each plugin is used for specific functionality, and can be remotely installed, started, stopped, updated, and uninstalled without needing a reboot. OM2M is built as an Eclipse product using Maven, a build automation tool for Java projects and Tycho, a set of maven plugins used in building eclipse products.

#### 2.2.1.2 Resource Tree Creation

The Internet of Things (IoT) system is a collection of interrelated computing devices, mechanical and digital machines, objects, animals or people that are provided with a unique identifier and the ability to transfer data over a network without requiring human-to-human or human-to-computer interaction [11]. In the oneM2M standards [1], every component of the IoT system is referred as a resource, and the structured arrangement of these resources is referred as a resource tree. Figure 2.2 shows the illustration of Smart Campus, IIIT-H resource tree. The root or parent for all the verticals in the resource tree is the IN-CSE, where IN stands for Infrastructure Node and CSE for Common Service Entity. Each CSE has its unique name, Ex. In the below diagram, the name of in-cse is in-name. The IN-CSE has multiple Application Entities (AEs) and each AE here corresponds to one IoT vertical. Each IoT node corresponds to one container describing data attributes and child resources. Every node container (CNT) contains two more containers, the Descriptor and the Data CNTs. Figure 2.3 illustrates the weather monitoring application entity's Node, Data and Descriptor containers. The Descriptor container contains a content instance (CIN) with a detailed description of the devices used for data collection and the monitored parameters. The Data container contains CINs that store data transmitted from the IoT node. Figure
2.4 (a) and (b) illustrate the descriptor and data content instance of the water flow node deployed at Pump-House 1 inside the IIIT-H campus.

[1]https://member.onem2m.org/Application/documentapp/downloadLatestRevision/default.aspx?docID=26462

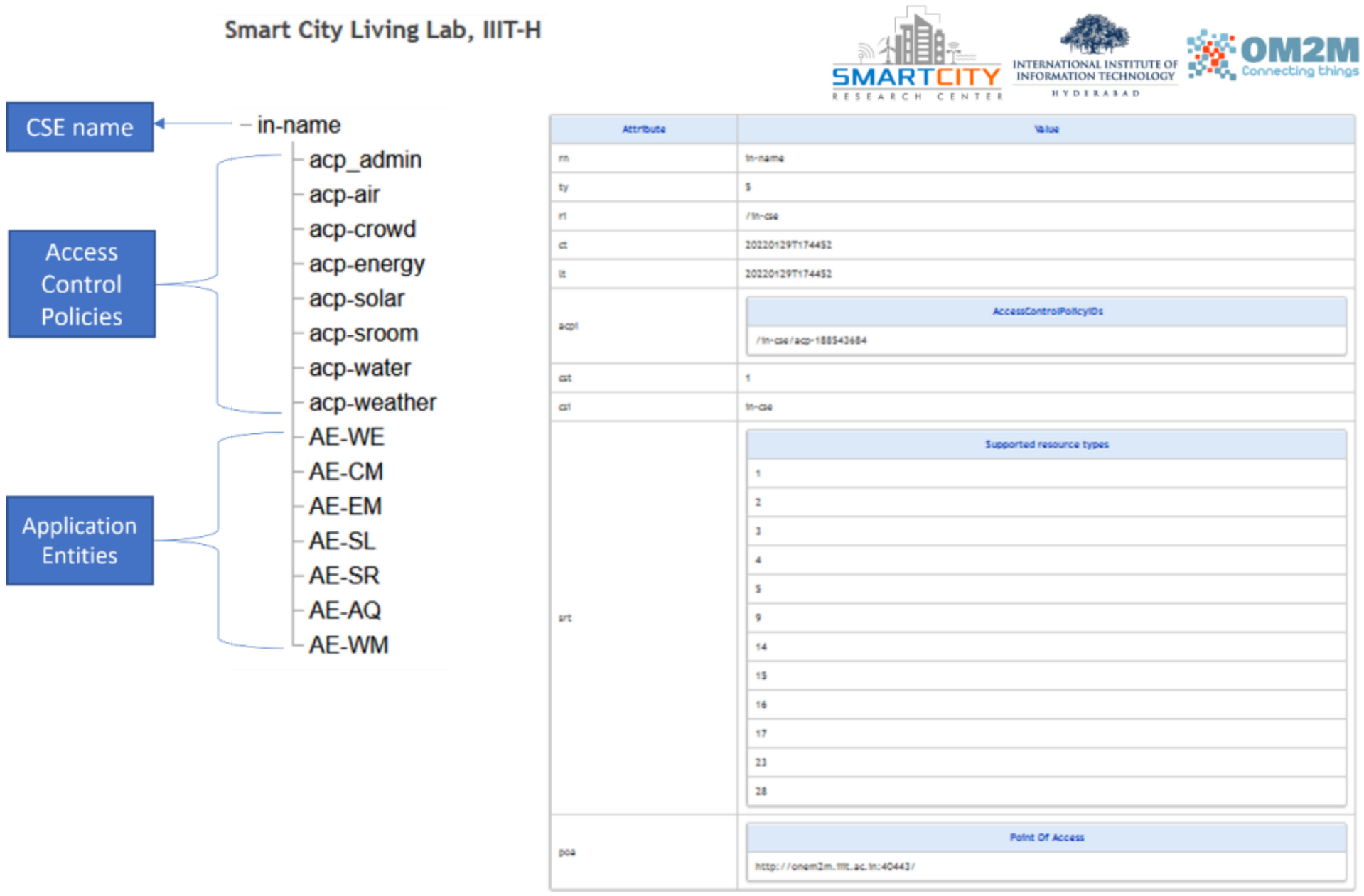

Figure 2.2: Illustration of the Smart Campus, IIIT-H resource tree.

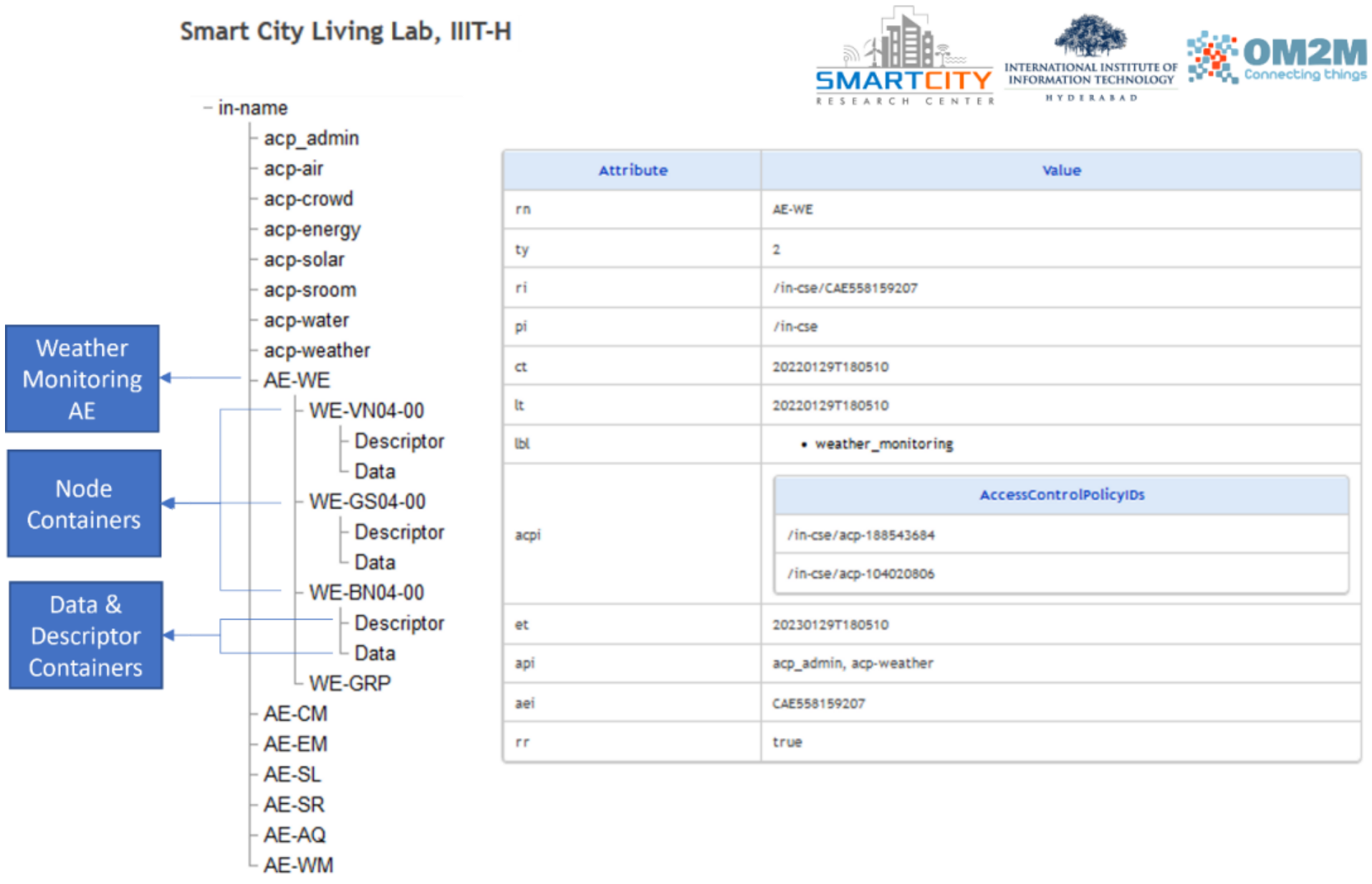

Figure 2.3: Illustration of the Node, Data and Descriptor containers of the weather monitoring application entity.

| Attribute | Value |
|---|---|
| Node ID | WM-WF-PH01-00 |
| Node Location | {'Latitude': 17.445793, 'Longitude': 78.351444} |
| Device Model | {'Controller': 'ESP 32, id=1.0', 'Device': 'Water Flow node with Flow Rate, Total Flow and Pressure', 'Sensors': ['(Timestamp)', '(Flowrate = Wprime Ultrasonic Water Meter, id=1.0)', '(Total Flow = Wprime Ultrasonic Water Meter, id=1.0)', '(Pressure = Danfoss MBS 3000, id=1.0)', '(Pressure Voltage = Danfoss MBS 3000, id=1.0)']} |
| Version History | [{'ver': 'V6.0.0', 'dt_start': '26-04-2021 00-00-00', 'dt_end': '31-12-9999 23-59-59'}] |
| Data String Parameters | ['Timestamp', 'Flowrate', 'Total Flow', 'Pressure', 'Pressure Voltage'] |
| Parameters Description | Data Description, [datatype], [Units], [Resolution], [Accuracy] |
| Timestamp | The number of seconds that have elapsed since Thursday, 1970 Jan 1 00:00:00 UTC, [int], [s], [60 s], [n/a] |
| Flowrate | The instantaneous value of Flow rate, [float], [$m^3/h$], [0.001], [± 2 % in [Q2,Q4), ± 5 % in [Q1,Q2]] |
| Total Flow | The instantaneous value of Total flow, [float], [$m^3$], [0.01], [± 2 % in [Q2,Q4), ± 5 % in [Q1,Q2]] |
| Pressure | The instantaneous value of Pressure, [float], [bar], [1], [± 1 %] |
| Pressure Voltage | The instantaneous value of pressure voltage, [float], [V], [0.000001 V], [N/A] |

(a)

| Attribute | Value |
|---|---|
| rn | cin_938437420 |
| ty | 4 |
| ri | /in-cse/cin-938437420 |
| pi | /in-cse/cnt-198060182 |
| ct | 20220219T123355 |
| lt | 20220219T123355 |
| lbl | • WM-WF-V6.0.0<br>• AE-WM-WF<br>• V6.0.0<br>• WM-WF-PH01-00 |
| st | 0 |
| cnf | text |
| cs | 50 |
| con | [1645254204, 867.00, 3091168.00, 260.00, 0.006418] |

(b)

Figure 2.4: Illustration of (a) Descriptor and (b) Data content instance, of the waterflow monitoring node deployed at Pump-House 1 inside IIIT-H campus.

#### 2.2.1.3 Access Control Policies

Access control policy (ACP) is a set of rules that decides the actions permittable to the user. The CSE uses ACPs to control access to resources. Every resource always has at least one ACP attached to it. The rule of an ACP defines,

- **WHO** can access the Resource (e.g.,Identifiers of authorized AE/CSE)
- For **WHAT** operation (CREATE / RETRIEVE / UPDATE / DELETE/NOTIFY/DISCOVERY)
- Under **WHICH** contextual circumstances (Time, Location, IP address)

In OM2M, each ACP has a unique identification number (ACPI) and set of usernames and passwords to define the access for the required operation. Table 2.1 illustrates two ACPs, acp-admin and acp-guest.

Each username and password pair has a number associated with it; the number is a binary combination of actions that can be performed. Figure 2.5 illustrates the relationship between the access types and their associated bits. The system administrator must select a combination of access rules and calculate the decimal number accordingly. For example, in Table 1, admin ACP has access control operation parameters (ACOP) as 63, which is the binary equivalent of 111111. From Figure 2.5, it can be deduced that the admin user has access to create, retrieve, update, delete, notify, and discover any resource to which the ACP is attached. Whereas the guest ACP has ACOP as 34, whose binary equivalent would be 100010, means the guest user can only discover and retrieve the resource to which the ACP is attached.

Table 2.1: Access control policies for an admin and guest user

| ACP | Credentials (username:password) | ACOP | ACP ID | Description |
|---|---|---|---|---|
| acp-admin | admin:admin | 63 | /in-cse/acp-348780115 | This ACP contains the admin rights which are common to all AEs residing under the CSE. |
| acp-guest | guest:guest | 34 | /in-cse/acp-644134496 | This ACP contains the guest rights which allows only to discover and retrieve |

| Enumeration | Discovery | Notify | Delete | Update | Retrieve | Create |
|---|---|---|---|---|---|---|
| 1 | 0 | 0 | 0 | 0 | 0 | 1 |
| 2 | 0 | 0 | 0 | 0 | 1 | 0 |
| 3 | 0 | 0 | 0 | 0 | 1 | 1 |
| ... | ... | ... | ... | ... | ... | ... |
| 63 | 1 | 1 | 1 | 1 | 1 | 1 |

Figure 2.5: Illustration of access control rule and its associated bit representation.

The OM2M platform allows the administrator to create and update an ACP by sending an HTTP POST request. The body of the request should contain privileges (PV) and self-privileges (PVS), which further include access control rules (ACR) through access control originator (ACOR) and access control operation (ACOP) procedures. Figure 2.6 (a) illustrates the XML formatted representation of an ACP and Figure 2.6 (b) illustrates the created ACP on the OM2M platform.

#### 2.2.1.4 Data Storage and Latency Improvement

The OM2M platform uses the H2 database, a Java-based, open source, relational structured query language (SQL) database. H2 supports embedded and server modes for data storage. The embedded mode is used as an in-memory database as part of the OM2M platform and is helpful for storing lesser amounts of data and having another database instance for larger data storage. The embedded mode

also

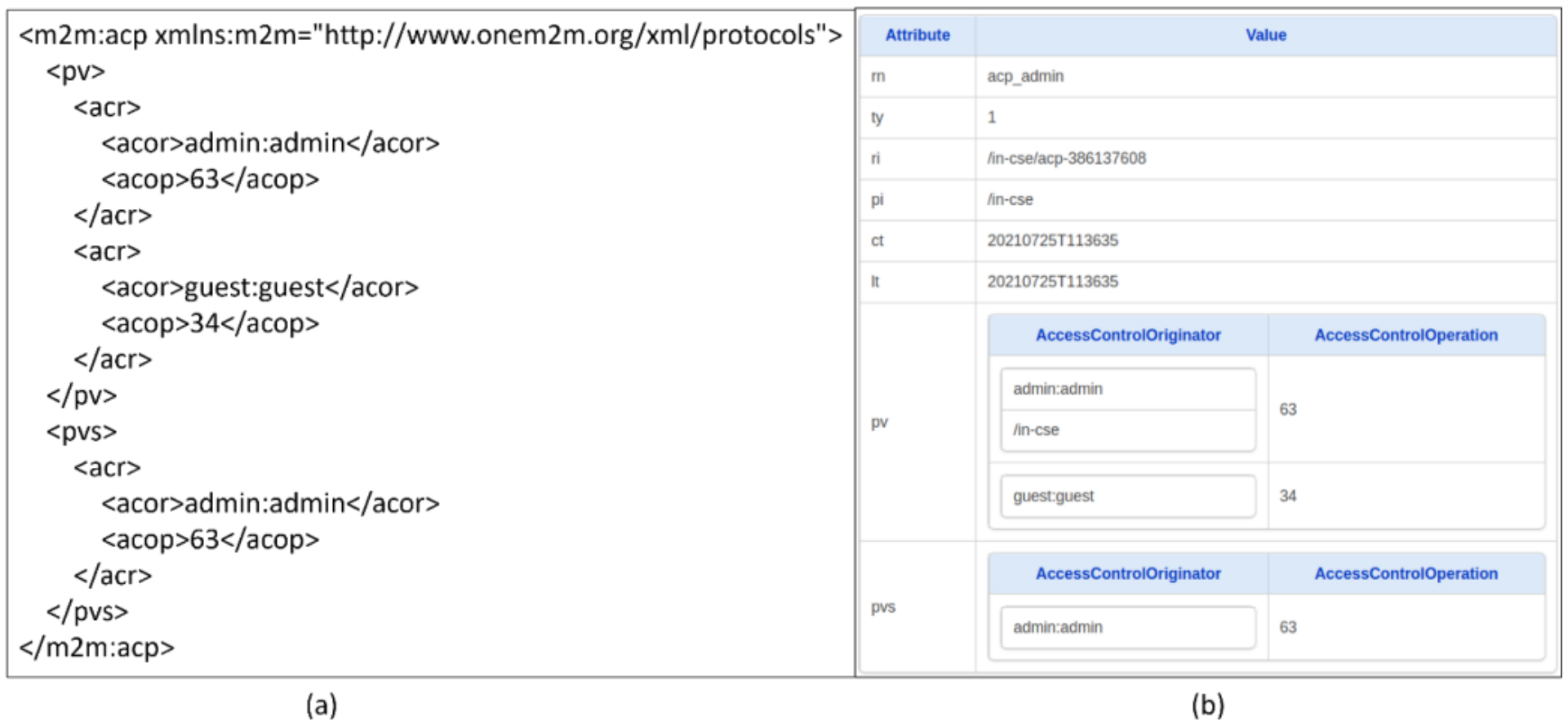

Figure 2.6: (a) XML formatted representation of an ACP b) Illustration of the created ACP on OM2M platform.

helps in faster interaction. For server mode, the H2 database can be hosted on the same machine as OM2M or on a separate dedicated machine. In the H2 server, we can perform actions such as managing databases, creating users, storing substantial amounts of data, etc.

With the help of POSTMAN, a REpresentational State Transfer (REST) client software, both modes of H2 database were tested by measuring data retrieval time from the database. Ten thousand content instances per container in 10 containers were stored under an AE. As a part of the data retrieval, the latest data content instance from each data container was retrieved using the latest data API provided by the OM2M platform. The test results are summarised in Table 2.2.

Due to the larger retrieval time for server mode compared to embedded mode, it was decided to use embedded mode for our application. Currently, the IIIT-H IN-CSE contains 150 nodes where the number of instances in each node's data container is restricted to 120, as illustrated in Figure 2.7. The reduction in the number of instances helped to further reduce the latest data retrieval time in the embedded mode to around 130-140 milliseconds. Later, a data lake system is developed to store data beyond 120 instances which is explained in section 2.2.2.

#### 2.2.1.5 oneM2M APIs for Data Retrieval

oneM2M REST APIs handle CRUD+N (Create, Retrieve, Update, Delete and Notification) operations. In this section the APIs for data retrieval are described.

- **Latest, Oldest and All Data Retrieval:** A user who requires the live / recently transmitted data through an IoT node can use the latest data API request. The standard also proposes another API where the user can retrieve the oldest data instance of a data container. If an application wants to analyze the entire data stored in a container, i.e., 120 instances in the current implementation, it can use all data retrieval API. Figure 2.8 illustrates the request format for retrieval of the latest,

| Attribute | Value |
|---|---|
| rn | Data |
| ty | 3 |
| ri | /in-cse/cnt-535825527 |
| pi | /in-cse/cnt-202656262 |
| ct | 20220129T182637 |
| lt | 20220129T182637 |
| lbl | •AQI<br>•AQI-MP (AQI Major Pollutant)<br>•AQL<br>•CO Concentration<br>•Data Interval<br>•NH3 Concentration<br>•NO2 Concentration<br>•PM10<br>•PM2.5<br>•Relative Humidity<br>•Temperature<br>•Timestamp |
| acpi | **AccessControlPolicyIDs**<br>/in-cse/acp-188543684<br>/in-cse/acp-898417228 |
| et | 20230129T182637 |
| st | 55482 |
| mni | 120 |
| mbs | 10000 |
| mia | 0 |
| cni | 120 |
| cbs | 8520 |
| ol | /in-cse/in-name/AE-AQ/AQ-MG00-00/Data/ol |
| la | /in-cse/in-name/AE-AQ/AQ-MG00-00/Data/la |

Labels useful for data discovery (→ lbl)

Maximum number of instances (→ mni)

Figure 2.7: Illustration of the data container of the air quality node deployed at the main gate of IIIT-H, with a maximum number of instances (mni) limited to 120

Table 2.2: Results of the latest data retrieval time measurement test

| **Number of instances in a container** | ***Latest data retrieval time (milliseconds)*** | |
|---|---|---|
| | ***Embedded mode*** | ***(Server mode)/n*** |
| 1000 | 198 | 786 |
| 2000 | 376 | 1088 |
| 3000 | 587 | 1484 |
| 4000 | 800 | 2289 |
| 5000 | 1158 | 3095 |
| 6000 | 1366 | 3787 |
| 7000 | 1550 | 4732 |
| 8000 | 1750 | 5590 |
| 9000 | 2002 | 6491 |
| 10000 | 2215 | 7163 |

oldest, and all data instances. The response format of the latest and oldest API requests is the same; only the data residing under the con attribute differs. The all-data API request response contains a list of individual CIN data responses. The responses of the oldest and all data API requests can be deduced from the latest data API response illustrated in Figure 2.9, i.e., the oldest API response contains the oldest data instance of a CNT, and all data API response contains all 120 instances of a data CNT.

- **Group Data Retrieval:** As described in the oneM2M standards, group data resource (GRP) is created by grouping the CNTs under an AE. Thus, in the Smart Campus Resource Tree, for applications that require data from a group of resources, such as data from all IoT nodes related to a specific vertical, GRP resources are created. Using such group resources, it becomes possible to retrieve the latest, oldest, and all content instances of all those data containers by sending a single API request. Figure 2.10 illustrates a POST request that is used to create a group resource, AQ-GRP, by grouping the data containers of air quality vertical. Figure 2.11 illustrates the request format for the retrieval of the latest, oldest, and all data instances of an AQ-GRP resource. **URL** is the resource identifier to locate the particular resource, **Header** contains the credentials that are required to authenticate the API request and **Method** tells the type of REST-API Ex. GET, POST, PUT, DELETE.

Latest data instance retrieval API request

| Field | Value |
| --- | --- |
| URL | https://<cse-address>/~/in-cse/in-name/AE-AQ/AQ-AN00-00/Data/la |
| Method | GET |
| Header | X-M2M-Origin: username:password<br>Accept: application/json |

Oldest data instance retrieval API request

| Field | Value |
| --- | --- |
| URL | https://<cse-address>/~/in-cse/in-name/AE-AQ/AQ-AN00-00/Data/ol |
| Method | GET |
| Header | X-M2M-Origin: username:password<br>Accept: application/json |

All data instances retrieval API request format

| Field | Value |
| --- | --- |
| URL | https://<cse-address>/~/in-cse/in-name/AE-AQ/AQ-AN00-00/Data?rcn=4 |
| Method | GET |
| Header | X-M2M-Origin: username:password<br>Accept: application/json |

Figure 2.8: Illustration of the response of the latest data API request.

**Response:**

```
{
  "m2m:cin": {
    "rn": "cin_755434944",
    "ty": 4,
    "ri": "/in-cse/cin-755434944",
    "pi": "/in-cse/cnt-799370622",
    "ct": "20220305T201834",
    "lt": "20220305T201834",
    "lbl": [
      "AE-AQ",
      "AQ-AN00-00",
      "V3.0.02",
      "AQ-V3.0.02"
    ],
    "st": 0,
    "cnf": "text",
    "cs": 71,
    "con": "[1646491691, 23.50, 42.80, 31.49, 32.25, nan, nan, nan, 42.80, 0, 1, 0]"
  }
}
```

Figure 2.9: Illustration of the request format for the latest, oldest and all data retrieval API requests.

**HTTP Request:**

| Field | Value |
|---|---|
| URL | https://<cse-address>/~/in-cse/in-name/AE-AQ |
| Method | POST |
| Header | X-M2M-Origin: username:password<br>Content-Type : application/json;ty=9 |

**Body:**

```json
{
  "m2m:grp":{
    "rn":"AQ-GRP",
    "mt":3,
    "mid": [
    "/in-cse/in-name/AE-AQ/AQ-AN00-00/Data",
    "/in-cse/in-name/AE-AQ/AQ-MG00-00/Data",
    "/in-cse/in-name/AE-AQ/AQ-KH00-00/Data",
    "/in-cse/in-name/AE-AQ/AQ-KN00-00/Data",
    "/in-cse/in-name/AE-AQ/AQ-VN90-00/Data",
    "/in-cse/in-name/AE-AQ/AQ-PH03-00/Data",
    "/in-cse/in-name/AE-AQ/AQ-PL00-00/Data",
    "/in-cse/in-name/AE-AQ/AQ-FG00-00/Data",
    "/in-cse/in-name/AE-AQ/AQ-SN00-00/Data",
    "/in-cse/in-name/AE-AQ/AQ-BN00-00/Data"
  ],
    "mnm":10
  }
}
```

Figure 2.10: Illustration of a POST request that is used to create a group resource of air quality vertical.

Latest data instance retrieval API request

| Field | Value |
|---|---|
| URL | https://<cse-address>/~/in-cse/in-name/AE-AQ/AQ-GRP/fopt/la |
| Method | GET |
| Header | X-M2M-Origin: username:password<br>Accept: application/json |

Oldest data instance retrieval API request

| Field | Value |
|---|---|
| URL | https://<cse-address>/~/in-cse/in-name/AE-AQ/AQ-GRP/fopt/ol |
| Method | GET |
| Header | X-M2M-Origin: username:password<br>Accept: application/json |

All data instances retrieval API request format

| Field | Value |
|---|---|
| URL | https://<cse-address>/~/in-cse/in-name/AE-AQ/AQ-GRP/fopt?rcn=4 |
| Method | GET |
| Header | X-M2M-Origin: username:password<br>Accept: application/json |

Figure 2.11: Illustration of the request format for the latest, oldest and all data retrieval API requests for group resource.

- **Discovery based data retrieval:** The Smart Campus DMS, IIIT-H, has several IoT nodes that monitor parameters such as energy consumption, air quality, weather condition, water quality and more. Some parameters are common in some of these nodes; for example, temperature and relative humidity are the common monitoring parameters in air quality and weather condition monitoring nodes. The data discovery API is proposed in oneM2M standards to ease the data retrieval for application that need data of nodes with common parameters. A Data CNT's label (LBL) attribute is used here for discovery purpose. Figure 2.12 illustrates the labels of an air quality data CNT, such as Temperature, Relative Humidity, etc. By using the labels, the data CNT's Uniform Resource Identifier List (URIL) can be retrieved. Later, using the URILs, the latest/oldest/whole data belonging to the respective data containers can be retrieved. Figure 2.13 illustrates the response of a data discovery API request to retrieve URILs of data CNTs with Temperature and Relative Humidity as common LBLs.

| Attribute | Value |
|---|---|
| rn | Data |
| ty | 3 |
| ri | /in-cse/cnt-432605719 |
| pi | /in-cse/cnt-378418196 |
| ct | 20210427T213107 |
| lt | 20210427T213107 |
| lbl | • AQI<br>• AQI-MP (AQI Major Pollutant)<br>• AQL<br>• CO concentration<br>• Data Interval<br>• NH3 concentration<br>• NO2 concentration<br>• PM10<br>• PM2.5<br>• Relative Humidity<br>• Temperature<br>• Timestamp |
| acpi | AccessControlPolicyIDs:<br>/in-cse/acp-292007947<br>/in-cse/acp-951301058 |
| et | 20220427T213107 |
| st | 0 |
| mni | 120 |
| mbs | 10000 |
| mia | 0 |
| cni | 0 |
| cbs | 0 |
| ol | /in-cse/in-name/AE-AQ/AQ-MG00-00/Data/ol |
| la | /in-cse/in-name/AE-AQ/AQ-MG00-00/Data/la |

Figure 2.12: Illustration of the attributes of an air quality monitoring node's data container.

**HTTP Request:**

| Field | Value |
|---|---|
| URL | https://<cse-address>/~/in-cse?fu=1&lbl=Relative Humidity&lbl=Temperature |
| Method | GET |
| Header | X-M2M-Origin: username:password<br>Accept: application/json |

**Response:**

```
{
  "m2m:uril": [
    "/in-cse/in-name/AE-SR/SR-OC/SR-OC-GW-KH03-01/Data",
    "/in-cse/in-name/AE-SR/SR-AQ/SR-AQ-KH03-03/Data",
    "/in-cse/in-name/AE-AQM/AQM-XX00-00/Data",
    "/in-cse/in-name/AE-SR/SR-AQ/SR-AQ-KH03-01/Data",
    "/in-cse/in-name/AE-AQ/AQ-AN00-00/Data",
    "/in-cse/in-name/AE-SR/SR-AQ/SR-AQ-KH00-00/Data",
    "/in-cse/in-name/AE-AQ/AQ-MG00-00/Data",
    "/in-cse/in-name/AE-WE/WE-GS04-00/Data",
    "/in-cse/in-name/AE-AQ/AQ-KH00-00/Data",
    "/in-cse/in-name/AE-SR/SR-AQ/SR-AQ-KH00-02/Data",
    "/in-cse/in-name/AE-SR/SR-OC/SR-OC-GW-KH01-00/Data",
    "/in-cse/in-name/AE-WE/WE-BN04-00/Data",
    "/in-cse/in-name/AE-SR/SR-OC/SR-OC-GW-KH03-00/Data",
    "/in-cse/in-name/AE-AQ/AQ-KN00-00/Data",
    "/in-cse/in-name/AE-AQ/AQ-VN90-00/Data",
    "/in-cse/in-name/AE-AQ/AQ-PH03-00/Data",
    "/in-cse/in-name/AE-SR/SR-OC/SR-OC-GW-KH03-02/Data",
    "/in-cse/in-name/AE-AQ/AQ-PL00-00/Data",
    "/in-cse/in-name/AE-SR/SR-AQ/SR-AQ-KH03-00/Data",
    "/in-cse/in-name/AE-SR/SR-AQ/SR-AQ-KH00-01/Data",
    "/in-cse/in-name/AE-SR/SR-OC/SR-OC-GW-KH95-00/Data",
    "/in-cse/in-name/AE-AQ/AQ-FG00-00/Data",
    "/in-cse/in-name/AE-WE/WE-VN04-00/Data",
    "/in-cse/in-name/AE-SR/SR-AQ/SR-AQ-KH95-00/Data",
    "/in-cse/in-name/AE-SR/SR-OC/SR-OC-GW-KH00-00/Data",
    "/in-cse/in-name/AE-AQ/AQ-SN00-00/Data",
    "/in-cse/in-name/AE-SR/SR-OC/SR-OC-GW-KH00-01/Data",
    "/in-cse/in-name/AE-SR/SR-AQ/SR-AQ-KH00-03/Data",
    "/in-cse/in-name/AE-SR/SR-AQ/SR-AQ-KH03-02/Data",
    "/in-cse/in-name/AE-AQ/AQ-BN00-00/Data"
  ]
}
```

Figure 2.13: Illustration API request and response of the label-based data discovery.

### 2.2.2 Data Lake System (DLS)

After developing a real-time DMS, there arise a need to store the data that can be accessed further to gain valuable insights. Therefore, the IIIT-H DLS has been proposed to store a large amount of sensor data by integrating it with the oneM2M platform.

#### 2.2.2.1 System Architecture

Figure 2.14 IIIT-H illustrates the integration of DMS and DLS, where the OM2M platform acts as a data ingestor and ingests all data collected from IoT nodes into a database hosted on DLS.

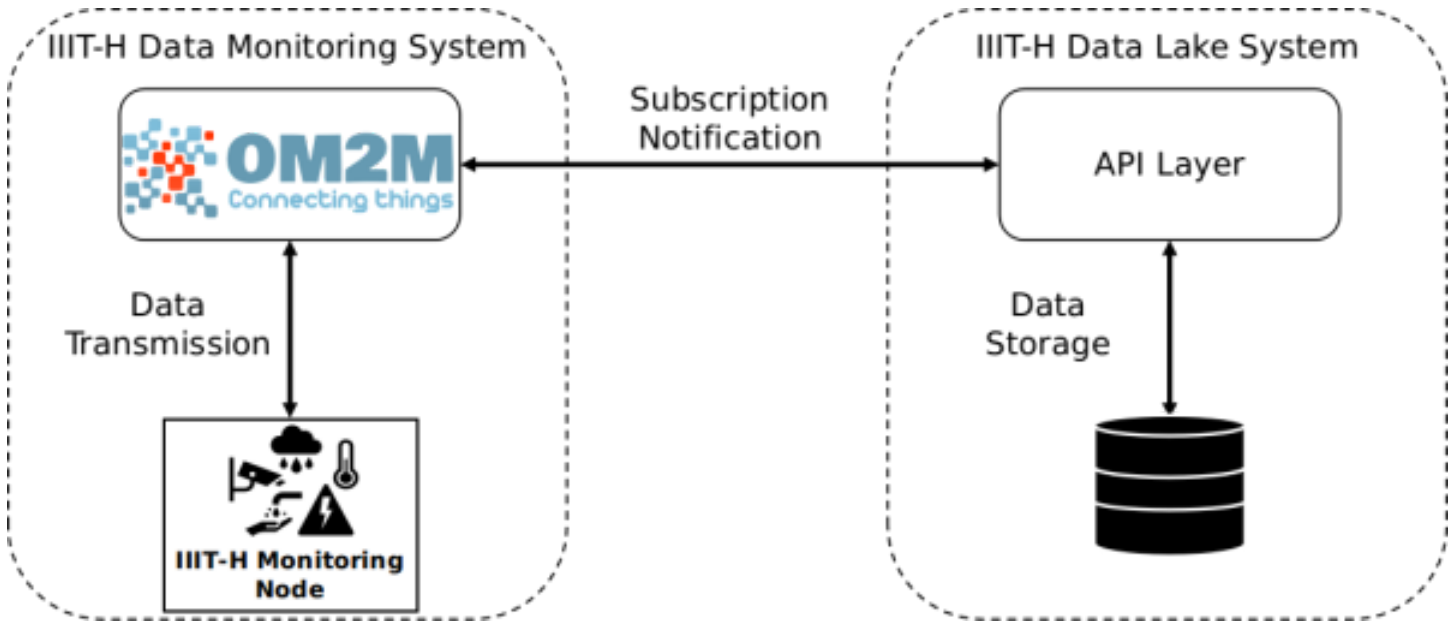

Figure 2.14: Illustration of the integration of IIIT-H Data Lake System with Data Monitoring System.

The oneM2M standards propose the use of a subscription resource based on the server-sent events (SSEs) technique for automatically forwarding newly created events (i.e., CINs) as notifications to the subscriber applications (i.e., DLS in this case). Therefore, on the OM2M platform, a subscription resource is created for every data CNT that forwards the newly created CINs to DLS. Currently, only the DMS administrator has access to create subscription resources to prevent unauthorized applications from subscribing to the data CNTs. It is happening with the help of ACP created on the platform. Once the CINs reach the DLS, they are parsed to extract information such as the value of the measured parameters, the node name from which the data is transmitted, the observation timestamp, and the data model version. The version specifies the current device model that contains the metadata, such as the description of monitored parameters, their resolution, accuracy, and the type of sensor used for the monitoring by the IoT node. Some modifications may happen in the nodes, such as the addition/removal of sensors, replacement of one sensor with another, and so on. Hence a unique number (version) is given to each device model to distinguish one device model from another. This information is stored in the persistent storage, which later helps the data users to understand which data is belonging to which version.

The main objective behind developing DLS is to enable efficient and flexible data storage of large amounts of heterogeneous IoT data. A Data Warehouse (DW) enables efficient collection and storage of historical and commutative data to analyze, report, and integrate transaction data from one or more sources. To design a DW, the data modeling (DM) technique is used that helps to optimize the database for faster data retrieval. The DM consists of the dimensional data model, where a data model

is a

collection of concepts used to describe the structure of the database Ex. relationships, data types, etc. In the case of the dimensional data model, concepts like entity, attribute and relationship are used. While designing a DM for DLS, the below steps are followed.

- **Identify Process:** This step is followed to identify the real-time process, i.e., in our case, the collection of data from different IoT nodes corresponding to other verticals. The process is illustrated in Figure 2.15.

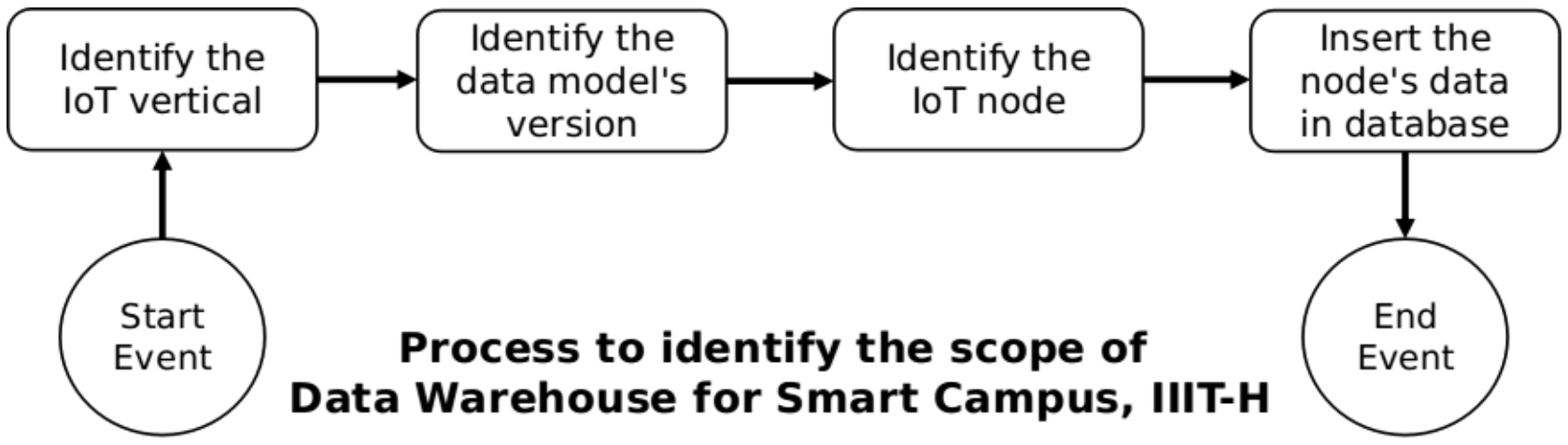

Figure 2.15: Illustration of identification of the use case covered by the Data Warehouse designed for the Data Lake System for Smart Campus, IIIT-H.

- **Identify Grain (level of detail):** This step is followed to identify the lowest level of information stored in the table in DW. In our case, the lowest level is the IoT node data, and the granularity of the node is the sampling period which is the period of collecting two consecutive data points by the node. Ex. AE-AQ node has a granularity of 15 seconds, AE-WM node has a granularity of 4 hours, etc.

- **Identify Dimensions:** This step is followed to identify the dimensions, i.e., measured parameters by an IoT node that need to be considered during data retrieval. In our case, the parameters are vertical name, node name, timestamp, geo-location and version. E.g., The user can send a request for information on nodes with air quality index (AQI) <= 100 at geographical location 14.00,13.00 (latitude, longitude) between 10 pm and 12 pm on May 3, 2021.

- **Identify Facts:** This step is mainly related to data storage and data users. It is followed while creating the fact table, a primary table in the DM. The current implementation contains two fact tables, Data Fact Table (DFT) and Parameters Fact Table (PFT). DFT includes facts such as node name, vertical name, sensors used for measurement, observation timestamp, values of the measured parameters and two foreign keys to the PFT. PFT contains facts such as parameter name, vertical name, sensor name, measurement's data type, unit, accuracy, resolution and a foreign key to the Vertical dimension table. The dimension tables are created to store information about the deployed nodes, the data model's version, monitoring vertical, sensors used for the measurement, and geolocation.

- **Build Schema:** This is the final step in implementing the dimension model by creating schemas, i.e., by properly designing the database tables. Of the three popular schemes, Star,

Snowflake and

Galaxy Schema, the Galaxy Schema has been selected and implemented for DW in the IIIT-H DLS. The Galaxy Schema is a collection of more than one-star schema and as the Star and Snowflake schema has only one fact table, therefore, Galaxy Schema has been chosen because of the need for two fact tables for DLS. The two fact tables in DLS are the data and the parameters table. The data table is associated with the data values obtained from the deployed nodes and the parameters table is associated with the parameters measured by the deployed nodes. Figure 2.16 illustrates the currently implemented Galaxy schema for DW in the IIIT-H DLS.

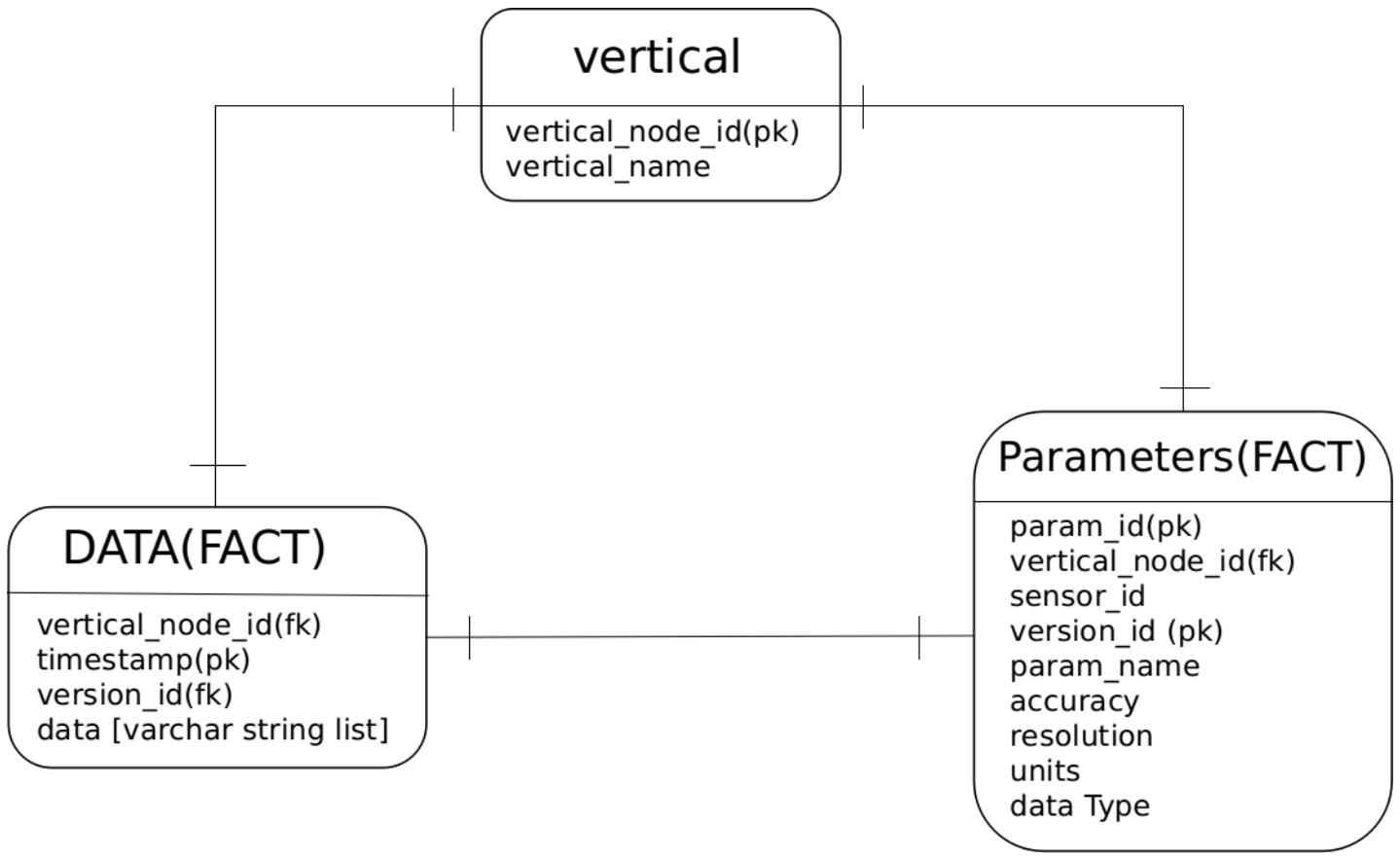

Figure 2.16: Illustration of the currently implemented Galaxy schema for the Data Warehouse in the IIIT-H Data Lake System.

In the case of a smart campus project, each vertical is maintained by different groups of students and professors, who deploy the IoT nodes and then use the data from those nodes to gain valuable insights. In this work a tenancy database architecture has been proposed and implemented to handle such scenario. A ”Tenant” is a part of the Software as a Service (SaaS) model where the service provider company does not sell licenses to the software, but customer makes rent payments to the company. Thus, making each customer a tenant of your company. In return tenant receives access to SaaS application components, and has its data stored in the SaaS system

Similarly, in case of the smart campus project, by considering a group of students and professors as a tenant, it was necessary to give separate access to each tenant by implementing a proper database access mechanism. Therefore, after finalizing the schema, in the next step, a multi-tenant database architecture is designed and implemented. In multi-tenant with multiple logical database architecture, the privacy and security of each tenant's data is ensured in three different ways.

- **Separate Database:** Allocating a separate logical database to each tenant
- **Separate Schema:** Allocating schema to multiple logical databases

- **Separate Rows:** Allocating same physical database and schema to store all the information within the same tables while separating the information by using the compound primary keys of node-id and timestamp that are allocated as part of the database design.

Table 2.3 illustrates the comparison between all three multi-tenant database architecture approaches by CloudExpo [43]. Depending upon various attributes, they have rated each approach as either Good, Fair or Poor.

Various issues may arise in the future such as database locking, the total database size may go up to terabytes (TB) or 100s of gigabytes (GB) causing concern in system scalability and moreover, frequent changes in the currently deployed vertical are less likely. Thus, it was decided to opt for a multi-tenant architecture with a separate logical database to store each vertical's data as shown in Figure 2.17.

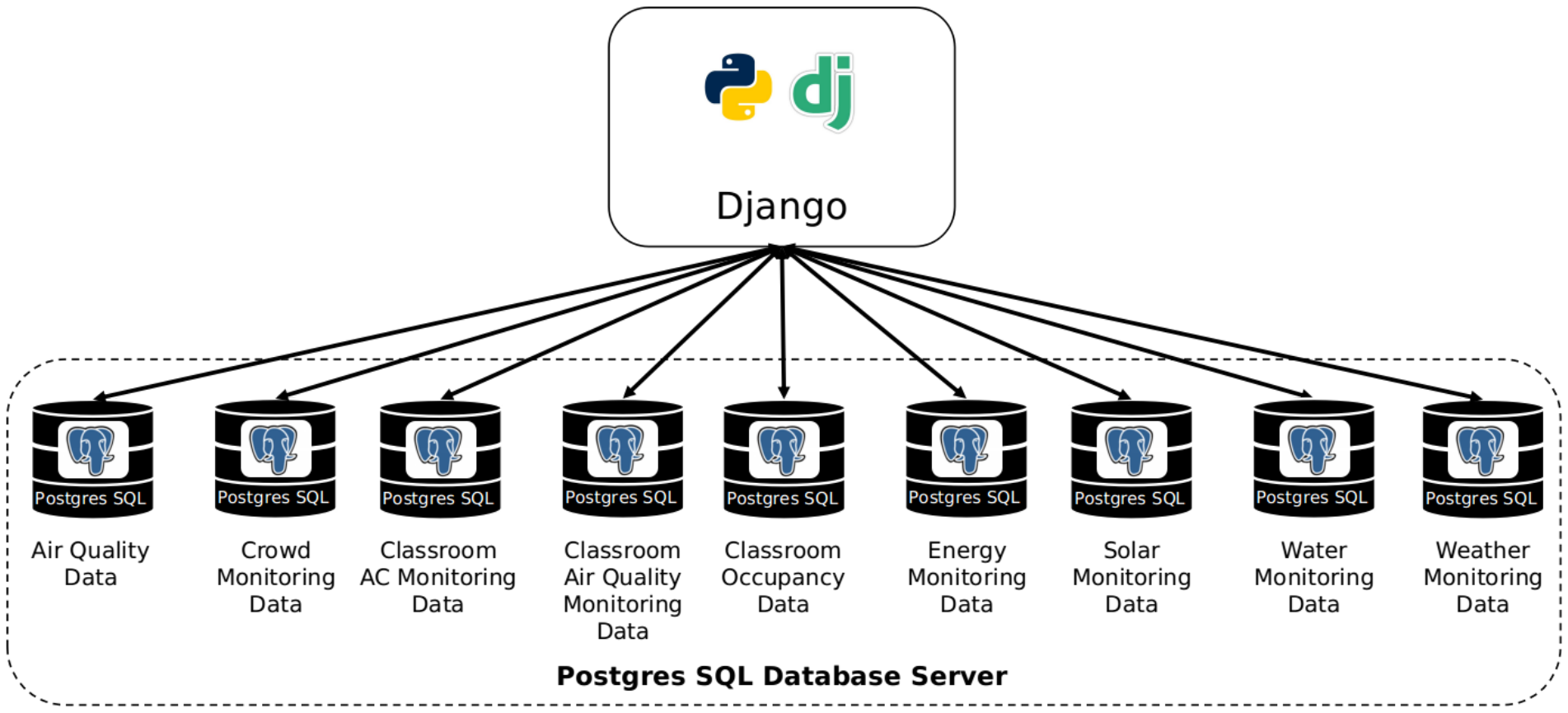

Figure 2.17: Illustration of multi-tenant separate database architecture implemented as part of the IIIT-H Data Lake System.

#### 2.2.2.2 Implementation

As per the oneM2M standards, for data forwarding, the data CNT's subscription resource must be associated with the unique resource locator (URL) of the client application (Notification URL (nu) in case of oneM2M standards). Moreover, after receiving the notification, the client application must send an acknowledgment response to the OM2M platform. Therefore, DLS is developed in such a way that it provides a URL to receive notifications and sends a ”200 OK” acknowledgment response as soon as a notification is received. Later, it processes the data and store into the database.

The URL creation happens using the REST API framework of Django. Further, to host a web framework like Django, the Apache web server is used that filters requests such as large static files like images, scripts retrieval which need not to be served by the Django application and hence acts as a shield by only allowing requests which are meant for Django. In addition, Apache server provides additional functionalities such as access control mechanism, web server security, and

high-performance

Table 2.3: Comparison between all three multi-tenant database architecture approaches

| Sr No | Attribute | Separate Database | Separate Schema | Separate Rows |
|---|---|---|---|---|
| 1 | Tenants are large enterprise customers who could store large amounts of data in TB or 100s of GB | Good | Fair | Poor |
| 2 | Tenants are individual consumers with low or moder- ate storage of data like a social networking site | Poor | Fair | Good |
| 3 | Rapid provisioning is the key, risk of losing business exists if the response is not quick, customers select the service on the fly | Fair | Fair | Good |
| 4 | Customers can opt out of (Cancel) service at a faster rate and space reclamation is not a concern | Good | Good | Fair |
| 5 | The business logic of your Software as a Service (SaaS) application is highly customized with respect to the Tenant | Good | Good | Poor |
| 6 | Application is prone to database locks | Good | Fair | Poor |
| 7 | Security and Legal Requirements require data sepa- ration, even if the application controls it | Good | Fair | Poor |
| 8 | Performance tuning is a concern and reports perfor- mance should be based on the volume of data | Good | Fair | Poor |
| 9 | Administration and maintenance of Schema and database code is a concern | Poor | Poor | Good |
| 10 | Scalability is a concern | Good | Fair | Poor |
| 11 | Frequent Changes to the Application Possible | Poor | Poor | Good |
| 12 | Data is mission critical, and Point in Time Recovery is needed in case of a crash | Good | Fair | Poor |

mechanism which is required for large scale systems. However, a standard interface is required for efficient communication between Django and Apache web server. Python Web Server Gateway Interface (WSGI), an interface that acts as a mediator between Django application and the web server, is used for this purpose. Finally, to store the data, Postgres database server is used. Figure 2.18 illustrates the communication between OM2M platform, Apache web server, WSGI, Django and Postgres database server.

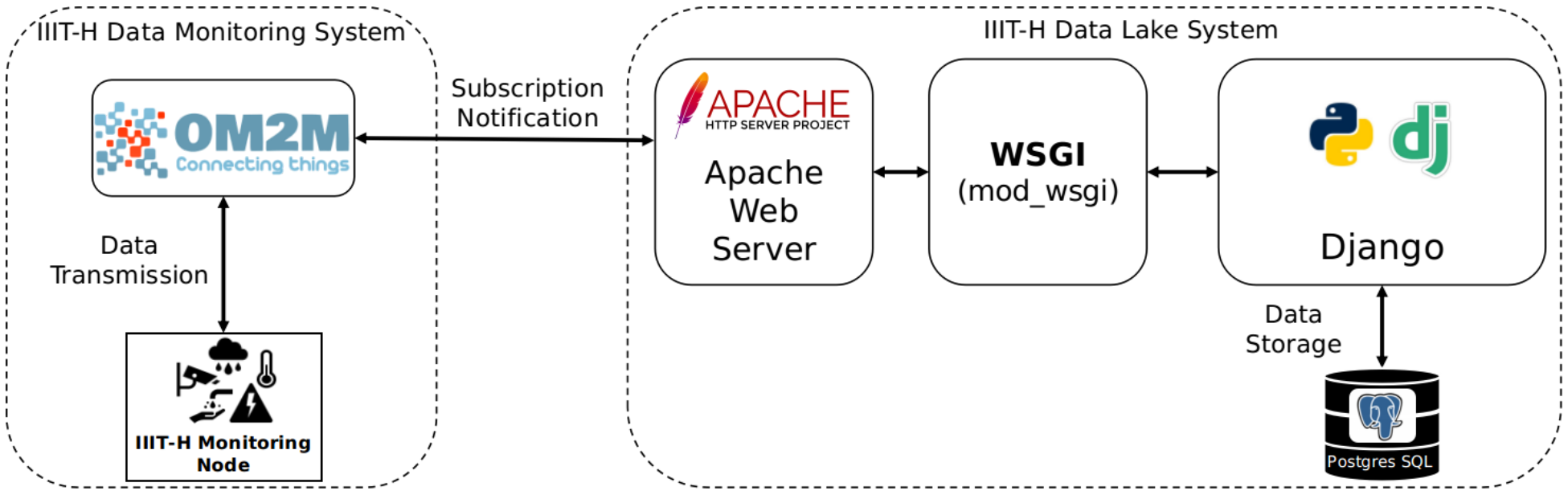

Figure 2.18: Illustration of the integration of IIIT-H Data Lake System with Data Monitoring System.

Figure 2.19 (a) illustrates a subscription resource created under one of the data CNT of AE-AQ and Figure 2.19 (b) illustrates the JSON object received by the DLS as a notification. The notification from OM2M towards DLS thus first reaches to Apache web server and then to Django (via WSGI) which sends the acknowledgement back to the web server (once again via WSGI) and eventually to the OM2M. After sending the acknowledgement, Django processes the received data and stores it in the corresponding database based on the vertical name obtained from the URI of received notification. The label from a received notification is used to obtain version number. Ex. Labels in Figure 2.19 (b) contains the vertical name (AE-name (AE-AQ)), node name (AQ-KH00-00), data model's version (V3.0.02) of the AQ-KH00-00 node from which the data is received. Based on the vertical name information, Django stores the data into AQ database. Similarly, data from other nodes is stored into respective databases.

The multiple databases helped to store the large volumes of data and retrieve it with minimum latency. Furthermore, if any database encounters a locking problem, it is easy to recover that database without disturbing other databases. The access management has also become easier because the administrator can allow or remove access to a database directly without interfering with the access policies of other verticals' database users.

Finally, the default firewall of the Ubuntu operating system called Uncomplicated Firewall (UFW) is also enabled which currently only allows data insertion requests from the OM2M platform i.e., sending notification requests to DLS and allows data retrieval from IIIT-H resource server (IIIT-H-RS), which is discussed in detail in Chapter 4.

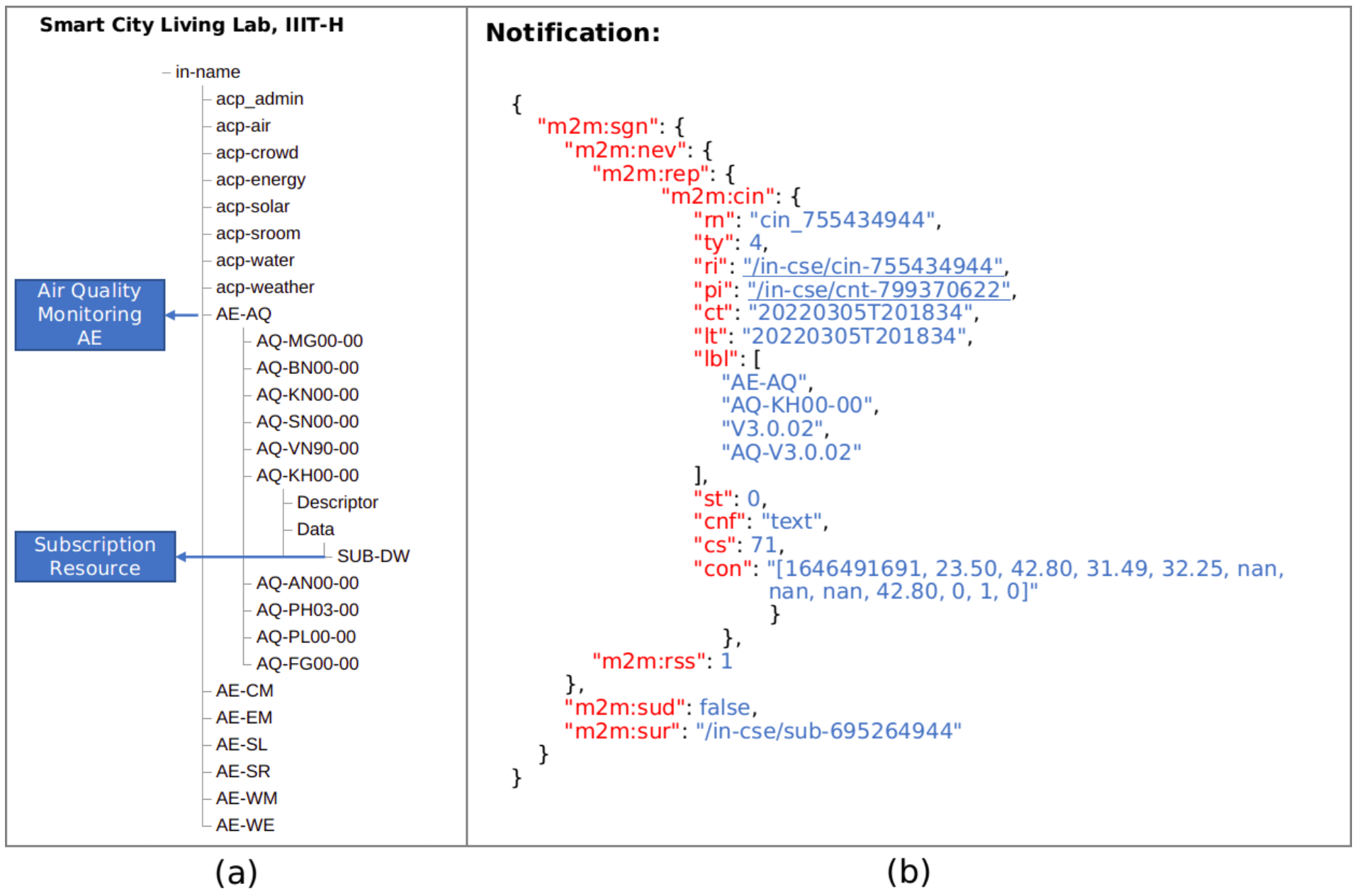

Figure 2.19: (a) Illustration of a subscription resource created under one of the data CNT of AQM vertical (b) Illustration of the JSON object received by the DLS as a notification.

## 2.3 Performance Assessment

### 2.3.1 Data Monitoring System

The primary purpose of this experiment was to mimic the DMS's real-time data insertion and retrieval scenarios. A 12- hour performance test has been conducted on the DMS, and the analysis revealed that it could handle 8 parallel users for the data-insertion scenario. As shown in Figure 2.20, it is observed that the retrieval latency increased proportionally with the increasing parallel users, and DMS can handle 600 parallel users with zero downtime for data-retrieval scenario.

### 2.3.2 Data Lake System

The impetus for this assessment was to highlight the ad- vantages of the proposed DSL architecture for a multi-sensor network scenario. This is reflected by the increased throughput in the standard temporal API scenario. As shown in Figure 2.21, the analysis is conducted with two variable parameters: the parallel user count and data points per request. The parallel user count is varied in steps of 25 users, and within each parallel user set, the requested data points are varied from 200 to 600 in steps of 200. The retrieval performance test for all such permutations are conducted on a single database and proposed DSL, where each request effectively retrieves data of three sensor networks simultaneously using the standard temporal API. Clearly, the proposed multi-tenant architecture with multiple logical databases has higher throughput when compared to a single database, as seen in Figure 2.21.

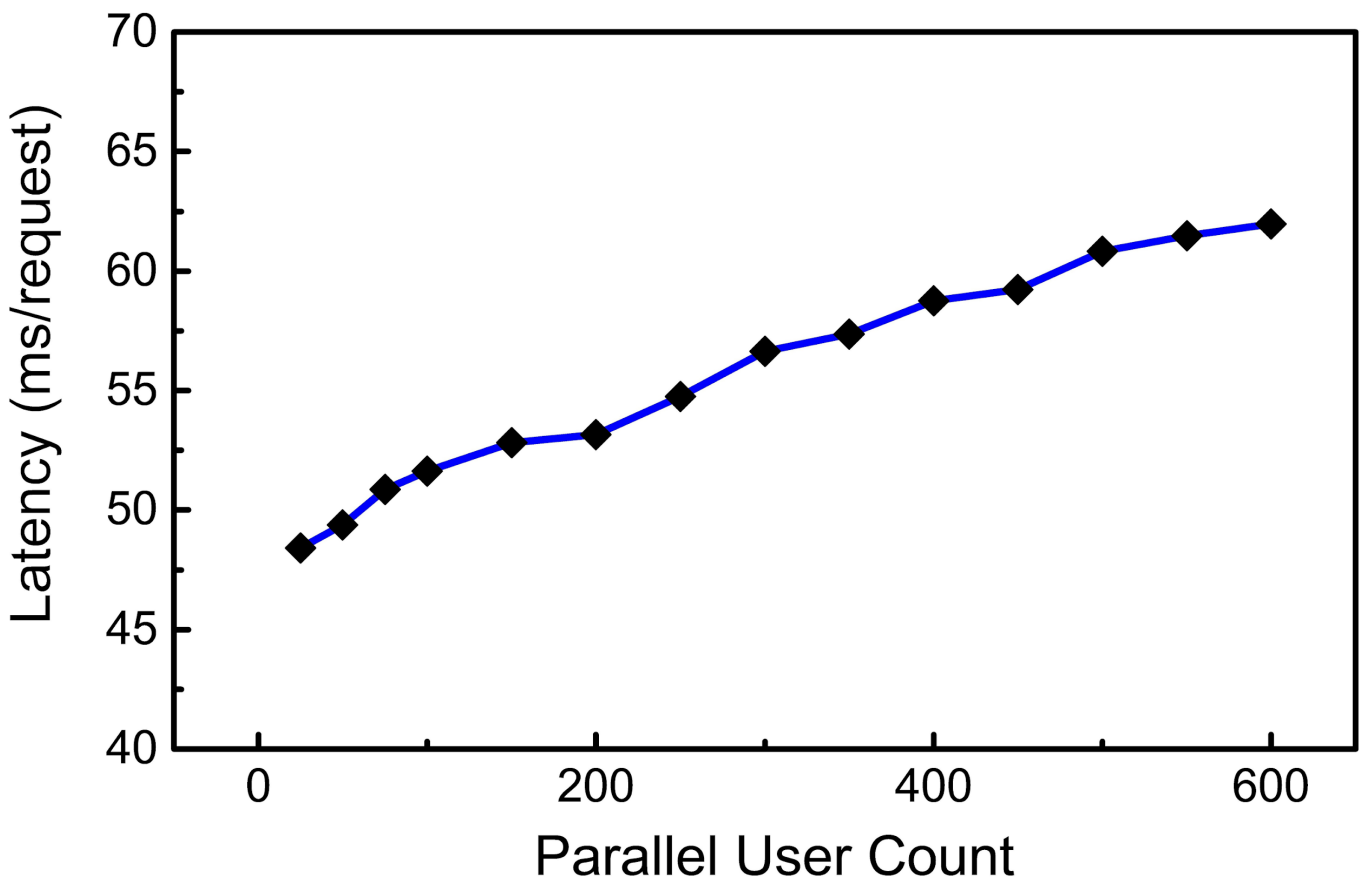

Figure 2.20: OM2M retrieval performance characteristic.

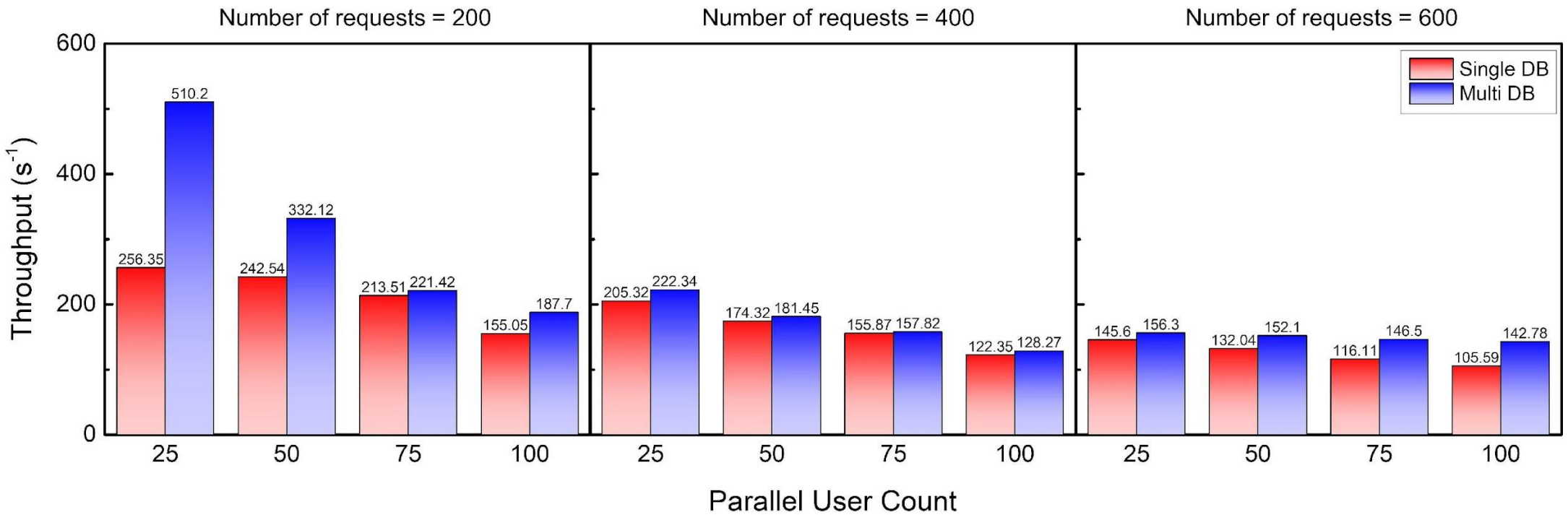

Figure 2.21: Throughput analysis for Single and Multiple DBs.

## 2.4 Summary

Smart cities play a vital role in limiting the ill-effects of rapid urbanization on the environment without compromising on benefits such as improving infrastructure, standard of living, and productivity. However, the collection, storage, and sharing of data from the plethora of sensor networks in a typical smart city deployment warrants a well-defined data platform architecture. Therefore, in this work, a novel architecture is proposed compliant with the oneM2M standards. The proposed architecture consists of DMS and DLS. oneM2M acts as the middleware in the DMS to enable IoT interoperability. A multi-tenant architecture with multiple logical databases enables efficient and reliable data management in the DLS.

*Chapter 3*

# Applications of oneM2M standards

Chapter 2 discussed the proposed DMS and DLS and their implementation as a part of the Smart Campus IIIT-H system. It discussed the use of CSFs of the oneM2M standards to develop a system for collecting data from multiple IoT verticals and storing it in a common location using a multi- tenant database architecture. This chapter discusses two IoT applications developed in this work to present the oneM2M standard's ability to integrate various applications independently of the underlying communication technology. The first application is developed to introduce the integration of LoRAWAN- enabled smart energy meters with the oneM2M platform, and the second one is developed to introduce the integration of OCPP-based energy meters with the oneM2M platform. Thus it shows that an application can use any communication protocol based on its requirements and then can be easily integrated into the system without any changes to the existing architecture.

## 3.1 LoRaWAN Enabled Smart Energy Meters Deployment

### 3.1.1 Background and Related Work

Climate change has brought about a paradigm shift in the global energy domain. There is now a constant effort to improve energy efficiency and provide clean distributed energy for sustainable global economic growth. IoT is pioneering this transformation, imparting capabilities such as real-time monitoring, surveillance [44], control, situational awareness, and intelligence, and cybersecurity [45] at various stages of energy generation and distribution. Real-time monitoring can assist in achieving better efficiency and minimizing load on the grids by analysing usage patterns, providing faster resolutions to power outages, lowering the cost for consumers using demand-based dynamic pricing, etc. and indeed, multiple IoT-based energy monitoring solutions have been proposed in recent years.

A comprehensive survey of smart electricity meters has been presented in [46], detailing aspects of the metering process, data analytics, existing technologies, and various stakeholders. In [47], a different perspective reviewing the development, deployment, functions including data recording and alarming, theft detection and security, pricing, etc. has been presented. An IoT-based energy

monitoring and

control system employing a Raspberry Pi and a MySQL database hosted on Amazon Web Services (AWS) has been presented in [48], to assess the energy consumption of air conditioning and improve efficiency. A Bluetooth Low Energy (BLE) based energy management method has been proposed in [49] using a BLE-enabled mobile device (any smartphone) to aggregate the energy consumption data. This was a modified architecture of their previous Zigbee, and PLC based system [50], primarily aimed to reduce the energy consumption. In general, IoT devices have gradually steered away from traditional protocols such as WiFi, BLE and cellular, since low power and long range are favoured in most IoT applications as opposed to higher bandwidth and data rate. The new protocols including LoRaWAN, Sigfox, NB-IoT and LTE-M are collectively termed as Low Power Wide Area Networks or LPWANs. Specifically, LoRaWAN is seen as a promising solution in sundry IoT applications [51] including smart cities [52], [53], localization [54, 55], utility metering [56, 57], etc.. owing to its long range and low power capabilities. Extensive research on signal propagation [58, 59], path loss modelling [60], and range evaluation [61], has been carried out in the recent years and LoRaWAN-based energy monitoring systems have already been proposed in the literature [62–64]. In [65], class A LoRaWAN-based smart meters were used to monitor energy parameters and a detailed analysis of package delivery rate was presented. Energy monitoring using an Arduino-based sensor, Raspberry Pi-based gateway using LoRa for communication, and a NoSQL-based MongoDB database for data storage has been presented in [66]. One of the major disadvantages of the rapid integration of IoT is the limitation of meaningful data exchange between different verticals. To combat this, a global standard, oneM2M, has been proposed. It acts as a horizontal layer between IoT end-nodes, communication networks and applications by defining CSFs, thus restoring the interoperability and scalability. In this work, a novel architecture is proposed employing LoRaWAN-enabled smart meters for sensing the energy parameters and an OM2M platform for collecting and analysing the data. OM2M is an open-source service platform that compliant to the oneM2M standards. A proprietary format is used for the LoRa Protocol Data Unit (PDU) to transmit electrical parameter information including individual phase currents and voltages, avg. Power Factor (PF), real, signed active, and signed reactive components of power and energy. The integration of oneM2M aids in realizing this energy monitoring vertical as part of a scalable monitoring system, including verticals such as air pollution, crowd monitoring, security, and water quality.

### 3.1.2 System Architecture and Implementation

A LoRaWAN-enabled three-phase smart meter, compliant with the IS 16444 standard [67] has been used for capturing energy data which is sent to multiple SenRa [68] gateways through LoRa modulation. This data is forwarded to the cloud based SenRa network server via an Ethernet backhaul, subsequently decoded in a proxy server, and forwarded to the oneM2M IoT platform. The live dashboard visualises this data, accessible to any device through a HTTP-based web browser. A schematic of this architecture is illustrated in Figure 3.1.

As part of a scalable monitoring system, energy meters were deployed at two locations inside the campus of IIIT Hyderabad, India. One of them is shown in Figure 3.2, along with the geolocation information of the deployed meters. We are in the process of deploying an additional 78 smart meters throughout the campus, phasing out the traditional digital meters entirely.

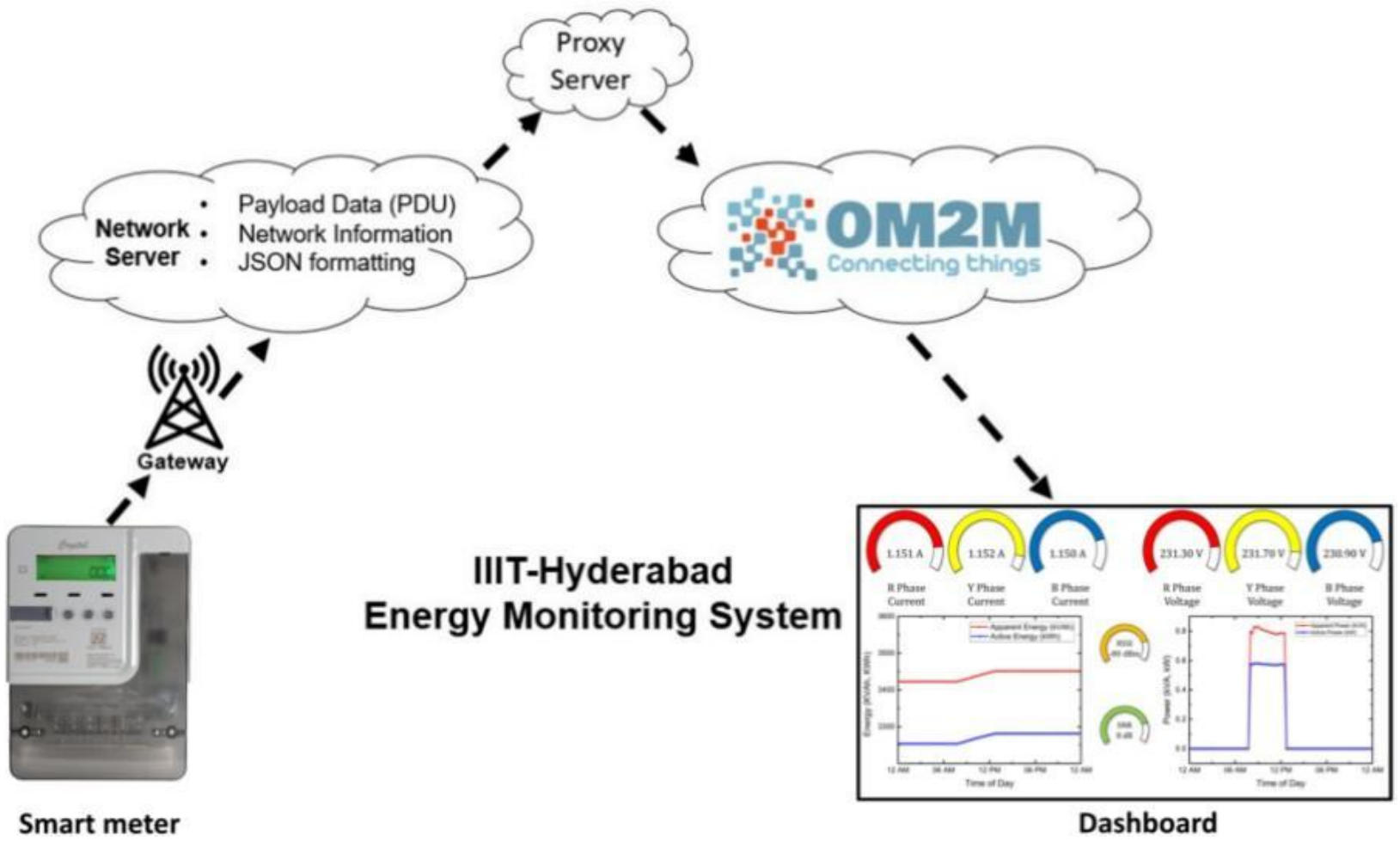

Figure 3.1: Illustration of the energy monitoring system implemented at IIIT Hyderabad.

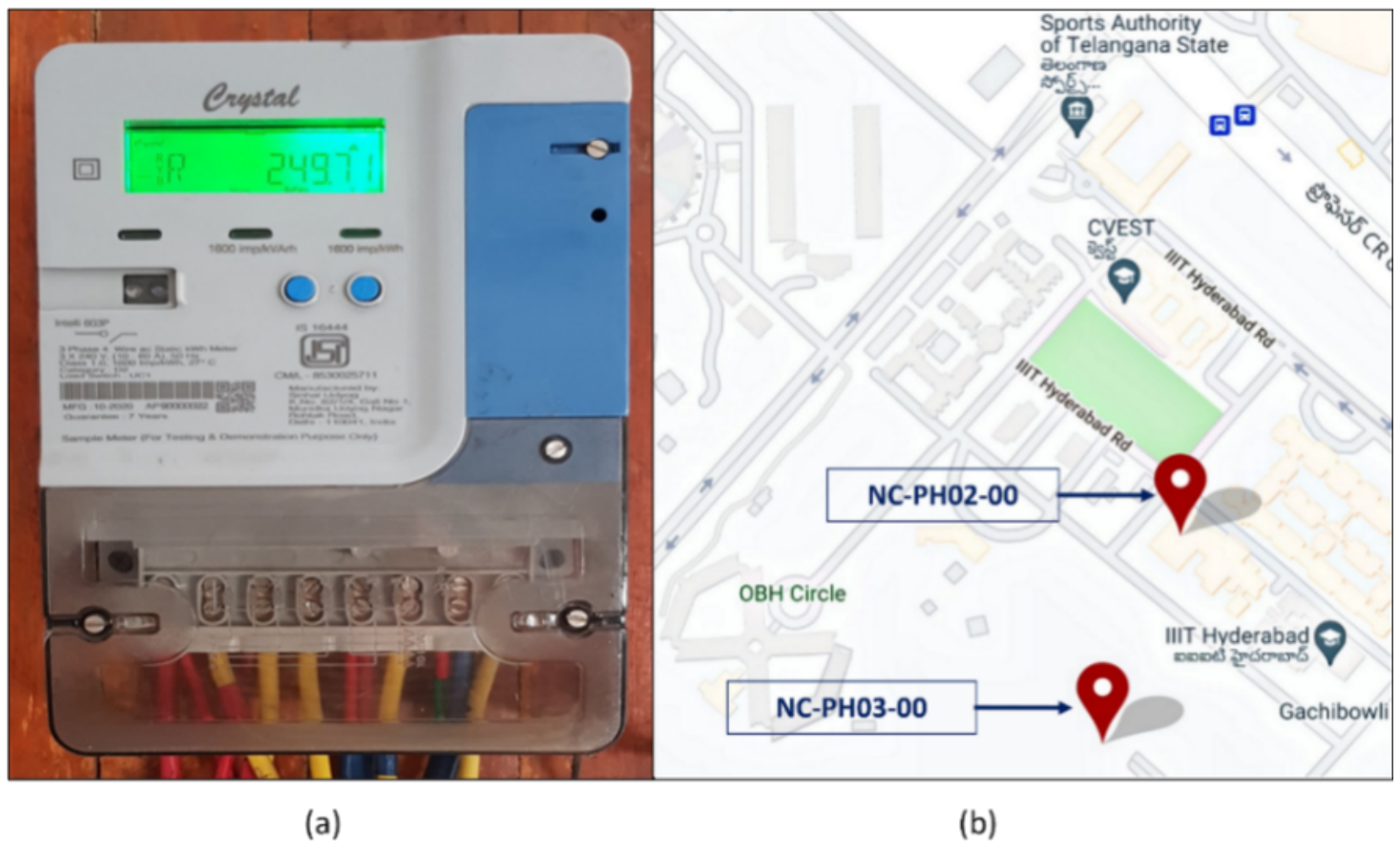

Figure 3.2: (a) Photograph of the deployed LoRaWAN-enabled smart energy meter. (b) Geolocation information of the two meters deployed.

The first step in establishing communication between the energy meter and the oneM2M platform is to register and activate the meter in the SenRa[1] network. Once activated, the meter starts transmitting an uplink message to multiple gateways that relay the message to the network server. On the network server, the data is formatted into a JSON object with multiple key-value pairs that hold the information such as the PDU or payload data, observation date and time, unique identifiers of the energy meters, transmit frequency, etc. Then, the packet with the best Signal to Noise Ratio (SNR) is forwarded to the proxy server. The hexadecimal bits of the PDU are converted into decimal values upon which the proxy server reformats the JSON object to comply with the oneM2M standards, and forwards it to OM2M. Finally, the dashboard application fetches the latest data from OM2M using standard REST APIs, for visualization. The proprietary format of the PDU contains a payload 46 bytes; each cluster of bytes representing an electrical parameter, as illustrated in Figure 3.3. The choice of 46 bytes depends on a plethora of factors including time-on-air (and hence duty cycle), the range provided, that further vary according to the values of Spreading Factor (SF), Coding Rate (CR), Data Rate (DR), etc. [69]. It also depends on the resolution of the phase currents, voltages, power factor (PF), power, and energy consumption, etc. as required by the IS 16444 standard. The conversion of the bytes in the PDU to the values of corresponding electrical parameters is detailed in Table 3.1.

[1]https://senraco.com/

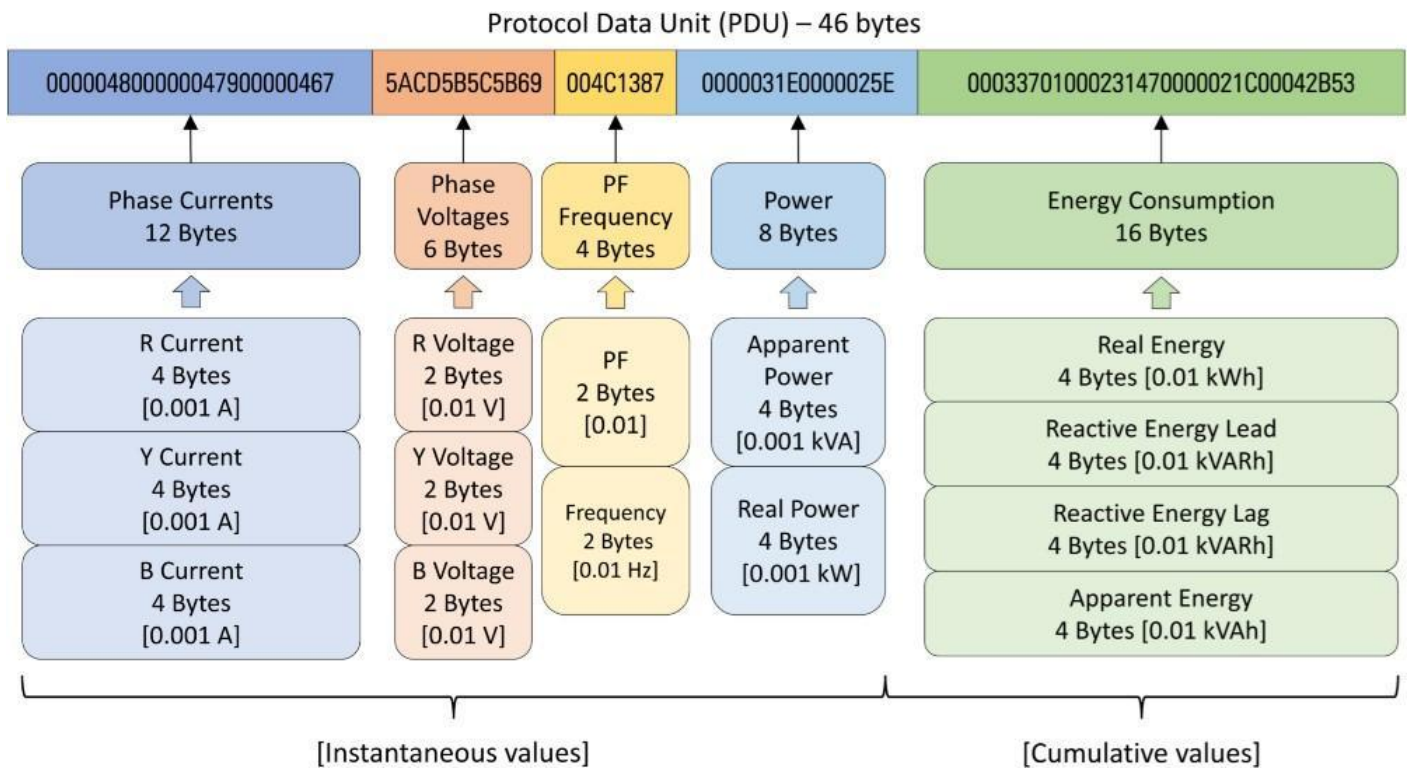

Figure 3.3: (The format followed for creating the LoRaWAN payload at the transmitter.

The data stored in the oneM2M platform can be subsequently used for various IoT solutions including graphic visualization in the form of a dashboard, real-time monitoring of energy consumption, designing algorithms for predictive maintenance of electrical appliances to improve performance, early fault detection to prevent failures, hence reducing the maintenance costs of the system.

### 3.1.3 Results and Discussion

The deployed meters have accumulated more than 10000 data points each, over a period of four months. The live data can be viewed in the dashboard illustrated in Figure 3.4. These real-time values of phase currents and voltages can help in immediate fault detection in the loads. The live plots of instantaneous power and cumulative energy assist in estimating the time during which energy was consumed in a day. The Received Signal Strength Indicator (RSSI) and Signal to Noise Ratio (SNR) values can be set to trigger a retransmission of payload if they fall below certain thresholds, minimizing data loss.

Table 3.1: PDU decoding process

| **Energy Parameters** | ***Hexadecimal to Decimal*** | | |
|---|---|---|---|
| | ***Hex value*** | ***(Hex-to-Dec)/n*** | ***Final Value*** |
| R Current (A) | 00000480 | 1152/1000 | 1.152 |
| Y Current (A) | 00000479 | 1145/1000 | 1.145 |
| B Current (A) | 00000467 | 1127/1000 | 1.127 |
| R Voltage (V) | 5ACD | 23245/100 | 232.45 |
| Y Voltage (V) | 5B5C | 23388/100 | 233.88 |
| B Voltage (V) | 5B69 | 23401/100 | 234.01 |
| Avg PF | 004C | 76/100 | 0.76 |
| Avg Freq (Hz) | 1387 | 4999/100 | 49.99 |
| Power (kVA) | 0000031E | 798/1000 | 0.798 |
| Power (kW) | 0000025E | 606/1000 | 0.606 |
| Energy (kWh) | 00033701 | 210689/100 | 2106.89 |
| kVRh Lead (kVRh) | 00023147 | 143687/100 | 1436.87 |
| kVRh Lag (kVRh) | 0000021C | 540/100 | 5.4 |
| Energy (kVAh) | 00042B53 | 273235/100 | 2732.35 |

RSSI values between -30 dBm and -120 dBm are considered ideal in LoRaWAN, with values below -120 dBm indicating potential packet loss. The RSSI values of all the transmissions from a meter over the four months of deployment have been analysed to conclude that approximately 83% of them were in this ideal range, as illustrated in Figure 3.5.

The access to daily energy consumption in both kilo Watt hour (kWh) and kilo Volt-Ampere hour (kVAh) is one of the pivotal advantages of smart meters over traditional energy meters. This can help eliminate monthly average-based pricing in favor of dynamic, demand-based pricing and effectively reduce costs incurred by end consumers. The flexibility of tracking both kWh and kVAh usage along with real-time monitoring of PF values is also a vital advantage in industrial deployments. We were able to perform a long-term analysis of the energy consumption data from one of the meters as shown in Figure 3.6 thus minimizing power loss and reducing bills.

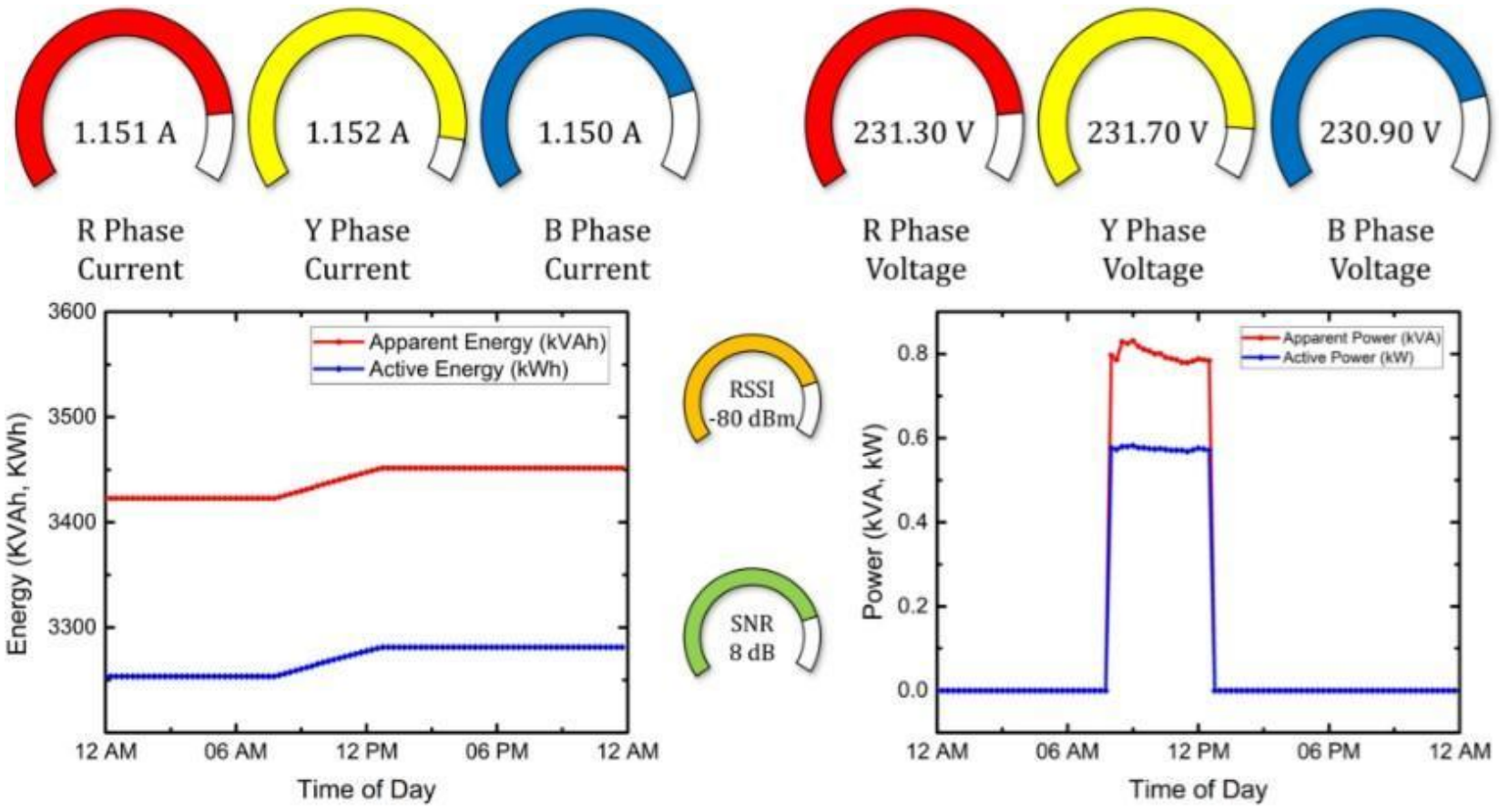

Figure 3.4: A live dashboard to monitor the electrical parameters and signal strength in real-time.

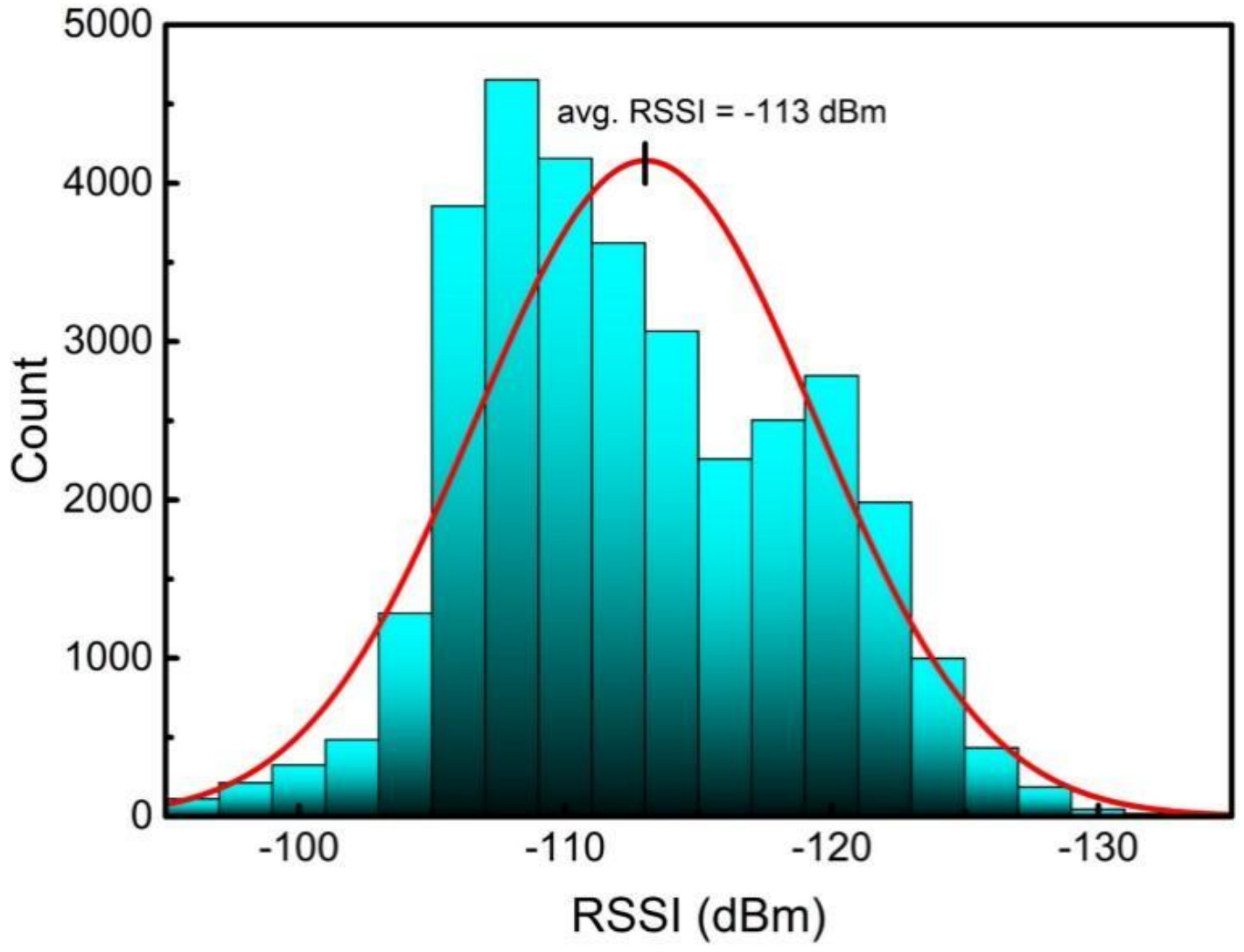

Figure 3.5: A histogram of all the RSSI values from energy meter NC-PH02-00, over the 4-month period of deployment.

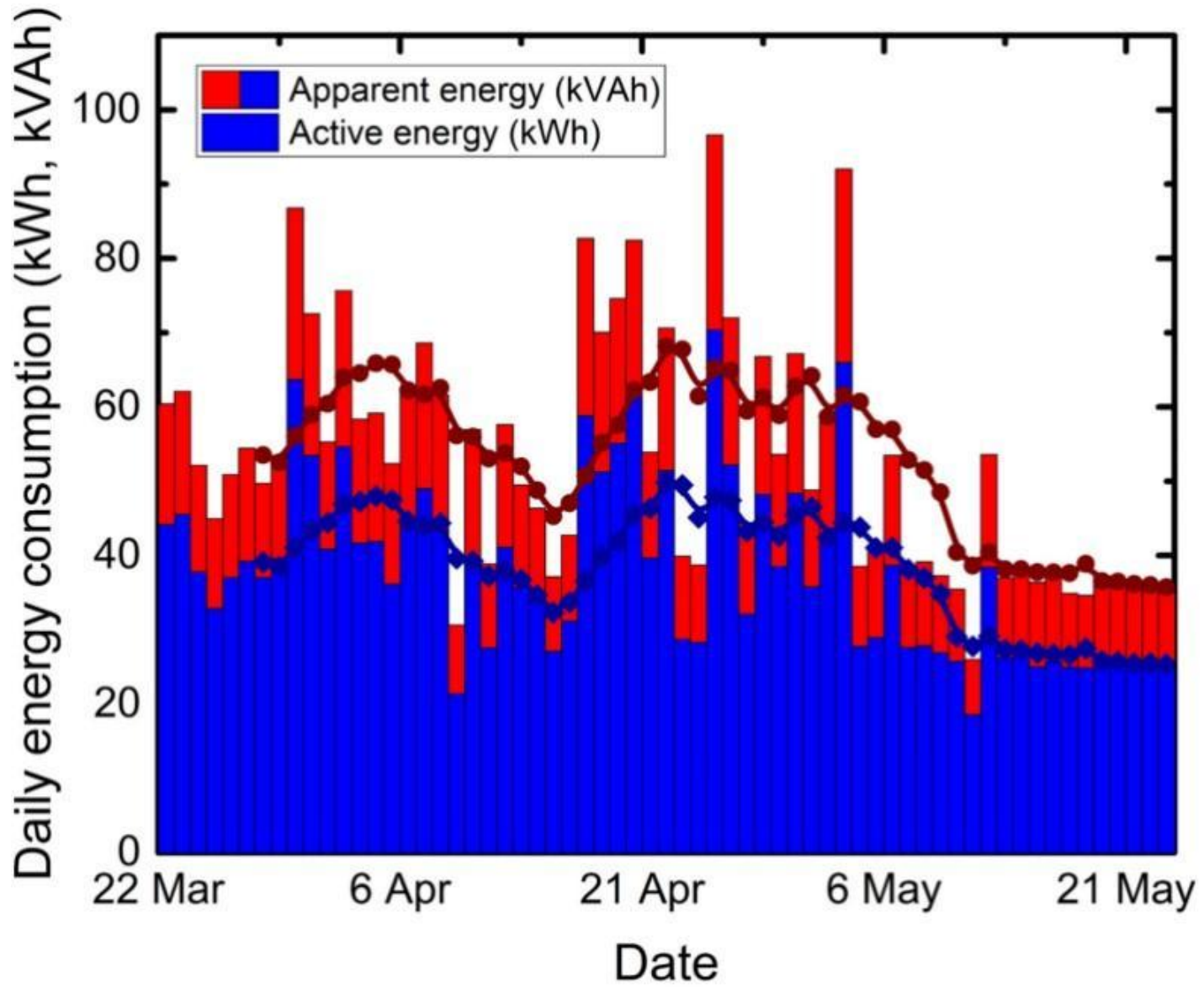

Figure 3.6: Daily energy consumption from 23rd March to 23rd May 2021. The line graphs indicate the 7-day moving average of active and apparent energy.

### 3.1.4 Summary

This work proposes a novel and scalable real time energy monitoring system employing LoRaWAN- enabled smart meters and a oneM2M service platform. A proprietary format was used for the LoRa PDU that contained 46 bytes of payload denoting values of phase currents and voltages, power, and energy consumption, etc. The data accumulated over the four months of deployment was analysed to detect faults, minimize power loss, and reduce bills. A dashboard is proposed for remote monitoring of live electrical parameters via a compatible HTTP browser.

## 3.2 Integration of OCPP based EV charger and oneM2M platform

### 3.2.1 Background and Related Work

Smart City is a technologically advanced urban area where a wide range of electronic and digital technologies with enhanced functional capabilities are used to improve the quality of life of the residents. The Government of India has already launched the Smart City Mission to develop 100 smart cities in the country. The current situation of Indian cities is improving day by day due to the addition of more and more capable smart devices. Electric vehicles (EVs) and their charging infrastructure are new entrants to the list of smart city technologies. However, with the growing population now preferring to adopt the various EVs available in the market, there is an urgent need for interoperable charging infrastructure capable of charging different EVs manufactured by different manufacturers. Developing such an interoperable charging station requires both reliable hardware design and efficient software implementation that will be possible with the help of standardization. The Open Charge Point Protocol (OCPP), an initiative by Open Charging Alliance (OCA), an international affiliate organization [70]. OCPP standard defines a method that enables charging stations to interact uniformly with a central system [71] and enables interoperability.

However, the heterogeneous devices in the Smart City have created the need for interoperability between EV chargers, their charging stations and other heterogeneous devices. As described in Chapter 1, oneM2M standards provide a common service layer that acts as a middleware between devices, com- munication protocols, and cloud and commercial applications that ultimately enable IoT interoperability. Therefore, we have developed a system that integrates OCPP based EV charging station [72] and OM2M platform based on oneM2M standards as a central management system for data storage and retrieval. This integration enables seamless data exchange between various IoT verticals at the city level with a clear semantic context of the data. OCPP establishes communication between the charging system (charge point) and the central management system (CMS) [73, 74]. The CMS is the cloud-based backend system that operates the charging process [75]. Thus, OCPP is used to interface between the OM2M platform and the charge point (charger).

### 3.2.2 Integration Workflow

The charging process is divided into three sub-processes: user authentication, electricity consumption based billing and final status update.

- **User Authentication Process:** The authentication process begins by verifying whether the user is already registered on the system with the help of a unique Radio-frequency identification (RFID) provided by the service provider. To initiate the charging process the user must swipe the RFID and enter the amount required for charging. As soon as the charger at the charging station receives the RFID, it sends a GET request to the OM2M platform to retrieve the information associated with it. If it cannot retrieve information because either the user is not already registered or the RFID is invalid, the platform sends an error message to the charger saying, ”user not found” and the same will be displayed on the charger’s screen. In this case, the user must first register and get a valid RFID and only then can he/she proceed for charging. In the case of successful user authentication, i.e., if the user is already registered and has used a valid RFID, the charger retrieves information and compares the user’s e-wallet balance. If it exceeds the required amount, the charging process begins and the power supply to the electric vehicle is initiated; Otherwise, an “Insufficient amount in the account” message will be displayed on the screen of the charger. However, if authentication is not approved, and the user does not proceed for registration, the charger returns to the default welcome screen.

- **Electricity Consumption Based Billing Process:** The billing process begins when the user is successfully authenticated. The cost per unit of power consumption differs according to the region and time of the day and hence the charger first retrieves and stores the current cost information to avoid incorrect billing and then it starts the charging process. Once charging has stopped, the charger first checks to see if the full amount has been charged. E.g. If the user enters Rs 100 and the vehicle consumes only Rs 90 equivalent electricity, the charger deducts Rs 90 from the user's wallet even if he initially enters Rs 100. Whereas, if the user enters Rs.100 and the vehicle consume Rs.110 then the charger deducts Rs.110 from the user's wallet. In the latter case if the user has only 100 rupees in the wallet, then the extra 10 rupees are stored as negative balance i.e., whenever user recharges the wallet, those 10 rupees will be automatically deducted. However, in the event of an emergency or sudden power outage, the charging station's processor works on an in-built battery backup to prevent data loss. Later, the information of the amount of completed charging is updated on the OM2M server according to usage.

- **Status Update Process:** When the charging process is completed, the charger generates two JSON objects using the Python3 JSON library's dumps () method which contain transaction related information. One JSON object with the user details is sent to the user's TRANSACTION container and another JSON object with the charger details is then sent to the charger's TRANSACTION con- tainer of the AE-EV-Chargers application entity on the OM2M platform. This way the information is updated on the cloud for further use.

Figure 3.7 illustrates the workflow of the integration of OM2M (oneM2M platform) with the EV charging station.

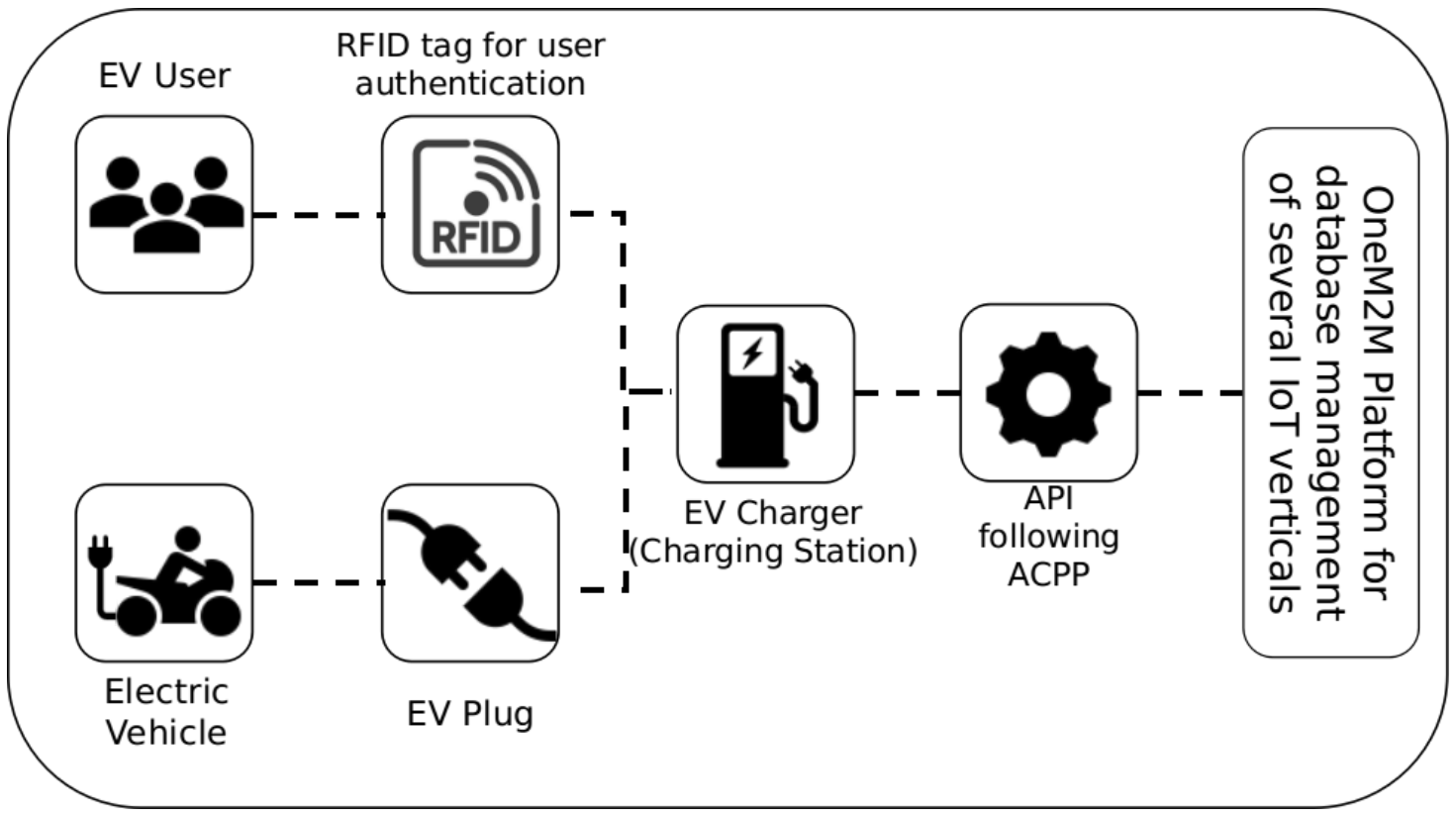

Figure 3.7: (Illustration of workflow of the integration of EV charging station and oneM2M platform.

### 3.2.3 Implementation

Figure 3.8 illustrates the structure of the EV charger AE on the OM2M platform's webpage, as one of the AEs among various smart city AEs.

In the first layer of AE, two containers (USER-DATA and CHARGER-DATA) are created to separately store the users' and chargers' (charging station) information. The second layer of containers is created to add the users' and chargers' identity information. i.e., a container WXYZ (where WXYZ is the number obtained from RFID tag) would be created under the USER-DATA container during user registration step and a container CHARGER-X (X is the charger number, Ex. CHARGER-1) would be created whenever a charging station id added to the system. The RFID tag fetched during the user authentication process is used to verify the already created container on the OM2M platform i.e., If container is already created then it means user is already registered otherwise the platform throws an error that the "user not found" (HTTP error 404).

In the final layer, user and charger information is stored along with the transactions. i.e., whenever a user finish charging, then the charging amount is subtracted from the previous balance at the charging station end. Further the updated balance is sent to the OM2M platform where the platform creates new content instances under user and charger TRANSACTIONS containers. The charger will then use the updated transaction's information during the next transaction cycle to check if the user has enough balance to charge the vehicle.

Figures 3.8 3.9 and 3.10 illustrate the screenshots of AE-EV-Charger, user's and charger's identity and transactions information respectively that are implemented on the OM2M platform.

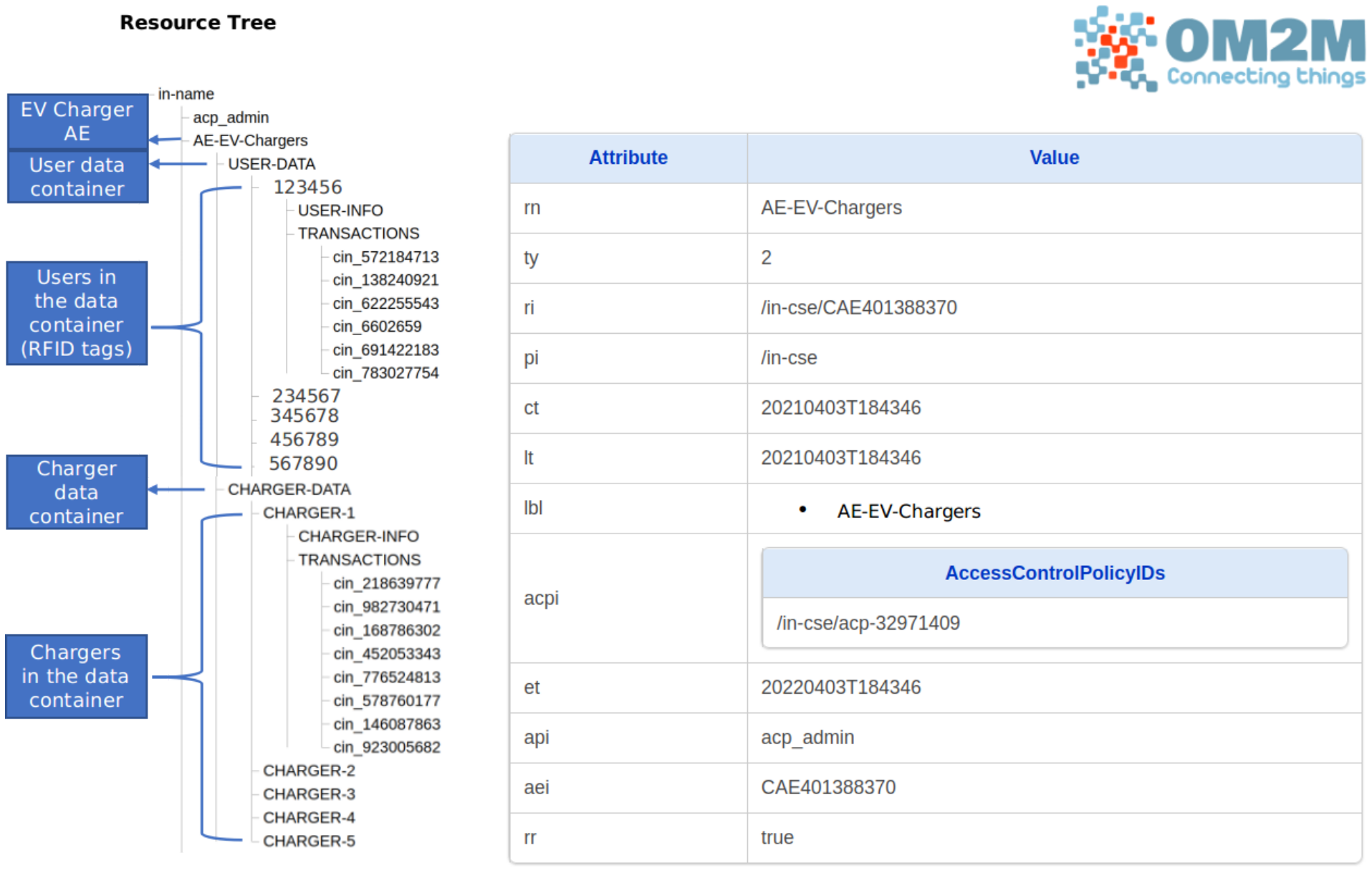

Figure 3.8: Illustration of the of the EV charger AE implemented on the OM2M platform.

**User Information**

| Attribute | Value |
|---|---|
| NAME | John Doe |
| EMAIL ID | abc@xyz.com |
| PHONE NUMBER | (+91)-1234567890 |

**Charger Information**

| Attribute | Value |
|---|---|
| CHARGER ID | Charger - 1 |
| GEO-LOCATION | {'Latitude': 17.44599, 'Longitude': 78.35142} |

Figure 3.9: Illustration of the of the User and Charger information created as a CINs under USER-INFO and CHARGER-INFO containers respectively.

**User's Transaction Details**

| Attribute | Value |
|---|---|
| USER ID | test-user |
| METER ID | test-charger |
| TRANSACTION DATE-TIME | 2021-04-03 23:15:34.229987 |
| TRANSACTION AMOUNT (IN RS) | 1000 |
| CURRENT AMOUNT IN USER'S ACCOUNT (IN RS) | 9000 |

**Charger's Transaction Details**

| Attribute | Value |
|---|---|
| USER ID | test-user |
| METER ID | test-charger |
| TRANSACTION DATE-TIME | 2021-04-03 23:15:34.064004 |
| TRANSACTION AMOUNT IN RS | 1000 |

Figure 3.10: Illustration of the of the User and Charger information created as a CINs under respective TRANSACTIONS containers.

### 3.2.4 Conclusion and Future Scope

The integration of an OCPP based EV charger with a oneM2M platform enables the charge point to integrate and exchange information with other smart city infrastructure. The concept of smart city becomes enormously successful only when substantial number of devices interact with each other and utilize the data gathered by other devices. Hence, the future course of action will be scaling up the approach to develop a system that can handle large number of charge points across the city along with the data generated from all the various infrastructure elements in the smart city.

### 3.2.5 Summary

Electric vehicles (EVs) are becoming increasingly popular around the world. These EVs need to be recharged to allow the EV owners to use the vehicles and for this purpose EV charge points are becoming the new entry to the Smart City devices. However, the interaction of an EV charge point with the smart city infrastructure requires seamless information exchange across various verticals. This horizontal flow of information is enabled by making the charge point compliant with the oneM2M platform. Thus, this work presents the design and fabrication of an EV charge point based on the OCPP communication standard and complaint with the oneM2M platform.

*Chapter 4*

# Indian Urban Data Exchange

Chapter 3 discussed two applications developed as part of this work to demonstrate the ability of the oneM2M standards to accommodate new heterogeneous applications without disrupting existing systems. This chapter discusses the need and necessity of IUDX data exchange framework, followed by the integration procedure to further enhance oneM2M based system. It discusses why data exchange is an essential part of any smart system, how it should be developed and how it can be integrated with previously proposed DMS and DLS. The proposed integration helps add data exchange to enhance oneM2M-based systems rather than becoming a separate individual component. The proposed DES consists of Catalogue Server (CS), Authorization Server (AS), and Resource Server (RS), where CS and AS are centralized common servers maintained by the IUDX organization, and the data provider maintains RS. In this work, Smart Campus, the IIIT-H system acts as a data provider; thus, RS is proposed, developed and integrated as part of this work.

## 4.1 Background and Related Work

The Indian Ministry of Housing and Urban Affairs (MoHUA) launched the Smart City Mission in 2015. In the initial phase, the ministry had planned to convert 100 cities into smart cities. The aim of the mission is to build the cities of the future through several domains such as by improving the security, public transport, protecting the environment through proper disposal of waste and so on. To work on such a big Indian cities, combined efforts are required from government, industry, research institutes and dedicated young entrepreneurs who are willing to contribute. Moreover, each of these organizations operates differently and therefore requires standardization and open innovation to bring uniformity in the city development process. Each city has its own issues that need to be understood first. Therefore, the first step should be the collection of the data of current state of the cities and their residents in terms of transportation facilities, weather conditions and daily use of resources such as electricity, water, natural gas, etc. Various government departments have been collecting such data for many years; however the heterogeneous data collection procedures by different departments inevitably leads to data silos which creates a difficulty for analysis. which created a data silo problem and made it difficult for data analysts to access and analyse the data. Thus, the proposed approach in Chapter 1 to develop a DMS and DLS based on oneM2M standards will ease the data collection process. The collected data can be sent to data analysts to develop applications through a common data exchange. This leads to the emergence of a data exchange. Various cities such as Columbus [76], Copenhagen [77] and Manchester [78] have developed their own data exchanges in the form of software platforms. Those platforms provide data in a common uniform format that can be easily shared among various departments as well as private data analytics firms as per the requirements. While collecting the data related to the city residents, the data privacy must be ensured and hence a data exchange allows the data owners to define a controlled access of the data to only authorised data users. In a similar way, the IUDX framework is emerged that is comprising of a set of NGSI-LD based APIs, data models, data privacy and security mechanisms. As a part of the Smart Campus, IIIT-H system, a RS has been proposed and developed to enable seamless data exchange between IIIT-H data storage and data users across the country. Moreover, the IIITH-RS is linked with the IUDX CS where the information of the data sources of all the India cities are stored commonly and linked with the IUDX AS that allows data providers to provide limited access control policies to registered data users. Further, five different data models have been proposed and developed with the help of the IUDX vocabulary to standardised data models for five different Smart City applications. Figure 4.1 illustrates the integration of the DMS, DLS with the RS, IUDX CS and AS.

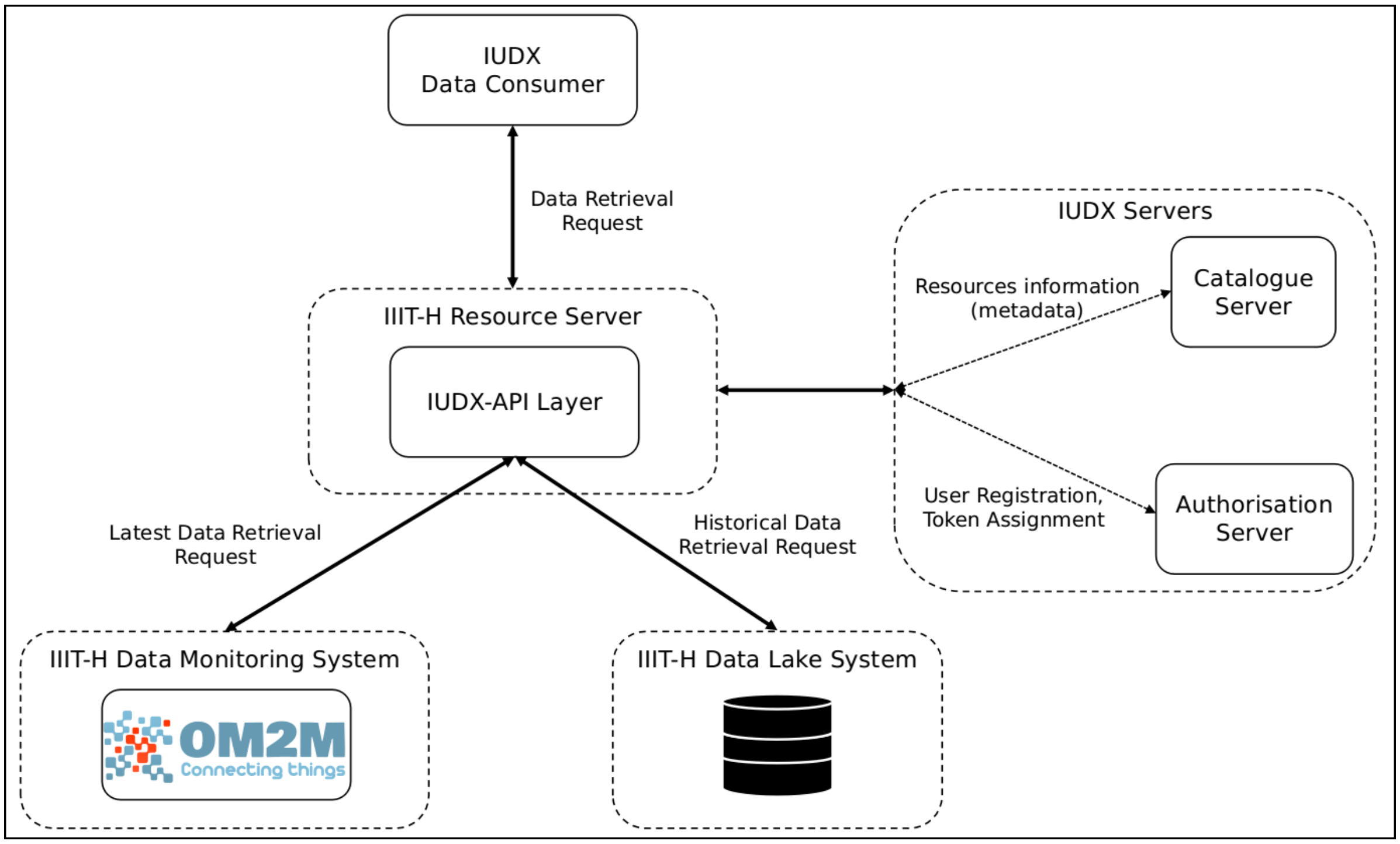

Figure 4.1: Illustration of IIIT-H Resource Server System Architecture.

## 4.2 System Architecture

### 4.2.1 IUDX Servers

As explained earlier, the IUDX organization maintains the CS and AS and thus, this section describes the working of the CS and AS.

#### 4.2.1.1 Catalogue Server (CS):

The CS[1] is used to store the metadata associated with the IUDX data sources. It uses the IUDX vocabulary [79], which is a shared vocabulary that describes data schemas for data exchange through IUDX data exchange APIs. The schema used to define this vocabulary is derived from RDF, a standard model for exchanging data across the web. This vocabulary uses the JSON for Linked Data (JSON-LD) 2, a format to encode linked data using JSON.

Various uniformly standardised data models have been published by IUDX [80] to enable data interoperability among city segments across multiple cities. With the help of standardised data models, all the resources belonging to the data model are described in JSON-LD format. The description is verified by the IUDX standardization body, and after successful verification it is onboarded onto the CS. The onboarded information is useful for the data users to understand the meaning of the data and also for the RS for resource ID verification where it verifies the information associated with a resource, such as the type of resource (open, secure), resource ID, resource group during the data retrieval process. A resource group is a collection of all the resource items that are associated with it, where a resource item refers to an IoT node and resource ID is the unique ID for each resource.

As a part of IIIT-H-RS development, five data models associated with five resource groups have been developed. A resource group is a collection of all the resource items associated to it, where a resource item refers to an IoT node. The resource groups associated with IIIT-H-RS are named as iiith-env-aqm (air quality monitoring), iiith-energy-meter (energy consumption monitoring), iiith-env-weather (weather monitoring), iiith-water-monitoring (water monitoring) and iiith-solar-panel (solar energy monitoring). Table 4.1 presents the catalogue webpage URLs of five resource groups. Figure 4.2 illustrates the description of parameters such as type of resource, resource ID, name, label, tags (parameters), detailed description, and geo-location co-ordinates of the resource item AQ-SN00-00 which belongs to the iiith-env-aqm resource group. The remaining data sources can be found CS [3] webpage.

#### 4.2.1.2 Authorization server (AS):

The AS is responsible for allowing data customers to access data authorised by the data provider in accordance with the data policy. This enables the data sharing in a controlled manner. With the help of a set of APIs provided by the AS, the data provider can create and modify access-controlled policies associated with the secure data resources. Similarly, by using another set of APIs provided by the AS, the data users can get the access tokens that are necessary to retrieve data from a RS.

[1]https://catalogue.iudx.org.in/
[2]https://json-ld.org/
[3]https://catalogue.iudx.org.in/datasets

Table 4.1: Catalogue webpage URLs of five resource groups

| Resource Group | Webpage URL |
|---|---|
| iiith-env-aqm | https://catalogue.iudx.org.in/dataset/research.iiit.ac.in-4786f10afbf48ed5c8c7be9b4d38b33ca16c1d9a-iudx-rs-onem2m.iiit.ac.in-iiith-env-aqm |
| iiith-energy-meter | https://catalogue.iudx.org.in/dataset/research.iiit.ac.in-4786f10afbf48ed5c8c7be9b4d38b33ca16c1d9a-iudx-rs-onem2m.iiit.ac.in-iiith-energy-meter |
| iiith-water-monitoring | https://catalogue.iudx.org.in/dataset/research.iiit.ac.in-4786f10afbf48ed5c8c7be9b4d38b33ca16c1d9a-iudx-rs-onem2m.iiit.ac.in-iiith-water-monitoring |
| iiith-solar-panel | https://catalogue.iudx.org.in/dataset/research.iiit.ac.in-4786f10afbf48ed5c8c7be9b4d38b33ca16c1d9a-iudx-rs-onem2m.iiit.ac.in-iiith-solar-panel |

IUDX divides resources into two types, open and secure resources. An open resource is a resource whose data doesn't affect any individual's privacy such as an air quality node deployed in an open ground. A JWT required for this resource's data retrieval can be retrieved by any registered data user. A secure resource is a resource that collects data which may belong to an individual or a group of individuals, for example, an energy meter that monitors the daily energy consumption of a government office.

The data retrieval process of these resources is based on the OAuth 2.0 framework. It is a framework that enables third-party applications to obtain limited access to a resource. The access mechanism for open and secure resources is different. Initially the data user needs to be registered on IUDX AS as a data consumer. Upon successful registration, the user is allowed to get a token for open resource's data retrieval. Whereas, in case of a secure resource, the registered user must first make the authorization request to the data provider for data retrieval access. Once the data provider sets the required policy on AS to grant the access to the user, the user can then get a token and retrieve the data. Figure 4.3 illustrates the format of JWT request's body for open and secure resources belong to

IIIT-H-RS. As illustrated in the

```
{
    "@context": "https://voc.iudx.org.in/",
    "type": [
      "iudx:Resource",
      "iudx:EnvAQM"
    ],
    "id": "research.iiit.ac.in/4786f10afbf48ed5c8c7be9b4d38b33ca16c1d9a/iudx-rs-onem2m.iiit.ac.in/iiith-env-aqm/AQ-SN00-00",
    "name": "AQ-SN00-00",
    "label": "Air Quality node 1 at Ground Floor of Sahana Atidhi Nivas",
    "description": "Air Quality node 1 at Ground Floor of Sahana Atidhi Nivas IIIT Hyderabad, publishing instantaneous values of air pollutant measures every fifteen seconds",
    "tags": [
      "environment", "air quality", "air", "aqi", "aqm", "pollution", "amonia", "carbon", "carbon monoxide",
      "nitrogen dioxide", "co", "no2", "pm2.5", "pm10", "humidity", "temperature", "nh3"
    ],
    "location": {
      "geometry": {
        "coordinates": [
          78.347483,
          17.445604
        ],
        "type": "Point"
      },
      "type": "Place",
      "address": "Ground Floor of Sahana Atidhi Nivas, IIIT Hyderabad"
    },
    "provider": "research.iiit.ac.in/4786f10afbf48ed5c8c7be9b4d38b33ca16c1d9a",
    "resourceGroup": "research.iiit.ac.in/4786f10afbf48ed5c8c7be9b4d38b33ca16c1d9a/iudx-rs-onem2m.iiit.ac.in/iiith-env-aqm",
    "itemStatus": "ACTIVE",
    "itemCreatedAt": "2021-08-03T07:06:42+0530"
}
```

Figure 4.2: Illustration of the description of the AQ-SN00-00 resource item that belongs to the iiith-env- aqm resource group.

Figure 4.3 (a), the registered user only needs to get JWT for the resource server and the same JWT can be used to retrieve data for all open resources of a specific RS. And the data user needs to get a separate token for each secure resource as illustrated in Figure 4.3 (b). In case of IIIT-H-RS, three resource groups are set as secure and two as open and Figure 4.4 illustrates the entire flow of API requests in the data retrieval process.

| Token request body for open resource | Token request body for secure resource |
|---|---|
| { "itemId": "iudx-rs-onem2m.iiit.ac.in", "itemType": "resource_server", "role": "consumer" } | { "itemId": "research.iiit.ac.in/4786f10afbf48ed5c8c7be9b4d38b33ca16c1d9a/iudx-rs-onem2m.iiit.ac.in/iiith-energy-meter", "itemType": "resource_group", "role": "consumer" } |
| (a) | (b) |

Figure 4.3: Illustration of the format of a get JWT request's body for (a) open and (b) secure resource.

A valid JWT received from IUDX AS contains information such as whom the token refers to (sub), token issuer (iss), for whom the token is intended for (aud), assign (iat) and expiration (exp) time, resource id (iid), token requester's role (role) and request type access (cons). An encoded and decoded JWT according to IUDX format is shown in Figure 4.5.

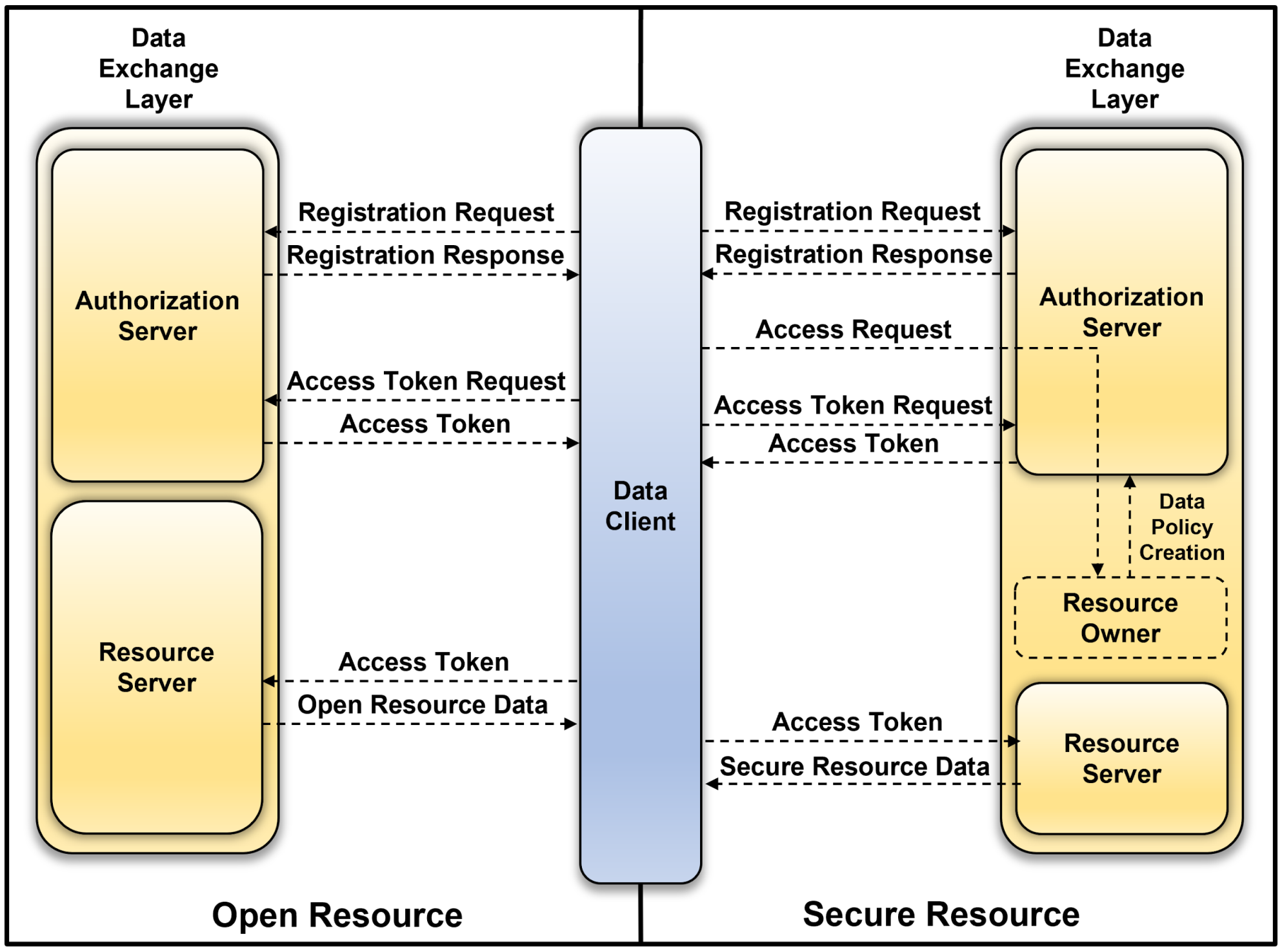

Figure 4.4: Illustration of the API requests' flow in the Oauth 2.0 security framework.

- **Encoded Information:**

"token":
"eyJ0eXAiOiJKV1QiLCJhbGciOiJFUzI1NiJ9.eyJzdWIiOiI1Yjk0NWNmZi0xOTk0LTQ1MjEtODgyYS05NmQ
wZDI1MGNjMGEiLCJpc3MiOiJhdXRodmVydHguaXVkeC5pbyIsImF1ZCI6Iml1ZHgtcnMtb25lbTJtLmlp
aXQuYWMuaW4iLCJleHAiOjE2NDc3NDE2NDAsImlhdCI6MTY0NzY5ODQ0MCwiaWlkIjoicnM6aXVkeC
1ycy1vbmVtMm0uaWlpdC5hYy5pbiIsInJvbGUiOiJjb25zdW1lciIsImNvbnMiOnt9fQ.vGWUtR-
kHvUTBxl2q_HtJyz1oxhOrZTcWWhn3hSfIWGoK0V_OUgMCrlL99JSvVOrxQTXIXUZdEUEZs7dLdMdiw"

- **Decoded Information:**

{
"sub": "5b945cff-1994-4521-882a-96d0d250cc0a",
"iss": "authvertx.iudx.io",
"aud": "iudx-rs-onem2m.iiit.ac.in",
"exp": 1647719267,
"iat": 1647676067,
"iid": "rg:research.iiit.ac.in/4786f10afbf48ed5c8c7be9b4d38b33ca16c1d9a/iudx-rs-onem2m.iiit.ac.in/iiith-energy-meter",
"role": "consumer",
"cons": {
"access": [
"apis"
]
}
}

Figure 4.5: Illustration of a JSON Web Token (JWT) in encoded and decoded format.

#### 4.2.1.3 IIIT-H Resource Server:

The RS is proposed as a part of Smart Campus, IIIT-H system to uniformly provide the data to the data users across the country. A detailed description of this procedure is given below.

As illustrated in Figure 4.4, when a HTTP GET request for data retrieval is sent by the data user reaches the IIIT-H-RS, it first passes through the authentication check mechanism and then depending on the type of request it is redirected to either the OM2M server or data lake. The user must attach a valid token with the GET request and if the token is not present or is invalid, the request will be discarded, and the following response will be returned.

```
{
   "type": "urn:dx:rs:InvalidAuthorizationToken",
   "title": "Not Authorized",
   "detail": "Token is invalid"
}
```

Figure 4.6: Response received by the user for invalid JWT.

Once the token is successfully verified, depending on the type of API request, the data is retrieved. In case of IIIT-H-RS, for data retrieval, four types of APIs are proposed and developed, they are, metadata API, latest data API, temporal data API and a revoke token API.

- **Metadata API:** It is proposed to provide the metadata associated with an IoT node such as type of sensors used, node location, units of the measured parameters, etc. Such metadata helps data users to properly understand the meaning of the data and leads to the easy application development process. Figure 4.7 illustrates the metadata APIs response of the air quality node, AQ-MG00-00, received by a user.

- **Latest data API:** It is proposed for applications that need latest observations made by the sensors. If the request is for the latest data retrieval, it is sent to the OM2M server to preserve data freshness. The data is then retrieved, reformatted as per the IUDX data model and sent back to the requestor. Figure 4.8 illustrates the successful latest data response of the air quality node, AQ-MG00-00, received by a user.

- **Temporal data API:** It is proposed for applications that require n days of data. Three types of keywords are used in temporal API requests, before, to retrieve data points before a certain date-time, after, to retrieve data points after a certain date-time, and during, to retrieve data points from a start time to a certain end-time.

  As illustrated in the Figure 4.9, the temporal data response is an array of JSON objects of the individual data points, where one data point response is similar to the latest data response as shown in Figure 4.8.

```
{
  "title": "Successful operation",
  "type": "urn:dx:rs:success",
  "results": [
    {
      "id": "research.iiit.ac.in/4786f10afbf48ed5c8c7be9b4d38b33ca16c1d9a/iudx-rs-onem2m.iiit.ac.in/iiith-env-aqm-version/version-info",
      "deviceInfo": {
        "deviceID": "AQ-MG00-00",
        "deviceName": "Air Quality node 1 at Main Gate"
      },
      "versionInfo": [
        {
          "versionName": "V2.01.33",
          "startDateTime": "2020-10-10T10:00:00+05:30",
          "endDateTime": "2020-12-31T10:00:00+05:30",
          "versionSpec": {
            "pm2p5": "SDS011",
            "pm10": "SDS011",
            "airTemperature": "DHT22",
            "relativeHumidity": "DHT22",
            "co": "Multichannel Grove Gas Sensor",
            "no2": "Multichannel Grove Gas Sensor",
            "nh3": "Multichannel Grove Gas Sensor",
            "controller": "ESP8266"
          },
          "comments": "comment on version change"
        },
        {
          "versionName": "V3.00.02",
          "startDateTime": "2020-12-31T10:00:00+05:30",
          "endDateTime": "9999-12-31T23:59:59+05:30",
          "versionSpec": {
            "pm2p5": "SDS011",
            "pm10": "SDS011",
            "airTemperature": "DHT22",
            "relativeHumidity": "DHT22",
            "co": "Multichannel Grove Gas Sensor",
            "no2": "Multichannel Grove Gas Sensor",
            "nh3": "Multichannel Grove Gas Sensor",
            "controller": "ESP8266"
          },
          "comments": "comment on version change"
        }
      ]
    }
  ]
}
```

Figure 4.7: Illustration of the metadata APIs response of the air quality node, AQ-MG00-00.

Sometimes, depending on the requirement, a user might ask for data of past one year and the response may contain millions of data points. This type of request suddenly increases the load on the RS and may result into a server crash. To avoid such situation, two types of limits are proposed in temporal data response, first is the maximum date difference between start date and end date as 10 and maximum data points limit per temporal data request as 2000. This limit may be an issue if there are more than 2000 data points during the mentioned period and hence the offset and totalHits are added to the response. The offset helps to shift the starting point of data retrieval and totalHits helps to explain the total number of data points between the mentioned period. Ex. If there are 3000 data points between the mentioned period, then from the first request, the requester will receive 2000 data points and the totalHits as 3000. totalHits explains the requester that the user must send another request with offset = 2000, where the second request's response will contain 1000 data points and in this way the requester will be able to retrieve all 3000 data points as well as prevent sudden increase of load on the server.

```
{
  "title": "Successful Operation",
  "type": "urn:dx:rs:success",
  "results": [
    {
      "id": "research.iiit.ac.in/4786f10afbf48ed5c8c7be9b4d38b33ca16c1d9a/iudx-rs-onem2m.iiit.ac.in/iiith-env-aqm/AQ-MG00-00",
      "pm2p5": {
        "instValue": 24.2
      },
      "pm10": {
        "instValue": 103.4
      },
      "airTemperature": {
        "instValue": 45.8
      },
      "relativeHumidity": {
        "instValue": 14.74
      },
      "co": {
        "instValue": "nan"
      },
      "no2": {
        "instValue": "nan"
      },
      "nh3": {
        "instValue": "nan"
      },
      "airQualityIndex": "102.27",
      "airQualityLevel": "POOR",
      "aqiMajorPollutant": "PM10",
      "observationDateTime": "2022-03-21T00:25:06+05:30",
      "versionInfo": {
        "versionName": "V3.0.02"
      }
    }
  ]
}
```

Figure 4.8: Illustration of the latest data response of the air quality node, AQ-MG00-00.

```
{
  "title": "Successful Operation",
  "type": "urn:dx:rs:success",
  "results": [
      {
      },
      {
      },
      {
      }
  ],
  "limit": 2000,
  "offset": 0,
  "totalHits": 3
}
```

Figure 4.9: Illustration of the temporal response with an array of JSON objects of individual data points.

Apart from the limit and offset, two more filtering techniques are proposed which are, attribute-based filtering where the requester can retrieve data of only one attribute for example only pm10 values in case of an air quality node, and value-based filtering where a requester can retrieve all the data points where temperature is above 40 degrees. This type of filtering reduces the data parsing efforts on the requester's side by allowing the requester to retrieve the data as required. Moreover, it saves the bandwidth by reducing the response size. Table 4.2 illustrates a list of the attributes of the filtering-based temporal data retrieval request's parameters and the Figure 4.10 illustrate its response.

Table 4.2: Access control policies for an admin and guest user

| **Attribute** | **Value** |
|---|---|
| time relation | during |
| start time | 2022-01-12T00:00:00Z |
| end time | 2022-01-12T00:01:00Z |
| attributes | pm2p5, observationDateTime |
| resource id | research.iiit.ac.in/4786f10afbf48ed5c8c7be9b4d38b33ca16c1d9a/iudx-rs-onem2m.iiit.ac.in/iiith-env-aqm/AQ-AN00-00 |
| query | pm2p5>30.00 |

```
{
   "title": "Successful Operation",
   "type": "urn:dx:rs:success",
   "results": [
        {
         "pm2p5": {
            "instValue": 31.2
         },
         "observationDateTime": "2022-01-12T11:24:43+05:30"
      },
      {
         "pm2p5": {
            "instValue": 30.6
         },
         "observationDateTime": "2022-01-12T11:22:43+05:30"
      },
      {
         "pm2p5": {
            "instValue": 30.7
         },
         "observationDateTime": "2022-01-12T11:22:28+05:30"
      },
      {
         "pm2p5": {
            "instValue": 31.3
         },
         "observationDateTime": "2022-01-12T11:22:12+05:30"
      }
   ],
   "limit": 2000,
   "offset": 0,
   "totalHits": 4
}
```

Figure 4.10: Illustration of the filtering-based temporal data retrieval request's response.

- **JWT Revoke API:** It is proposed to revoke all the JWTs that are generated before the time of revoke API request. It is useful in a scenario where, a user generates a token but somehow loses it or if some unauthorised user tries to use it without a prior permission, then in such case the revoke token API allows the user to revoke such token's access to prevent its misuse. The user must send a request to the IUDX authorization server to revoke the token. First, the authorization server verifies the request. Upon successful verification, the authorization server automatically sends the request to IIIT-H-RS with the information of the requester's ID. At IIIT-H-RS, the request is first verified and after successful verification, the ID of the requestor and the time of request information is stored. Then onwards, if a token belongs to the same user that was generated before the time of revoke token API request is received by IIIT-H-RS, it is automatically discarded and thus it prevents the reuse of revoked token.

## 4.3 Implementation

The IIIT-H-RS is programmed in Python and implemented using Django's REST API framework. Apache, an open-source, cross platform web server is used to host Django along with the mod wsgi, a python WSGI package that provides an Apache module for the communication between Django and Apache. Figure 4.11 illustrates the communication between data user (requester), Apache web server, WSGI and Django. The GZIP, a module provided by the Apache web server, is also integrated into the IIIT-H-RS implementation. The module compresses the response before sending it to the user. Such compression helps in reducing the response size which decreases API data retrieval latency. After the integration of the module, the data retrieval time is reduced to less than 300 milliseconds for the latest data and less than 1 minute for temporary data retrieval (2000 data points).

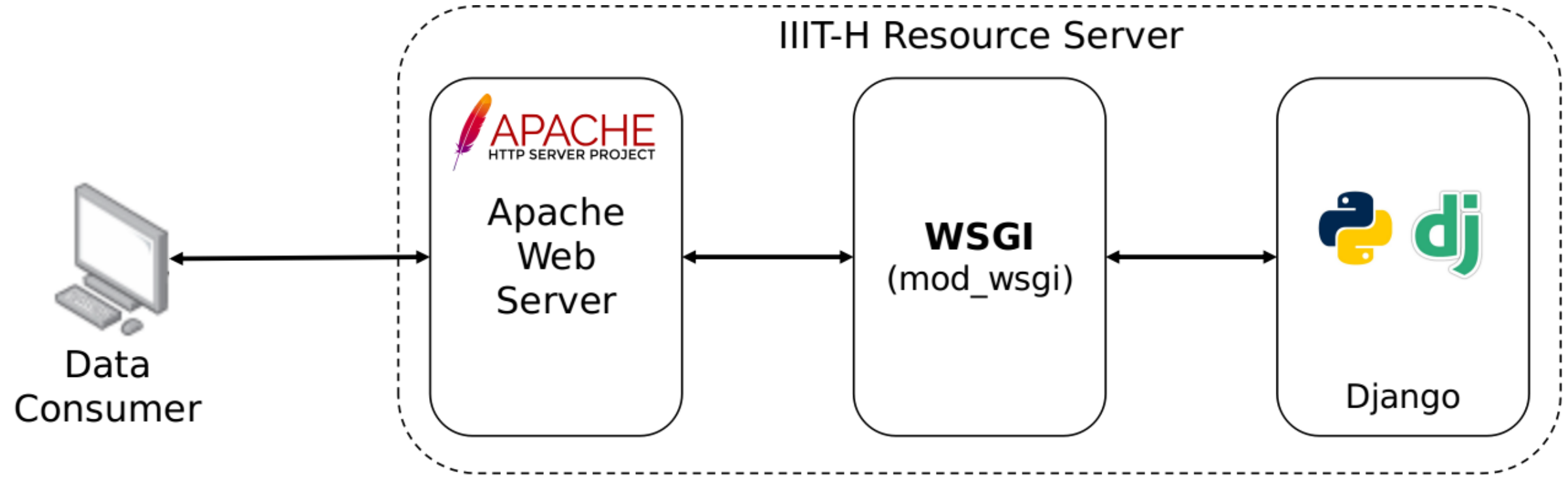

Figure 4.11: Block diagram of the implementation of IIIT-H-RS.

## 4.4 Performance Assessment

Performance analysis has been conducted on the individual layers to evaluate the feasibility and sustainability of the proposed multi-layer architecture. The implementation is assessed based on the number of parallel users supported and latency in milliseconds. The experiments have been conducted on a computer with Intel Core-i5-8400 2.80 GHz processor, 8 GB of RAM, and a 64-bit Ubuntu 18.04 operating system.

As described in Section 4.2, the IIIT-H RS currently supports three different APIs for data exchange in the NGSI-LD standard. As shown in the Table 4.3, the performance analysis describes the efficiency under a twelve-hour test with a constant hundred parallel users. The temporal data with parameter filtering API outperformed the other APIs with much higher throughput.

Table 4.3: Throughput analysis of Data Exchange Layer

| | Type of API | | | | |
|---|---|---|---|---|---|
| | Latest | Temporal | Temporal with Parameter Filtering | Temporal with Limit-Offset | Meta-Info |
| Throughput | 3.37 | 1.91 | 4.46 | 4.79 | 82.17 |

## 4.5 Summary

The emerging needs of the smart city can be aligned not only to improve the quality of life but also to ensure sustainability and optimal utilization of resources by in-depth analysis of real-time data collected from a wide range of sensor networks. While the deployment of sensors on a scale of a smart city to monitor these parameters of interest is an enormous task, the subsequent phases of data sharing is equally challenging. Hence, a well-defined data exchange framework to perform the analysis is a vital requirement in any smart city. The framework should define standardized data models and structured APIs, thus avoiding data silo problems and allowing interoperability. Therefore, the DES is proposed which currently shares data from heterogeneous sensing nodes. The DES has been evaluated in terms of throughput, and the evaluation results show the feasibility and sustainability of the system.

*Chapter 5*

# 5D-IoT Framework

Chapter 4 discussed the proposed DES and its integration with DMS and DLS. This chapter discusses the proposed data quality assessment framework as integration to DMS, DLS and DES. In smart city systems, most of the sensing nodes are located in remote locations that can be damaged by harsh weather, mishandling, or malicious cyber attacks, thereby generating inaccurate data. Moreover, inconsistencies arise in a smart city due to numerous data sources transmitting data using different formats and objects. Therefore, the issue of data quality must be addressed before deploying any smart city system. Thus, this work proposes a 5D-IoT framework to assess data quality and help users understand data clearly.

## 5.1 Background and Related Work

Data is now openly available, thanks to the emerging IoT projects that provide data with the help of Web APIs. However, often open data from multiple data streams becomes difficult to process and analyse because of the heterogeneity in the data models and data descriptions. Such diversity leads to the problem of data interoperability, which leads to IoT fragmentation. It also becomes a significant concern for application developers who must understand the various data formats and the underlying communication network before developing any application. Various works have been proposed in the past to unify data models for multiple systems to standardize model creation [81]. However, every user might not be allowed to change such data models once standardized. After identifying this need, the World Wide Web Consortium (W3C) started a Web of Things (WoT) working group in February 2017. Their approach consists of a standardized data model called the RDF to characterize or model the information. In this model, every resource has a unique resource identifier (URI), and the data description is in a linked data format that consists of logical assertions of the form; Subject, Predicate, and object – known as a triple. A triple capture both entity attributes and relationships between entities, and a set of triples (triplestore) can describe anything describable. Each triple expresses a fact about a subject. For example, one of the parameters observed by a DHT11 sensor is a temperature value, which can be written in a triple format as:

| ⟨*Subject*⟩ | ⟨*Predicate*⟩ | ⟨*Object*⟩ |
| --- | --- | --- |
| http://my.data/DHT11 | http://my.data/measures | http://my.data/temperature |

In this way, the RDF model solves the data interoperability problem by developing, linking, and integrating IoT data from various IoT devices into a standardized framework. Moreover, the Semantic Web (SW) technologies and principles provides expressive vocabularies to describe data and manipulate resources. The domain at the interface between the SW and the IoT called the Semantic Web of Things (SWoT), where the smart agents interact with devices in the user's environment, has shown some practical achievements in recent years [82]. However, IoT systems have further problems such as additional data transmission delay, duplicate data generation, erroneous and null data transmission [83]. Application developers can avoid these problems by filtering each time they use the data which needs better understanding of entire IoT deployment, and the factors that affect the data quality; once they understand, then only, they can filter and use the data. Even for the data provider, it becomes difficult to check and improve the quality of data as the system grows. Therefore, to ensure the IoT data quality, above mentioned factors need to be assessed. W3C recommends SHACL, a standard language for RDF data validation. SHACL provides shapes, a collection of constraints that are applied on specific RDF resources and based on those constraints, a validation report is generated that describes whether the data graph conforms to the shapes graph, if not, it describes the reason of failures. In [84], authors identified this issue and developed a platform that maps each run-time monitoring observation from heterogeneous data streams with precise semantics of context, observed quantities, units, and provenance. Further, as an extension of the previously proposed framework, an annotation engine called LSane is proposed in [85], which enriches and validates the semantically mapped data before reaching the customers. The enrichment part was added primarily as part of the data transformation as required by the metadata repository. The engine uses SHACL constraints to identify and filter out invalid observations. However, the authors did not describe how the engine applies these constraints by considering the heterogeneity in IoT application; hence further research is needed. Later, in the paper [86], authors described data quality as a measure of trust when using heterogeneous data sources. Authors used phase index (PI) while assessing the quality, and based on the PI, it will be decided whether to terminate or continue the process to improve the trust. However, they did not mention how do they define the weight of each parameter and how to calculate the PI, and therefore it was not clear how they assess and improve the quality of the IoT data. The authors in [87] proposed an approach for data quality evaluation in smart city applications, where they consider data outliers and missing data issues that typically occur during data collection from IoT systems. However, they did not discuss other factors such as hardware failure, duplicate data transmission and additional delays that need to be evaluated. Therefore, a five-layered 5-D IoT framework is proposed, namely data enrichment, data duplicacy, data delay, data validation, and data storage. An overview of the proposed framework is illustrated in Figure 5.1. In the first layer, the raw data from heterogeneous IoT sources is uniformly enriched and converted into graph data based on RDF defined by semantic web principles. Such enriched data makes it easy for users or application developers to understand the meaning of data which leads to faster application development. The next three layers represent the novel approach of the data quality assessment using SHACL shapes, which assesses duplicate data transmission, transmission delays, and inaccurate data generation, respectively. The assessment results are then added to the data graph using the SHACL inference rule and our proposed vocabulary. Finally, the fifth layer consists of the RDF Triple Store, which stores the processed data and makes it available as a SPARQL endpoint to the users.

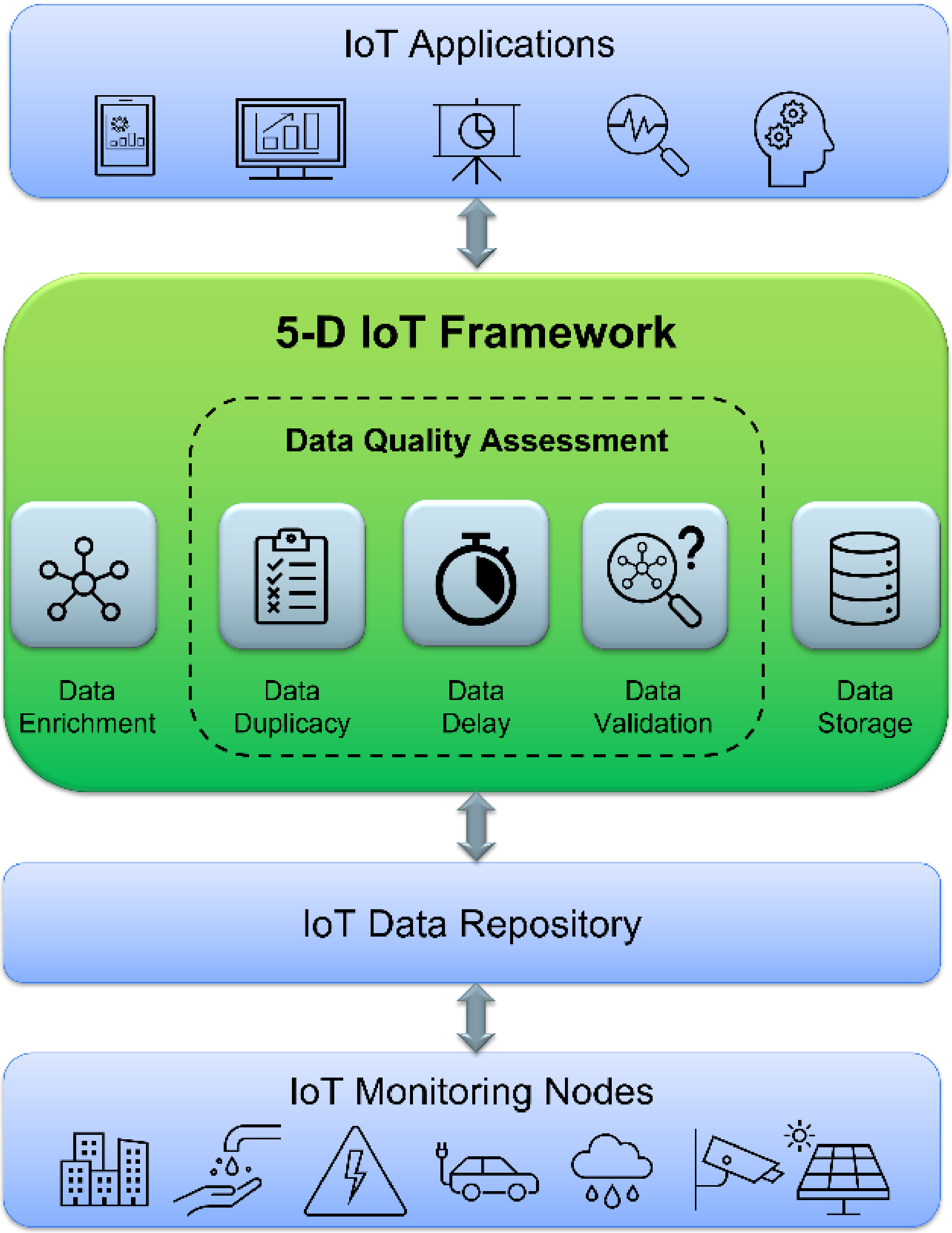

Figure 5.1: Illustration of the five-layered 5-D IoT framework that assesses heterogeneous IoT data and provides uniform description of the assessment results.

## 5.2 Factors affecting IoT data quality

Devices in the IoT system generate a tremendous amount of data that can be used further for scientific or industrial analysis. However, due to the factors such as the resource-constrained nature of devices, network failures, security vulnerabilities, vandalism, faulty devices, and harsh weather, the quality of data in the IoT context is hugely affected. Models built on such data to gain some valuable insights give incorrect results and affect the entire analysis process.

Various researchers have published studies in the past that discuss and categorize factors affecting data quality. In one such study by [88], authors discussed the factors that affect data quality at the physical, network, and application layer and then provided a description of data quality enhancement techniques proposed by other researchers while focusing primarily on data cleaning techniques. Although the quality of data in the IoT context depends on the intended use-case, there are some attributes that are common to all use-cases. Therefore, in [89], authors discussed user-defined attributes and classified them into four main categories, which are internal, accessibility, context, and representation. Based on these studies, the factors affecting quality are summarized and classified into the following five categories:

- **Completeness:** Lack of context or semantics, such as the unit of observation, datatype, timestamp, device type, and device name, can lead to misinterpretation of data. E.g., A correct analysis of the value received from the temperature sensor is impossible until the user understands whether it is in Celsius, Kelvin, or Fahrenheit. It can be avoided by providing the required semantics.

- **Accuracy:** The term accuracy refers to the correctness of the data. In the case of IoT data, noise, errors, and outliers [90] affect accuracy. Generally, IoT data collected from a sensor follows a specific pattern, but sometimes harsh weather affects sensors' working, resulting in erroneous observation values that differ from that pattern contribute to the noise in data. Such deviated observed values from the expected range of values are considered as outliers.

- **Timeliness:** Real-time applications require data generated by the IoT system with very minimal delay. For example, if the data obtained by a doctor using a remote health monitoring system is delayed by a few minutes, the patient's life may be endangered. Therefore, processing delays should be minimized for faster transmission of new or latest data.

- **Trustworthiness:** Security attacks such as Man in the Middle Attack cause communication

infiltration to alter data transmission from sensing nodes and data storage servers. Therefore, a reliable communication channel with a secure and correct information flow is essential to prevent data loss or false data storage and maintain data quality.

- **Confidence:** It is another crucial aspect of IoT data quality. The data provided by the IoT system can be evaluated based on its completeness, reliability, timeliness, and accuracy. Such an assessment will help the user to use the data according to the quality of the data received.

## 5.3 5D-IoT Framework

As described in the section 4.1, it is essential to provide explicit semantics with quality assessment information to facilitate the use of data. And in this section, the proposed framework, its implementation and obtained results by evaluating real-time data are discussed in detail.

### 5.3.1 The Proposed Methodology

This section provides a detailed description of the proposed 5D-IoT framework.

#### 5.3.1.1 Data Enrichment:

In the first layer of the framework, i.e., the data enrichment layer, the SOSA-SSN ontology [91–93] recommended by W3C has been used as the primary ontology to describe the sensor data. In the example illustrated in Figure 4.11, a description of a temperature observation is shown. The observation description describes the observed property (sosa:observedproperty), the entity whose property is being observed (sosa:FeatureOfInterest), observed value (sosa:hasResult), its datatype, unit (qudt-unit-1-1:unit), and time of observation (sosa:resultTime). For this description, a template is used that maps all new observations to their associated semantics and converts them into an RDF graph. The structure of the template can be modified by the user based on the application's needs while configuring the framework. The sensor description describes the sensor that made the observation (sosa:madeBySensor), its normal operating range (ssn-system: hasOperatingRange) and capabilities (ssn-system: hasSystemCapability) such as accuracy, frequency, sensitivity, and precision. The sensor description is static information that does not change until the sensor is replaced by a new sensor. For this purpose, a Knowledge Base (KB) is used that stores such static information. The KB is created using a form and is updated every time a new node is added to the system. Later the sensor description is linked with every new observation using URIs as shown in Figure 5.2. In this way, the framework enriches the data and forwards it to the next layer for quality assessment.

#### 5.3.1.2 Data Quality Assessment

In this framework, SHACL shapes use is proposed for the quality assessment of semantically enriched data and a vocabulary to provide a detailed description of assessment results. W3C specifies

two types

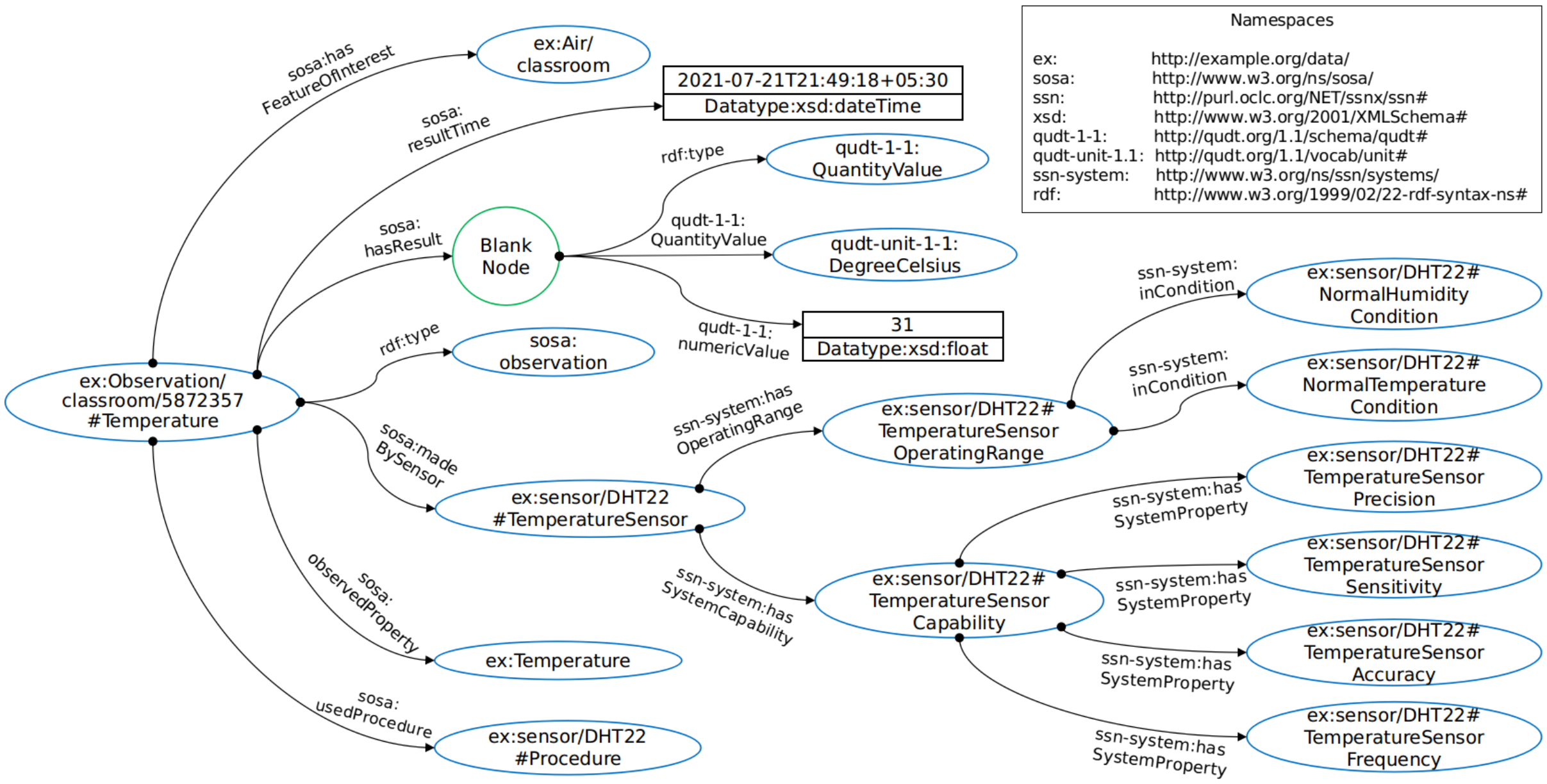

Figure 5.2: RDF graph of a temperature sensor observation created by linking the properties and their associated values to it.

of shapes, a node shape that declares constraints on a specific node and a property shape that declares constraints on the values associated with that node through a path. Collectively these shapes are helpful to validate the data based on the observed property associated with an IoT node.

- **IDQA Vocabulary:** The current data quality ontologies such as DQV [94], and DaQ [95], either focus on assessing the quality of a dataset or focus on a specific domain [96]. These ontologies are helpful to assess the quality of the dataset formed by the 5-D IoT framework, but as the primary goal is to assess each data observation for micro-level quality assessment, any of those ontologies could not be used. Therefore, the IoT Data Quality Assessment (IDQA) vocabulary is proposed which is combined with sosa-ssn ontology's ssn-system: MeasurementRange to enhance the description of the quality assessment results of each observation. The IDQA vocabulary describes two aspects: the factors that should be considered while assessing the quality and the quality assessment result.

  For the first aspect, three classes and six properties are defined. The class idqa: QualityFactor represents an observation quality factor that depends on a given observed property at a given interval of time for a given Feature of Interest (FOI). To represent such information the data property idqa: forInterval and the two object properties, idqa: forFeatureOfInterest and idqa: forObservedProperty are proposed. Moreover, two types of quality factors are defined with the subClasses of idqa: QualityFactor, idqa: RangeValue and idqa: ExpectedDelay. For a idqa: RangeValue factor, two data properties are defined: a minimum expected observation value for the results of such observations that is expressed through the data property idqa: minValue, and a maximum expected observation value expressed through the data property idqa: maxValue. For a idqa: ExpectedDelay factor, the expected delay between two successive observations for a given feature of interest and a given property is defined through the data property idqa: delayValue.

As mentioned before, this representation is complementary to sosa-ssn ontology's ssn-system: MeasurementRange, and considers constraints on value independently on the sensor observing the property. Assessing the quality is thus possible for different observations dealing with the same FOI but produced by different sensors observing different properties.

For the second aspect, which represents the quality assessment result, the object property system:qualityOfObservation from the sosa-ssn ontology is reused, and three classes and the corresponding three data properties are defined. The idqa: numOfDuplicates property of the class idqa: DuplicacyAssessment describes the number of duplicate packets received, the idqa: timeDelay property of the class idqa: DelayAssessment describes the added delay than the expected delay between two successive observations. Finally, the idqa: isOutofRange property of the class idqa: RangeValueAssessment describes whether the observation value is within the idqa: minValue and idqa: maxValue.

These quality types will be added to the observation graph, which will help data users to understand the assessment results.

- **Assessment Techniques:** This section describes the use of shapes and IDQA vocabulary for assessing the data and semantically describing the assessment result, respectively.

  IoT systems receive data from heterogeneous devices deployed in multiple locations that monitor numerous properties, and the values of these properties change over time; hence, it becomes challenging to evaluate the quality of IoT data. For this purpose, a dynamic mechanism is proposed by using the SHACL-SPARQL shapes that assess the data by collectively considering the FOI, observed property, sensor capabilities, and time of observation. Every shape is built using a template designed based on the type of assessment. A shape can be activated or deactivated based on the framework's configuration. It provides the flexibility to the framework administrator to apply various constraints without modifying the complete framework. Once the data is assessed, the assessment results are added to the observation using IDQA vocabulary.

- **Data Duplicacy:**

  As mentioned in [97], one of the most common problems in IoT systems is the transmission of duplicate data, i.e., multiple transmissions of the same measurement value. The node must wake up repeatedly for such multiple transmissions, which results in battery power wastage. The duplicate observations also increase the size of data storage and consume larger bandwidth while retrieving the data from data storage. Discarding such data is also not a desirable choice because by analysing it, a data provider can understand the node’s functionality and reduce such duplicate transmission in the future. Hence, there must be a way to keep the information without storing duplicate observations so that, without any additional processing, the user can get meaningful information while preserving the bandwidth as well as data storage.

Duplicate data can be identified by comparing the timestamps of two successive non-duplicate observations. Therefore, in the proposed framework, a shape using the SHACL-SPARQL constraint is applied to every observation. The shape validates if the timestamp of the newly received observation to new is less than or equal to that of the previous non-duplicate observation to last from the same FOI and observed property. If yes, it means the observation has already been received and is a duplicate observation; otherwise, it is a non-duplicate observation. Later, based on the validation result, an inferred triple will be added to the observation graph using the CONSTRUCT- based SHACL advanced rules. So, if it is a non-duplicate observation, then 0 is inferred and added as an object value to the data property, idqa: numOfDuplicates otherwise, n, the number of times observation is received will be inferred and added. For example, if the same observation is received three times, then three will be inferred and added. The non-duplicate observation graph along with inferred triple is forwarded to the next layer for further quality assessment, and only the inferred triple of duplicate observation is forwarded and stored in the triple store. The reason behind this is to avoid the repetitive processing of the same information. The flow chart illustrating this process is shown in Figure 5.3.

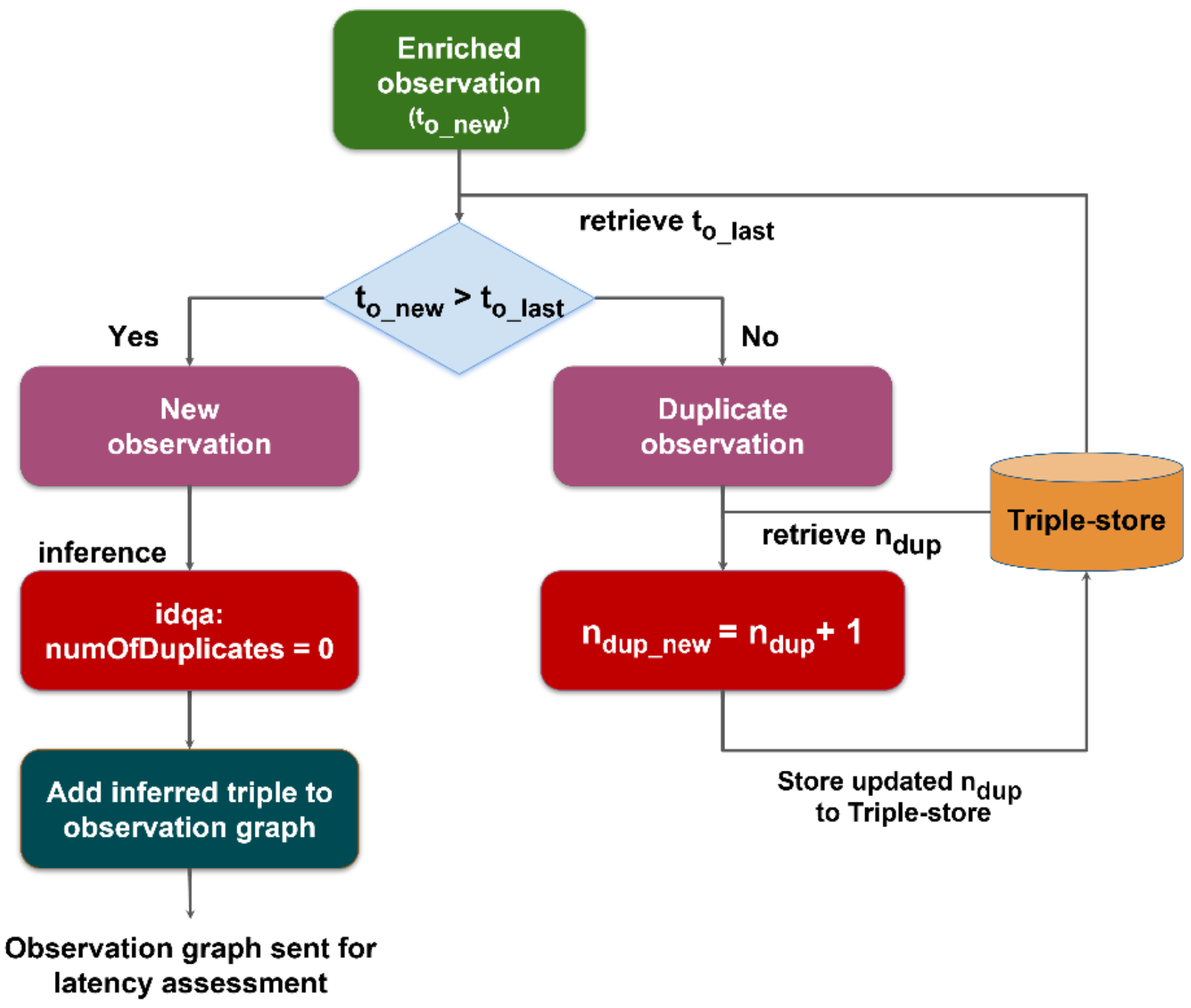

Figure 5.3: Flowchart of the process of duplicate observation assessment and inferred triple addition.

- **Data Delay:** Data collected from IoT nodes is a time-series data, where data can be either collected at a specific frequency defined by the data provider or when there is a change in the observed parameters, for example, an application like a theft detection where the data is transmitted only when someone enters a locked house. However, sometimes, network failures and hardware malfunctions can cause the node to send delayed data affecting the performance of applications

  that receive such delayed data. Hence, it becomes crucial to assess the latency to take necessary action to minimize it. However, as the system grows, its complexity increases, making it difficult for the data provider to assess the latency of each IoT node. Therefore a novel approach is proposed for evaluating delayed observations in which the delay is assessed using SHACL-SPARQL constraint, and a triple will be inferred that describes the added delay using a data property.

  In the first step, divide the delay problem is divided into two parts, transmission delay, and sampling delay. The transmission delay is caused during the transmission of data from node to the IoT platform, and sampling delay is caused during the collection of data by an IoT device (sosa: Sampler). The transmission delay is calculated by taking a difference of observation_time to new and the time when the observation is recorded in the data repository platform tr. To calculate the sampling delay, first, the timestamp of the last non-duplicate observation to last is queried from the RDF triple-store. Using the FILTER SPARQL query of SHACL-SPARQL constraint, that timestamp is compared with the latest observation's timestamp to new. If the difference between the timestamps exceeds the idqa: delayValue (T) stored for the instance of the class idqa: expectedDelay linked to the corresponding feature of interest and observed property, it means the received observation is delayed. Later, triples that contain the delay values are inferred and added to the observation graph using the data properties, idqa: transmissionDelay and idqa: timeDelay. The flow chart illustrating this process is shown in Figure 5.4.

  The reason for choosing a non-duplicate observation during sampling delay calculation is that the timestamp of any duplicate observation will be smaller than the lastly received non-duplicate observation. Therefore, the difference between the latest and any duplicate observation timestamp will always exceed the idqa: expectedDelay, resulting in inaccurate latency assessment.

- **Data Validation:** As mentioned before, noise, errors in measurements, faulty devices, or outliers affect data accuracy; hence, it becomes necessary to assess such factors as they might lead to inaccurate decision-making results. Devices in IoT systems observe various properties whose values differ over several factors like time, location, observed property, the feature of interest, device capabilities such as accuracy and precision. For example, on August 13, 2021, the daytime temperature in Gachibowli, Hyderabad, India, varies from 23°C to 30°C weather.com. If the sensor that is used to measure the temperature has an accuracy of +-2 °C, then due to the accuracy range, the measured temperature variation might be in between 21°C to 32°C. A faulty sensor may transmit inaccurate observation that is out of this range which may affect further analysis based on the inaccurate value. A user that is not aware of the valid range for August 13, 2021, for this observation might not understand that the observed value is inaccurate until such information is explicitly provided with the observation. Another example could be a real-time energy usage monitoring application for electric pumps used daily to fill water tanks in many parts of the city. Based on the energy consumption statistics of the pumps, it is possible to evaluate their health and fault prediction using machine learning algorithms. However, due to defects in the pump, if the data collected contains a large number of outliers, and if not correctly identified, it can lead to incorrect health assessments and erroneous estimates of faults. The problem is exacerbated when different vendors manufacture and deploy pumps, for which they use different data collection techniques. All this makes it difficult for developers to every time filter the data from multiple data sources, which ultimately delays the analysis process.

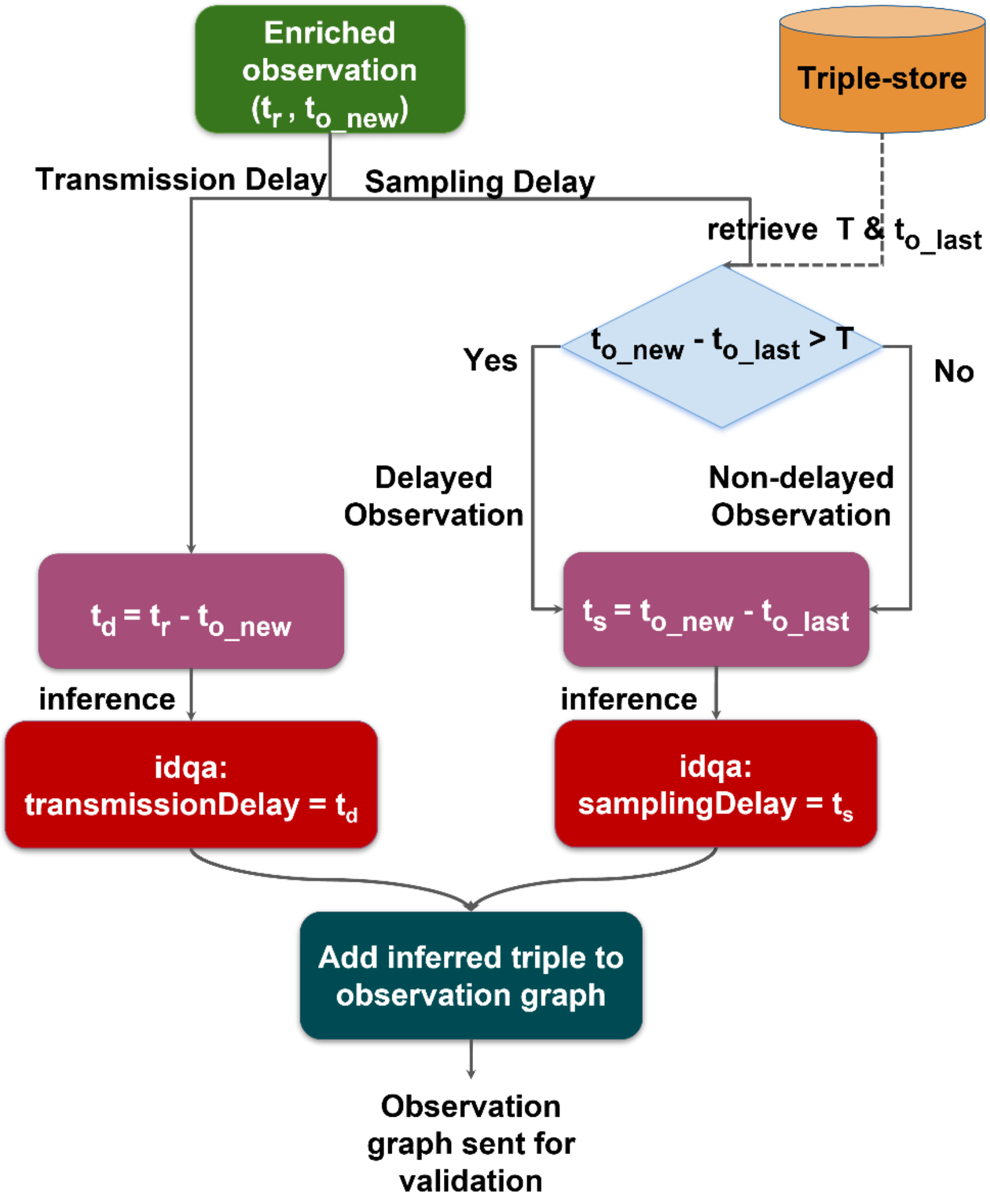

Figure 5.4: Flowchart illustrating the process of transmission and sampling delay assessment and inferred triple creation and addition.

Therefore, there arises a need for a technique that considers all the factors discussed above when assessing data accuracy and adds that information to the data for better understanding. For this purpose, a novel approach is proposed to use the SHACL shapes, which apply dynamic range value constraints on each observation for a more comprehensive accuracy assessment. When a new, semantically enriched, non-duplicate, delayed or non-delayed observation is received at the data validation layer; first, its FOI and observed property would be checked, i.e., in the above examples, the FOIs are an outdoor area and pump house and observed properties are temperature and energy consumption, respectively. This information would be provided by the data provider using the idqa vocabulary's object properties idqa: forFeatureOfInterest and idqa: forObservedProperty respectively. Then in the next step, the minimum Vmin and maximum Vmax values for an interval would be provided using the data property idqa: forInterval and the subClass idqa: RangeValue respectively. All these values will be used while applying the range value constraint to validate the outliers. So, if the observation value Vobservation is null or out of the allowed range, it is declared as inaccurate data, and an inferred triple is added to the observation graph using the data property idqa: isOutofRange, with value as TRUE. The flow chart illustrating this process is shown in Figure 5.5.

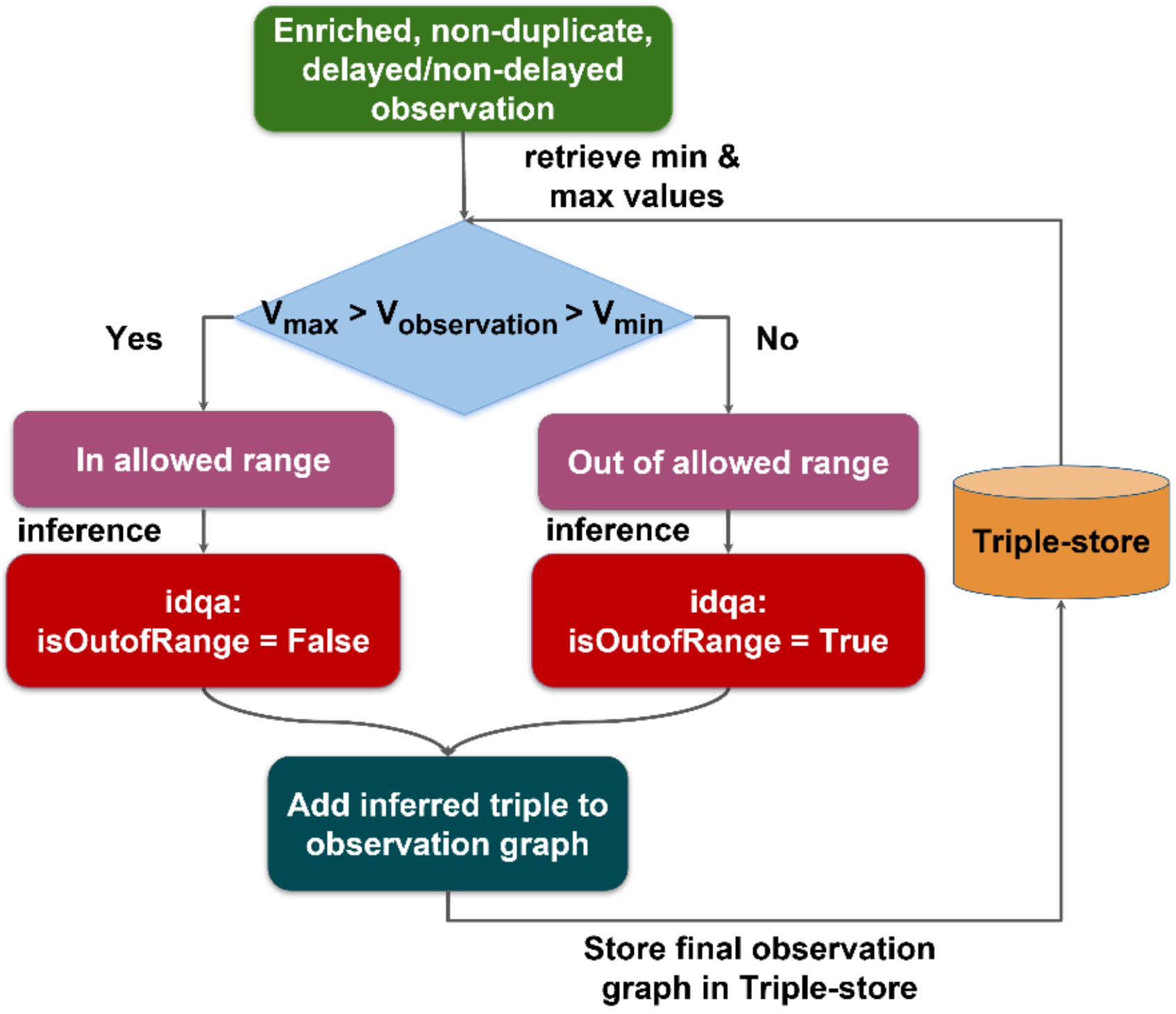

Figure 5.5: Flowchart illustrating the process of data validation and inferred triple creation and addition.

#### 5.3.1.3 Data Storage

The fifth and final layer consists of an RDF triple-store capable of storing the completely processed data. This stored data will benefit various applications such as quality assessed data archiving, descriptive data visualization, correct data analysis and many more.

### 5.3.2 Evaluation

This section provides evaluation results for the framework with the help of implementation by using real-time heterogeneous data from an IoT data repository.

### 5.3.3 Implementation

In the form of implementation, it is described that how quality assessment data can be provided using the proposed framework.

The proposed framework is implemented using Django. The REST APIs provided by the Django REST framework enable the data transmission from repositories to the framework and framework to end-users. During the implementation, apart from built-in libraries of Python, other libraries are used, such as RDFLib [98] to work with RDF in Python and pySHACL [99] to validate RDF data graphs using SHACL shapes. It is assumed that the data packet received by the platform would be a JSON object containing a set of key-value pairs with or without any semantics. The Apache Jena Fuseki [100] triple store is used to store the RDF data that stores the information of devices used and processed observations graphs.

In the first step, sensor graphs, FOIs, expected range value constraints, the expected delay between two successive observations, and templates for shapes are loaded and stored. Then, once the framework receives a new observation, its values are parsed and enriched by linking them with associated properties as shown in Figure 5.2 and converted into an RDF graph using the RDFLib library. Every new observation gets a unique resource identifier (URI) that helps while adding the inferred triples to the observation graph. The RDF graph is then validated against SHACL shapes using the pySHACL module. Based on the validation result, the data properties for sosa: Observation will be added to the observation graph to describe the quality assessment of an observation. Finally, the completely processed data is stored in Fuseki.

### 5.3.4 Results and Discussion

The results shown below are obtained by evaluating the real-time monitoring data received from the Smart city living lab, IIIT-H repository. The repository collects data from heterogeneous IoT verticals such as air [39], water [41], energy [40], and weather monitoring. 24-hour data from one node of each of the four IoT verticals is processed to evaluate the 5D-IoT framework. The devices of IoT nodes used for this evaluation transmit data at different frequencies according to application requirements. These devices are also deployed in different working conditions, as shown in Table 5.1. Such diversity increases the complexity for data providers as well as data users to use such heterogeneous data.

Table 5.1: IoT Devices Specifications

| **IoT Data Monitoring Applications** | ***IoT Application Characteristics*** | | | |
|---|---|---|---|---|
| | ***Sampling Period*** | ***Device Used*** | ***Device Location*** | ***No of Obs*** |
| Energy Consumption | 15 minutes | Energy Meter | Pump Room | 96 |
| Water Flow | 3 minutes | ESP8266 | Pump Room | 480 |
| Weather Condition | 1 minute | Weather Station | Rooftop | 1440 |
| Air Quality | 15 seconds | ESP32 | Classroom | 5760 |

- **Duplicacy Assessment:** With the help of 5D-IoT framework, observations are first enriched by linking them to their associated information, as described in section 5.3.1.1. Then during the duplicacy assessment, it is observed that duplicate data transmission for each device is different due to the difference in sampling period and working conditions. For example, in the case of the air quality monitoring (AQM) node, based on the sampling period, ideally, 5760 unique observations should be transmitted in one day from the node, whereas in reality, 3473 unique observations were received. Among those received observations, 1747 received once, 1196 received twice, 283 received thrice, and 247 received four times. Later after analysing further, it was found out that due to the high frequency, sometimes a node does not receive an acknowledgment from the IoT platform repository, and hence it tries to re-transmit the same observation until it receives the acknowledgment. Due to such repetitive transmission, the node misses transmitting new observations, which results in data loss. The count is identified and added to the number of duplicate observations in the observation graph. Similarly, assess the duplicate observation transmissions from 24 hours data of other nodes are assessed, which is illustrated in Figure 5.6.

- **Delay Assessment:** Based on the delay assessment results, it is found that the observations received over 24 hours from the AQM node produces a significant amount of delay, as can be seen in Figure 5.7 and 5.8. On an average, the AQM observation produces a combined transmission and sampling delay of around 30-40 seconds. For weather monitoring nodes, the delay was even higher, as can be seen in Figure 5.8 and similarly for the other two nodes. Such large delays are unacceptable in applications such as fire alarm systems, where the delay of few seconds may result in detrimental life-threatening situations.

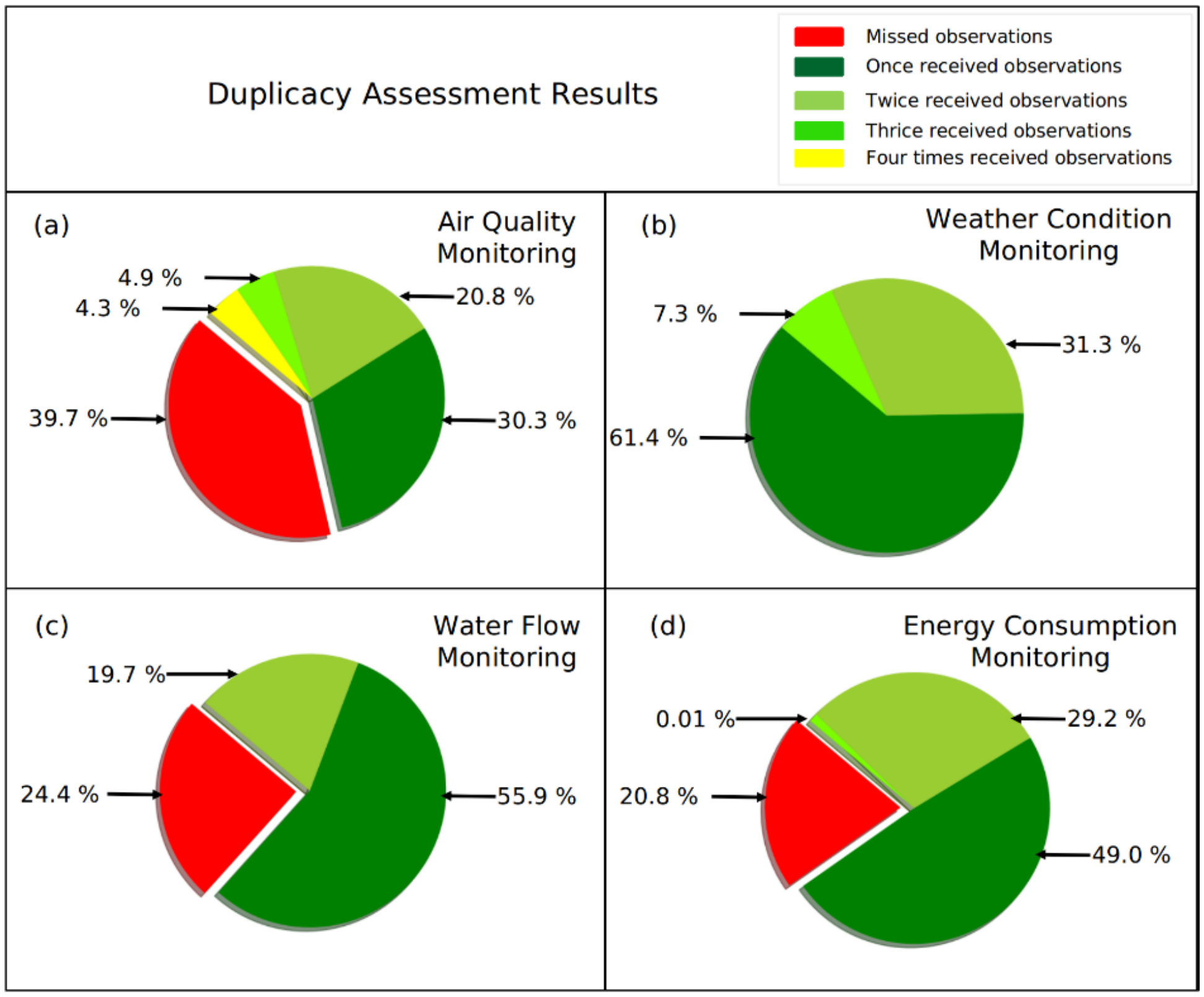

Figure 5.6: Pie charts of observations received from (a) air quality monitoring node (b) weather condition monitoring node (c) water flow monitoring node (d) energy consumption monitoring node.

- **Data Validation:** The assessment results due to the data validation layer helped data providers to assess the hardware failures and provide data users a detailed description of the inaccurate data. It is observed that the cause of inaccurate data transmission is different for every IoT node. For example, the water flow monitoring data provider recognized that ideally, the pressure voltage values should be between 0.03 & 1 Volt, but the voltage losses due to the esp32 boards during data collection result in inaccurate observation values in the range 0.015 to 0.027 Volts. The bar graph in Figure 5.9 illustrates the observed properties by each IoT node and the number of received observations that are in and out of the allowed range.

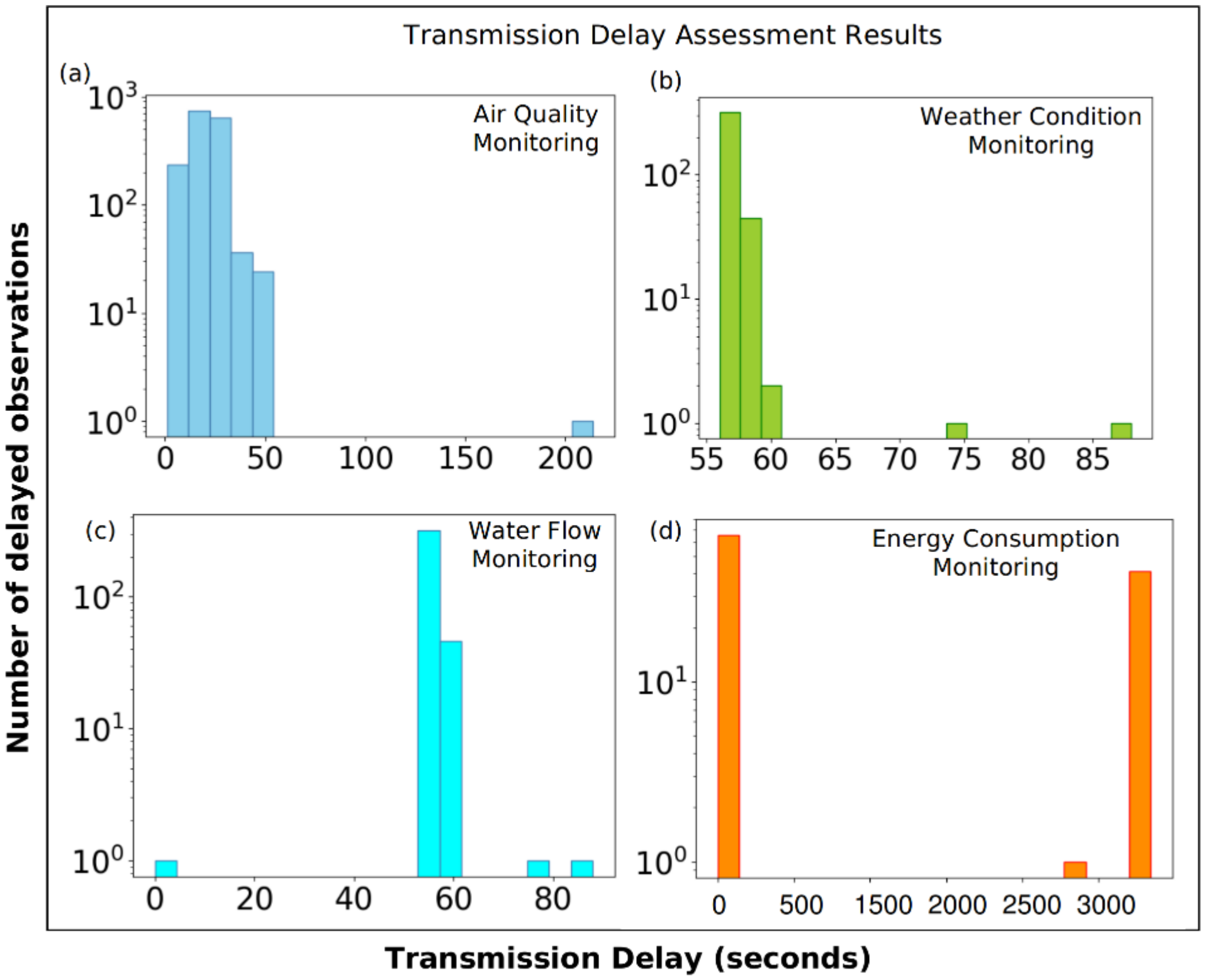

Figure 5.7: Histograms of transmission delays for observations received from (a) air quality monitoring node (b) weather condition monitoring node. (c) water flow monitoring node (d) energy consumption monitoring node.

## 5.4 Summary

Due to the increasing number of IoT devices, a large amount of data is being generated. This data helps analyse daily energy consumption, pollution, weather conditions, health problems, and many other factors. However, factors such as hardware malfunctions, network failures, or cyber-attacks affect data quality and result in inaccurate data generation. It becomes a severe problem in large-scale IoT applications, such as smart cities with numerous devices and different data models, increasing the complexity for developers while developing an application based on data obtained from such systems. Therefore, to facilitate data usage based on the data obtained from such systems. Therefore, to facilitate data usage, a novel 5D-IoT framework is proposed that provides uniform data quality assessment for heterogeneous IoT systems with meaningful data descriptions. Based on the quality assessment result, a data user can directly access data from any IoT source, which ultimately speeds up the analysis process and helps gain important insights in less time. The framework relies on semantic descriptions of sensor observations and SHACL shapes assessing the quality of such data. Evaluations carried out on real-time data show the added value of such a framework.

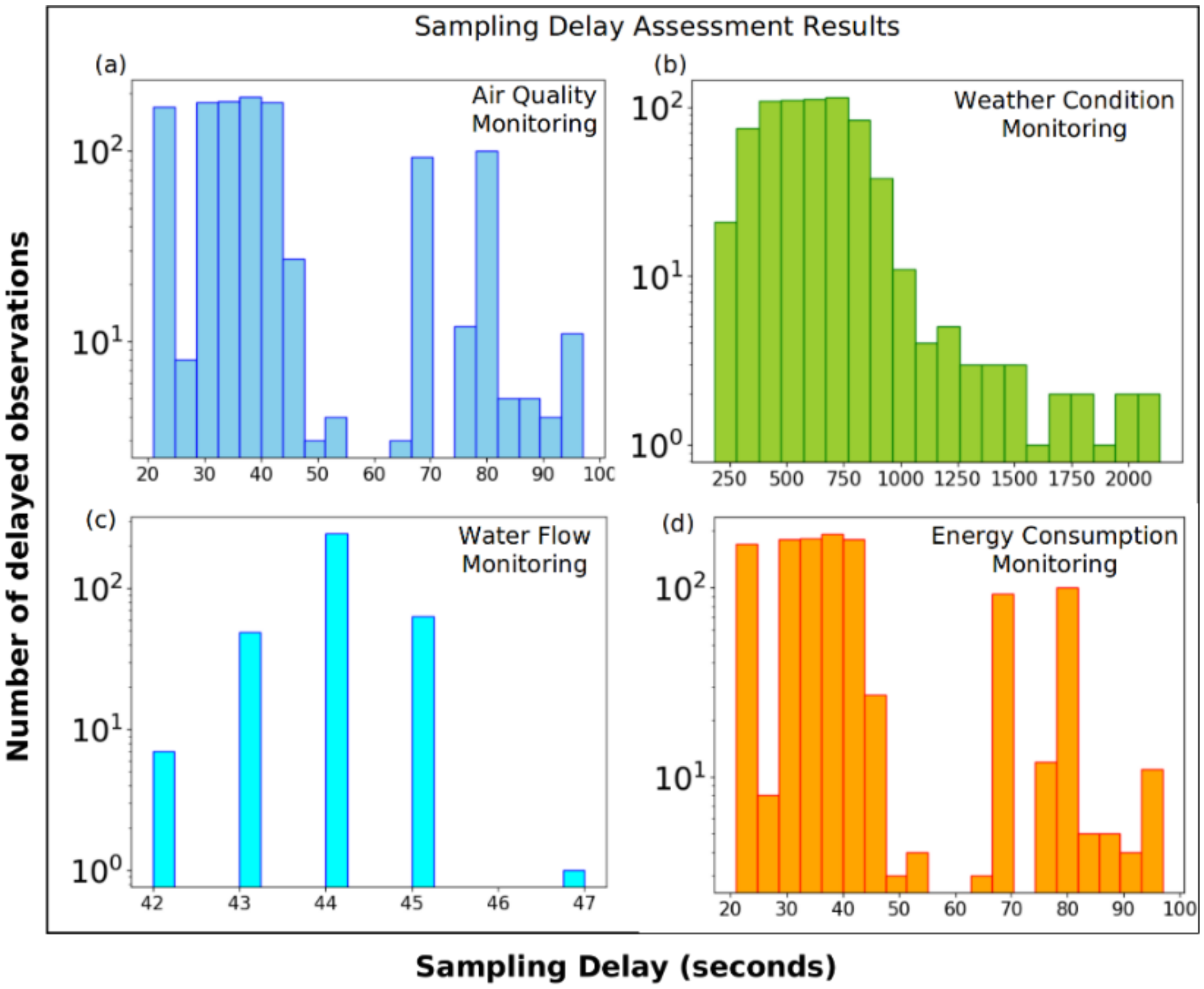

Figure 5.8: Histograms of sampling delays for observations received from (a) air quality monitoring node (b) weather condition monitoring node. (c) water flow monitoring node (d) energy consumption monitoring node.

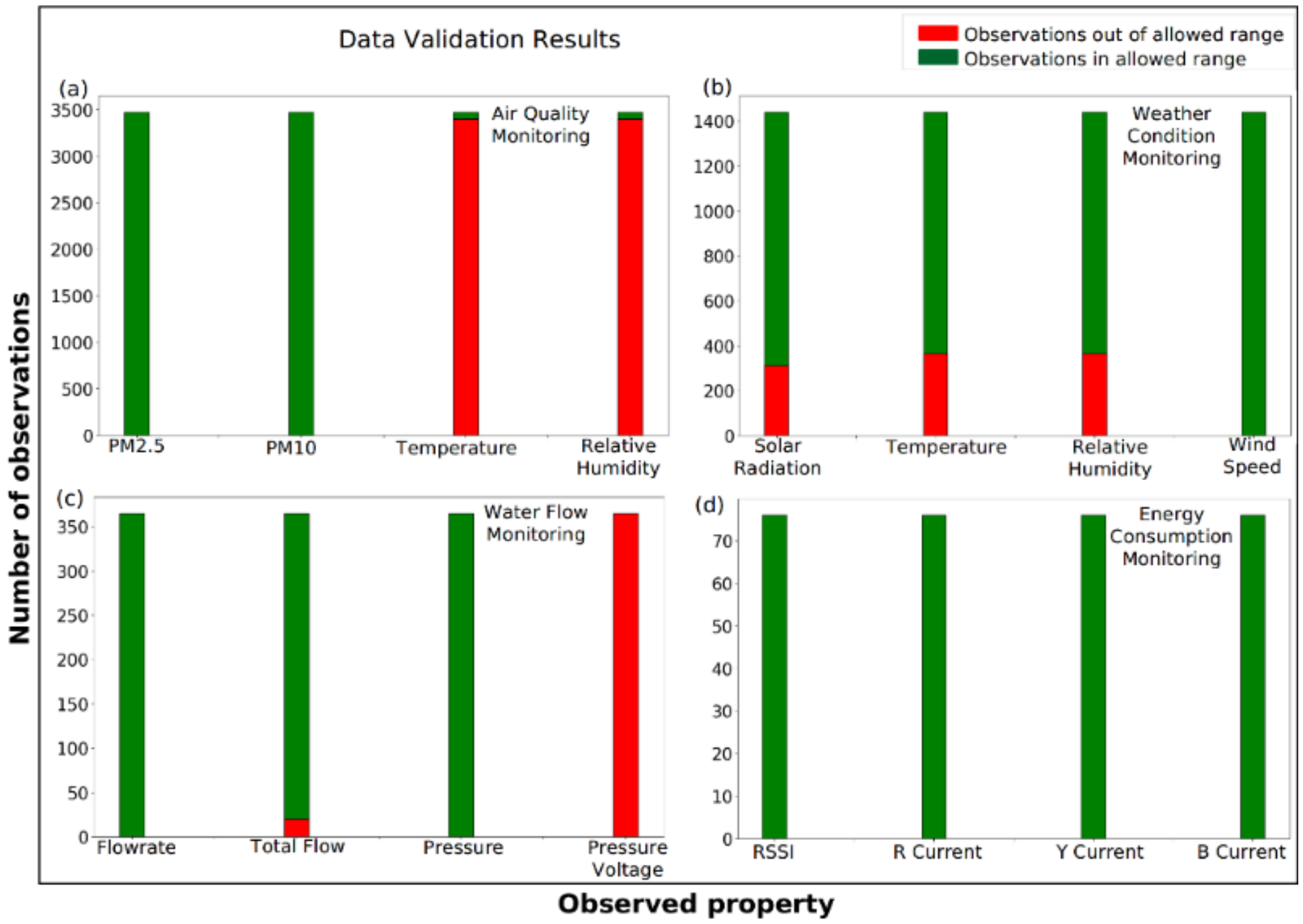

Figure 5.9: Bar graphs of accurate and inaccurate observations received from (a) air quality monitoring node (b) weather condition monitoring node. (c) water flow monitoring node (d) energy consumption monitoring node.

*Chapter 6*

# Conclusion and Future Work

The smart city development consists tasks such as the deployment of sensors to monitor various parameters and the subsequent phases of data collection, storage, secure sharing, visualization and the detailed analysis. Hence, there arises a need of a well-defined system architecture to enable efficient dataflow from the sensing nodes to various applications. Therefore, in this work, a novel system architecture compliant with various international standards is proposed that enables interoperability among various monitoring verticals.

The data collection happens via a common service layer provided by oneM2M standards. The service layer allows the interworking among various monitoring applications enabling IoT interoperability. Currently, the real-time data belongs to 7 different IoT verticals is being monitored using various sensing nodes and communication protocols. Once a data point from a sensing node reaches to the oneM2M platform, it is automatically forwarded to another system for data storage. The data can be then subsequently accessed by authorised data users using the NGSI-LD based data exchange APIs. Finally, the quality of the data is assessed, and the assessment results are added to the data using the 5D-IoT framework to simplify the data understanding process for data users.

As a future work, the schema for the DW can be improved to accomodate the changes in the Smart Campus system. Further, a mechanism can be integrated to allow automatic creation of all the required resources such as CNT, Data CNT, subscription resource and data storage API at DLS for easy addition of newly deployed nodes. Currently, the OM2M platform supports username:password based authentication, hence in future a dynamic authentication mechanism can be integrated to it as described in oneM2M standards. In case of the energy-meter application, an Inter-working Proxy Entity (IPE) can be incorporated instead of a proxy server. In future a layer can be added on top of DES to support data retrieval in other formats apart from APIs, Ex. a Comma Separated Value (CSV) file or Excel sheet for temporal data retrieval. Finally, the added value of data filtering according to assessed quality for data users deployed on the network and automatic generation of SHACL shapes like ASTREA [101] can be integrated to the 5D-IoT framework to eliminate the need for expert knowledge for shapes creation and to use more constraints for assessing the IoT data quality.

## Related Publications

- **S. Mante**, R. Muppala, D. Niteesh, and A. M. Hussain, ”Energy Monitoring Using LoRaWAN-based Smart Meters and oneM2M Platform.” In 2021 IEEE Sensors, pp. 1-4. IEEE, 2021.
- D. Devendra, **S. Mante**, D. Niteesh and A. M. Hussain, ”Electric Vehicle Charging Station using Open Charge Point Protocol (OCPP) and oneM2M Platform for Enhanced Functionality.” In TENCON 2021-2021 IEEE Region 10 Conference (TENCON), pp. 01-05. IEEE, 2021.
- **S. Mante**, N. Hernandez, A. M. Hussain, S. Chaudhari, D. Gangadharan and T. Monteil, ”5D-IoT, a semantic web based framework for assessing IoT data quality.” In Proceedings of the 37th ACM/SIGAPP Symposium on Applied Computing, pp. 1921-1924. 2022.
- **S. Mante**, S. S. S. Vaddhiparthy, R. Muppala, A. M. Hussain, D. Gangadharan and A. Vattem, “A Multi Layer Data Platform Architecture for Smart Cities using oneM2M and IUDX”, Accepted in 2022 IEEE Virtual World Forum on Internet of Things (WF-IoT), 2022.